\documentclass[aps,10pt,prd,reprint,showpacs,showkeys,floatfix,footinbib,superscriptaddress,longbibliography,a4paper]{revtex4-2}
\pdfoutput=1 
\usepackage{bbding}
\usepackage{amsmath}
\usepackage{amssymb}
\usepackage{latexsym}
\usepackage{graphics}
\usepackage{graphicx}
\usepackage{slashed}
\usepackage{color}
\usepackage{comment}
\usepackage{cancel}
\usepackage{subcaption}
\usepackage{multirow}
\usepackage{diagbox}
\usepackage{here}
\usepackage{lipsum}
\usepackage{booktabs}
\usepackage[normalem]{ulem}
\usepackage[disable,colorinlistoftodos,backgroundcolor=white,bordercolor=steelblue,linecolor=steelblue]{todonotes}
\usepackage{bm}
\usepackage[colorlinks=true,citecolor=cyan,urlcolor=blue,bookmarks=true,bookmarksopen=true,bookmarksnumbered=true,bookmarksopenlevel=3]{hyperref}
\pdfoutput=1
\newcommand{\orcid}[1]{\hspace{1mm}\href{https://orcid.org/#1}{\includegraphics[height=0.3cm,keepaspectratio]{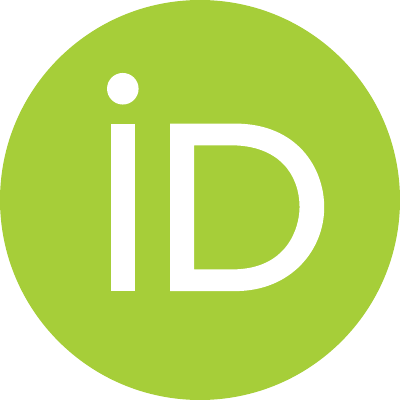}}}
\begin{document}
\title{Probing alternating-parity bands in even–even nuclei within an axial quadrupole-octupole-hexadecapole collective model}
\author{M. Chabab\orcid{0000-0002-2772-4290} }
\email{mchabab@uca.ac.ma}
\affiliation{Cadi Ayyad University (UCAM), FSSM, High Energy Physics, Astrophysics and Geosciences Laboratory, P.O.B 2390, Marrakesh 40000, Morocco.}
\affiliation{Cadi Ayyad University (UCAM), National School of Applied Science, P.O.B. 63 Safi 46000, Morocco.}
\author{A. El Batoul\orcid{0000-0003-4554-029X}}
\email{abdelwahed.elbatoul@uca.ac.ma} 
\affiliation{Cadi Ayyad University (UCAM), FSSM, High Energy Physics, Astrophysics and Geosciences Laboratory, P.O.B 2390, Marrakesh 40000, Morocco.}
\author{L. El Ouaourti}
\email{loubnaouaourti@gmail.com}
\affiliation{Cadi Ayyad University (UCAM), FSSM, High Energy Physics, Astrophysics and Geosciences Laboratory, P.O.B 2390, Marrakesh 40000, Morocco.}
\pacs{} 
\keywords{Collective model; quadrupole-octupole-hexadecapole; alternating parity bands; staggering; sextic potential.}
	\begin{abstract}
	An analytic quadrupole octupole hexadecapole (QOH) axially symmetric model is used to provide a coherent framework for describing both octupole vibrations and octupole deformations in even-even nuclei. The model offers analytical solutions for the energy spectra and electric E1, E2 and E3 transition probabilities. The spectroscopic features are governed by three physically meaningful parameters: the stiffness constant $c_2$, the hexadecapole deformation parameter $\phi_{02}$, and the sextic potential parameter $\alpha$, directly linked to $\phi_{01}^{\pi}$ which represents the interplay between the quadrupole and octupole deformations for both even and odd angular momentum states. The model is applied across a broad range of nuclei including Ra, Th, Sm, Ba, U, Gd, Nd, Rn, Pu, Xe, and Ce isotopes, effectively capturing the characteristic properties of alternating parity bands. Detailed calculations of E1, E2, and E3 transition matrix elements have been performed for $^{220,222}$Rn, $^{222,224,228}$Ra, and $^{144,146}$Ba yielding results in good agreement with available experimental data. 
	\end{abstract}
	
	\maketitle
\section{Introduction}
\par The study of nuclear shapes and collective excitations remains a central theme in modern nuclear structure physics. 
Low-energy properties of even--even nuclei are governed by the interplay of multipole degrees of freedom, dominated 
by quadrupole deformation ($\lambda = 2$), but significantly enriched by octupole ($\lambda = 3$) and hexadecapole 
($\lambda = 4$) components. Quadrupole deformation underlies the well-known rotational spectra of deformed nuclei, 
while octupole correlations give rise to reflection-asymmetric shapes, parity-doublet structures, enhanced E3 strengths, 
and characteristic alternating-parity sequences. Hexadecapole contributions, although less frequently addressed explicitly, 
influence energy staggering, band mixing, and electromagnetic transition properties in several mass regions, particularly 
in actinides and rare earths. Understanding the coupled dynamics of quadrupole, octupole, and hexadecapole modes is therefore 
essential for achieving a unified description of collective spectra and electromagnetic transition systematics across the 
nuclear chart\cite{b1,b2,b3}.
\par Experimental evidence accumulated over the last decade confirms strong octupole collectivity in several isotopic chains. 
Precision Coulomb-excitation measurements, most notably on neutron-rich Ba isotopes, have revealed enhanced octupole transition 
strengths, providing direct evidence for static or near-static octupole deformation in nuclei such as $^{144}$Ba 
\cite{b4}. Complementary observations in Ra and Th nuclei 
\cite{b2} further support the interpretation that octupole correlations emerge from the interaction between 
opposite-parity orbitals with $\Delta j = \Delta \ell = 3$. These findings motivate the development of theoretical models capable 
of simultaneously describing quadrupole, octupole, and higher-order multipole dynamics.
\par Modern microscopic methods such as HFB, symmetry restoration, and GCM mixing provide detailed insights into shape coexistence and multipole correlations across several isotopic chains. Large-scale HFB/GCM studies describe the evolution of quadrupole–octupole collectivity and parity splitting using Gogny or relativistic energy density functionals\cite{b5}(and references therein). Recent beyond-mean-field work also shows that hexadecapole deformation significantly impacts potential energy surfaces, alternating-parity bands, and transition strengths\cite{b6}.
 \par In addition to quadrupole ($\lambda = 2$) and octupole ($\lambda = 3$) degrees of freedom, increasing experimental and theoretical evidence suggests that hexadecapole ($\lambda = 4$) correlations may play a non-negligible role in nuclear structure, particularly in heavy and transitional nuclei. Recent Coulomb-excitation experiments and lifetime measurements\cite{add1,add2} indicate enhanced higher-order multipole contributions in selected isotopes.  From a microscopic perspective, beyond-mean-field calculations based on energy density functionals show that hexadecapole deformation can influence potential energy surfaces, band mixing, and electromagnetic transition rates\cite{add3}.
 \par It is well established that hexadecapole correlations are associated with low-lying $4^+$ excitations and enhanced $B(E4)$ transition probabilities. Although such observables are not explicitly treated in the present work, the inclusion of the $\lambda = 4$ degree of freedom effectively accounts for higher-order shape fluctuations and their impact on the collective dynamics.
 \par In this framework, the hexadecapole mode is introduced as an additional collective degree of freedom, described by small oscillations around an equilibrium value $\phi_{02}$, and accounting for higher-order shape contributions beyond the quadrupole–octupole dynamics.
\par On the other hand, Bohr-Mottelson Hamiltonian provides a versatile framework for exploring nuclear shape transitions and collective motion\cite{b7}. Of particular importance are the critical-point symmetries (CPS)\cite{b8,b9}, which offer analytic, parameter-independent (up to scale factors) predictions for nuclei at phase transitional points. The E(5) \cite{b8} and X(5) 
\cite{b9} symmetries correspond to the critical points of the spherical-to-$\gamma$-soft and spherical-to-axial quadrupole transitions, respectively. Both are obtained by appropriate separations of variables in the Bohr Hamiltonian, using an infinite square-well potential in the $\beta$ (quadrupole) degree of freedom. Their predictive power has been supported by experimental observations in a variety of transitional nuclei \cite{b10,b11,b12,b13}. The success of E(5) and X(5) has stimulated numerous efforts to extend analytic collective models by incorporating additional  degrees of freedom, particularly octupole deformation\cite{b14,b15,b16,b17,b18}. Also, an axially symmetric quadrupole-octupole model have been developed to  describe alternating-parity bands and yield analytic expressions for energies and electromagnetic transition probabilities under sextic potential\cite{b19}. However, these formulations typically consider only $\lambda = 2$ and $\lambda = 3$ modes, leaving the role of hexadecapole deformation largely unexplored within analytic frameworks.
\par Motivated by the experimental evidence for strong octupole correlations in selected even–even nuclei and by recent microscopic indications that hexadecapole degrees of freedom can noticeably affect structure, this article develops an axially symmetric Quadrupole–Octupole–Hexadecapole collective model for even–even nuclei. The model is designed to :
\begin{itemize}
\item [(i)]incorporate $\lambda=2,3,4$ deformation coordinates within a consistent collective Hamiltonian,
\item [(ii)] admit physically motivated potentials, including analytically tractable limits, that allow fits to energies and B(EL) values, and 
\item [(iii)]provide a framework that can be constrained by experimental matrix elements.
\end{itemize}
\par The structure of this study is as follows: Section I provides a general introduction, while Sections II and III outlines the theoretical framework of the model. Section IV presents and discusses the numerical results. Section V summarizes our conclusions. Finally, Appendix A provides the integrals required for the calculation of the normalization constants, whereas Appendix B details the explicit radial and angular integrals employed in the calculation of electric transition matrix elements.
\section{THEORETICAL FRAMEWORK OF THE MODEL}
The coexistence of quadrupole $\beta_2$, octupole $\beta_3$ and hexadecapole $\beta_4$ axially symmetric deformations within an even-even nucleus, is represented by the following Hamiltonian \cite{b20}:
\begin{widetext} 
	\begin{align}
		H = -\sum_{\lambda=2,3,4} {\hbar^2 \over 2 B_\lambda}{1\over \beta_\lambda^3}{\partial \over \partial \beta_\lambda} \beta_\lambda^3 {\partial \over \partial \beta_\lambda} + {\hbar^2 \hat {L^2} \over 6(B_2 \beta_2^2 + 2 B_3 \beta_3^2) + 20 B_4 \beta_4^2 } + V(\beta_2,\beta_3,\beta_4) 
		\label{Eq1} 
	\end{align}
\end{widetext} 
where $B_{2}$, $B_{3}$ and $B_{4}$ are the mass parameters.\\
The solutions of the Schrödinger equation read\cite{b20}:
\begin{widetext} 
\begin{equation}
	\Phi^{\pm}_L(\beta_2,\beta_3, \beta_4,\theta) = (\beta_2 \beta_3 \beta_4)^{-3/2} 
	\Psi^{\pm}_L(\beta_2,\beta_3,\beta_4) \vert LM_{L}K_{L},\pm\rangle,\label{Eq2}
\end{equation}
\end{widetext} 
in the case of an axially symmetric nucleus where the collective rotation is perpendicular to the intrinsic symmetry $z'$-axis; $K_{L}$=0, the function $\vert LM_{L}K_{L},\pm\rangle$ is expressed as follows\cite{b21} :
\begin{equation}
	\vert LM_{L}0, \pm\rangle = \sqrt{ 2L+1 \over 32 \pi^2} (1 \pm (-1)^L ) {\cal D}^L_{0,M_{L}}(\theta), 
	\label{Eq3}
\end{equation}  
$\theta$ are the three Euler angles, ${\cal D}_{L}(\theta)$ is the Wigner function of rotation. The positive and negative signs correspond respectively to symmetric states (L even) and antisymmetric states (L odd). Using the solutions of Eq. (\ref{Eq2}) for the Hamiltonian in Eq. (\ref{Eq1}), and introducing the reduced deformation parameters\cite{b20} $\tilde \beta_2 = \beta_2 \sqrt{B_2\over B},  \tilde \beta_3 = \beta_3 
\sqrt{B_3\over B},  \tilde \beta_4 = \beta_4 \sqrt{B_4\over B}$ with $B= {B_2+B_3+B_4 \over 3}$, as well as the reduced energy  $\epsilon_{L}=(2B/\hbar^2) E_{L}, v=(2B/\hbar^2) V$, the collective Schrödinger equation can be simplified to the form:
\begin{widetext} 
\begin{align}
	\left[-\frac{\partial^2}{\partial \tilde{\beta}_2^2}-\frac{\partial^2}{\partial \tilde{\beta}_3^2}-\frac{\partial^2}{\partial \tilde{\beta}_4^2}   +\frac{ L(L+1)}{3\left(\tilde{\beta}_2^2+2 \tilde{\beta}_3^2\right)+ 10 \tilde \beta_4^2} \right. \left.  +\frac{3}{4}\left(\frac{1}{\tilde{\beta}_2^2}+\frac{1}{\tilde{\beta}_3^2}+\frac{1}{\tilde{\beta}_4^2}\right)   +v\left(\tilde{\beta}_2, \tilde{\beta}_3, \tilde{\beta}_4\right) \right.  \left. -\epsilon_L \right] \Psi_L^{ \pm}\left(\tilde{\beta}_2, \tilde{\beta}_3, \tilde{\beta}_4\right)=0 .\label{Eq4}
\end{align} 
\end{widetext} 
We next introduce the spherical coordinates\cite{b20},
\begin{equation}
	\begin{aligned}
		\tilde\beta_{2} &= \tilde\beta\, \sin\phi_{2}\cos\phi_{1}, \qquad
		\tilde\beta_{3} = \tilde\beta\, \sin\phi_{2}\sin\phi_{1},\\[2mm]
		\tilde\beta_{4} &= \tilde\beta\, \cos\phi_{2}, \qquad
		\tilde\beta = \sqrt{\tilde\beta_{2}^{2} + \tilde\beta_{3}^{2} + \tilde\beta_{4}^{2}}.\label{Eq5}
	\end{aligned}
\end{equation}
where $0 \le \tilde\beta < \infty$ is the radial coordinate, and the angular variables 
$-\pi/2 \le \phi_{1} \le \pi/2$ and $0 \le \phi_{2} < \pi$ specify the relative contributions of the three deformation modes.
These angles highlight the interdependence among the nuclear shape degrees of freedom. In particular,
\begin{itemize}
	\item $\phi_{2} = 0$ corresponds to a purely hexadecapole deformation $(+\beta_{4})$;
	\item $\phi_{2} = \pi/2$ yields either:
	\begin{itemize}
		\item a pure octupole deformation $(\pm \beta_{3})$ when $\phi_{1} = \pm \pi/2$, or
		\item a pure quadrupole deformation $(+\beta_{2})$ when $\phi_{1} = 0$.
	\end{itemize}
\end{itemize}
With these definitions, the Schrödinger equation becomes:
\begin{widetext}
	\begin{align}
		\Bigg[ 
		-\frac{\partial^2}{\partial \tilde{\beta}^2}
		- \frac{2}{\tilde{\beta}} \frac{\partial}{\partial \tilde{\beta}}
		&- \frac{1}{\tilde{\beta}^2} \Bigg\{ 
		\frac{1}{\sin^2\phi_2} \frac{\partial^2}{\partial \phi_1^2}
		+ \frac{\partial^2}{\partial \phi_2^2}
		+ \frac{\cos\phi_2}{\sin\phi_2} \frac{\partial}{\partial \phi_2}
		\Bigg\} + \frac{1}{\tilde{\beta}^2}\frac{L(L+1)}{3\bigl(\sin^2\phi_1 + 1\bigr)\sin^2\phi_2 + 10\cos^2\phi_2}
		+ v(\tilde{\beta},\phi_1,\phi_2) - \epsilon_L \nonumber \\[2mm]
		&\; + \frac{3}{4} \frac{1}{\tilde{\beta}^2} \Bigg\{ 
		\frac{4}{\sin^2 2\phi_1\, \sin^2\phi_2} 
		+ \frac{1}{\cos^2\phi_2} 
		\Bigg\}
		\Bigg] \Psi_L^{\pm}(\tilde{\beta},\phi_1,\phi_2) = 0.\label{Eq6}
	\end{align}
\end{widetext}

The present model assumes an exactly separable form of the potential,
\begin{equation}
	v(\tilde{\beta},\phi_{1},\phi_{2})
	= v(\tilde{\beta})
	+ \frac{v(\phi_{1},\phi_{2})}{\tilde{\beta}^{2}\sin^{2}\phi_{2}},\label{Eq7}
\end{equation}
together with the factorized wave function 
$\Psi_{L}^{\pm}(\tilde{\beta},\phi_{1},\phi_{2})
= \Psi_{L}(\tilde{\beta})\,\Psi_{L}^{\pm}(\phi_{1},\phi_{2})$.
Considering small oscillations around the equilibrium values,
where $\phi_{1}$ remains close to $\pm \phi_{01}$ and
$\phi_{2}$ remains near $\phi_{02}$, the deformation ratios may be
treated as effectively constant. Under these assumptions, the
Schrödinger equation (\ref{Eq6}) becomes exactly separable, yielding:
\begin{equation}
	\left[
	-\frac{\partial^{2}}{\partial\tilde{\beta}^{2}}
	-\frac{2}{\tilde{\beta}}\frac{\partial}{\partial\tilde{\beta}}
	+ v(\tilde{\beta})
	- \epsilon_{L}
	+ \frac{\lambda_{L}}{\tilde{\beta}^{2}\sin^{2}\phi_{02}}
	\right]
	\Psi_{L}(\tilde{\beta}) = 0,\label{Eq8}
\end{equation}
and
\begin{widetext}
\begin{align}
	\Bigg[
	-\frac{\partial^{2}}{\partial \phi_{1}^{2}}
	- \sin^{2}\phi_{02}\,\frac{\partial^{2}}{\partial \phi_{2}^{2}}
	- \frac{\sin 2\phi_{02}}{2}\,\frac{\partial}{\partial \phi_{2}}
	+ \frac{3}{\sin^{2} 2\phi_{01}}
	+ \frac{3}{4\cot^{2}\phi_{02}}
	+ v(\phi_{1},\phi_{2})
	- \lambda_{L}  \nonumber\\
	\qquad
	+ \frac{L(L+1)}{3(\sin^{2}\phi_{01}+1) + 10\cot^{2}\phi_{02}}
	\Bigg]
	\Psi_{L}^{\pm}(\phi_{1},\phi_{2}) = 0.\label{Eq9}
\end{align}
\end{widetext}
Here the angular potential is written as
$v(\phi_{1},\phi_{2}) = v(\phi_{1}) + v(\phi_{2})$, and the separation
constant is $\lambda_{L} = \lambda_{L,\phi_{1}} + \lambda_{\phi_{2}}$.
Moreover, the angular wave function factorizes as
$\Psi_{L}^{\pm}(\phi_{1},\phi_{2}) = \chi_{L}(\phi_{1})\,\chi(\phi_{2})$,
allowing the angular equation (\ref{Eq9}) to be approximately separated into two independent equations:
\begin{widetext}
	\begin{align}
		&\Bigg[
		- \sin^{2}\phi_{02}\, \frac{\partial^{2}}{\partial \phi_{2}^{2}}
		- \frac{\sin 2\phi_{02}}{2}\, \frac{\partial}{\partial \phi_{2}}
		+ \frac{3}{\sin^{2} 2\phi_{01}}
		+ \frac{3}{4\cot^{2}\phi_{02}}
		+ v(\phi_{2})
		- \lambda_{\phi_{2}}
		\Bigg] \chi(\phi_{2}) = 0,\label{Eq10} \\[1em]
		&\Bigg[
		- \frac{\partial^{2}}{\partial \phi_{1}^{2}}
		+ \frac{L(L+1)}{3(\sin^{2}\phi_{01}+1) + 10 \cot^{2}\phi_{02}}
		+ v(\phi_{1})
		- \lambda_{L,\phi_{1}}
		\Bigg] \chi_{L}(\phi_{1}) = 0. \label{Eq11}
	\end{align}
\end{widetext}

\section{Analytical determination of the energy spectra}
In this section, we present the analytical determination of the collective energy spectra within the adopted separation scheme. Since the differential equations associated with the angular variables $\phi_{2}$ and $\phi_{1}$ have already been derived in the previous section, we only recall here the forms of the potentials employed for each collective coordinate. We then summarize the corresponding eigenvalues and eigenfunctions and combine the partial solutions to obtain the full expression of the total energy as well as the associated collective wave function, which is then employed to calculate the electromagnetic $B(EL)$ transition rates.
\subsection{The $\phi_{2}$ degree of freedom}
For the $\phi_{2}$ variable, we adopt a harmonic oscillator potential centered at the equilibrium value $\phi_{02}$, namely
\begin{equation}
	v(\phi_{2})=\frac{1}{2}c_{2}\,(\phi_{2}-\phi_{02})^{2}
	=\frac{1}{2}c_{2}\,\tilde{\phi}_{2}^{\,2},
	\qquad \tilde{\phi}_{2}=\phi_{2}-\phi_{02}.\label{Eq12}
\end{equation}

With the transformation
\begin{equation}
	\chi(\phi_{2})= \exp\!\left(-\frac{\tilde{\phi}_{2}}{2}\cot\phi_{02}\right)F(\tilde{\phi}_{2}),\label{Eq13}
\end{equation}
the $\phi_{2}$ equation reduces to a one-dimensional harmonic oscillator equation. Consequently, the eigenvalues take the form
$\lambda_{\phi_{2}}^{\prime}
	=\sqrt{2c_{2}\sin^{2}\phi_{02}}\left(n_{\phi_{2}}+\frac{1}{2}\right),\qquad n_{\phi_{2}}=0,1,2,\cdots.
$
This leads to the $\phi_{2}$ contribution to the separation constant
\begin{widetext}
\begin{align}
	\lambda_{\phi_{2}}^{\pi}
	=\sqrt{2c_{2}\sin^{2}\phi_{02}}\left(n_{\phi_{2}}+\frac{1}{2}\right)
	+\frac{3}{\sin^{2}2\phi_{01}^{\pi}}  +\frac{1}{4}\cos^{2}\phi_{02}
	+\frac{3}{4\cot^{2}\phi_{02}}.\label{Eq14}
\end{align}
\end{widetext}
The corresponding eigenfunctions are given by
\begin{equation}
	\chi_{n_{\phi_{2}}}(\phi_{2})
	= N_{n_{\phi_{2}}}\,
	H_{n_{\phi_{2}}}(a\tilde{\phi}_{2})\,
	\exp\!\left(-\frac{a^{2}\tilde{\phi}_{2}^{\,2}}{2}
	-\frac{\tilde{\phi}_{2}}{2}\cot\phi_{02}\right),\label{Eq15}
\end{equation}
where
\begin{equation}
	a=\left(\frac{c_{2}}{2\sin^{2}\phi_{02}}\right)^{1/4},\label{Eq16}
\end{equation}
$H_{n}$ denotes the Hermite polynomial of order $n$, and $N_{n_{\phi_{2}}}$ is a normalization constant.

\subsection{The $\phi_{1}$ degree of freedom}

For the $\phi_{1}$ variable, we employ a sextic potential of the form\cite{b22}:
\begin{equation}
	v(\phi_{1})=(\alpha^{2}-4c)\phi_{1}^{2}
	+2\alpha\,\phi_{1}^{4}
	+\phi_{1}^{6}.\label{Eq17}
\end{equation}
By comparing Eq.(\ref{Eq11}) for the present potential with the corresponding quasi-exactly solvable case discussed in Ref.~\cite{b22}, one obtains the following correspondence:
\begin{equation}
	L(L+1)=\left(2s-\frac{1}{2}\right)\left(2s-\frac{3}{2}\right),\label{Eq18}
\end{equation}
one readily finds\cite{b23}:
\begin{equation}
	s=\frac{L}{2}+\frac{3}{4}.\label{Eq19}
\end{equation}
The parameter $c$ entering the sextic potential then reads
\begin{equation}
	c=s+\frac{1}{2}+M
	=\frac{L}{2}+\frac{5}{4}+M.\label{Eq20}
\end{equation}

It is evident that the potential $v(\phi_{1})$ depends explicitly on the angular momentum $L$ through the parameter $c$. In order to ensure an $L$-independent potential, $c$ must be fixed to a constant value, leading to the constraint
\begin{equation}
	c=\frac{L}{2}+\frac{5}{4}+M=\text{const.},
	\qquad M=0,1,2,\ldots \label{Eq21}
\end{equation}
To solve the sextic potential equation for both positive and negative parity levels,
we introduce two distinct constants defining two families of angular-momentum states.
For even values of $L$, the allowed pairs
\[
(M,L)=(K,0),(K-1,2),(K-2,4),\ldots
\]
lead to
\begin{equation}
	c^{+}=K+\frac{5}{4}. \label{Eq22}
\end{equation}
For odd values of $L$, the sequence
\[
(M,L)=(K,1),(K-1,3),(K-2,5),\ldots
\]
is obtained, yielding
\begin{equation}
	c^{-}=K+\frac{7}{4}. \label{Eq23}
\end{equation}
Here, $K=M_{\mathrm{max}}$ is a fixed integer. Therefore, when $M$ increases by one unit, the angular momentum $L$ must decrease by two units in order to keep $K$ constant.
The condition of a \textit{constant potential} is ensured by supplementing $v(\phi_{1})$ with two parity-dependent constants $u_{0}^{\pi}(\alpha)$. As a result, the potential assumes the following general form :
\begin{equation}
	v^{\pi}(\phi_{1})
	=(\alpha^{2}-4c^{\pi})\phi_{1}^{2}
	+2\alpha\,\phi_{1}^{4}
	+\phi_{1}^{6}
	+u_{0}^{\pi}(\alpha),\ \pi=\pm.  \label{Eq24}
\end{equation}

To simplify the notation and to remain consistent with Refs.~\cite{b19,b23,b24}, we introduce the identification
\begin{equation}
	\phi_{1} \equiv y .\label{Eq25}
\end{equation}
The constants $u_{0}^{\pi}(\alpha)$ are fixed by requiring that the deformed minima 
$(\phi_{01}^{\pi} = y_{0}^{\pi} > 0)$ of the two parity-dependent potentials have the same energy. 
For $u_{0}^{+}=0$, one finds
\begin{widetext}
\begin{align}
	u_{0}^{-} ={} \left(\alpha^2 - 4 c^{+}\right) (y_{0}^{+})^2 
	- \left(\alpha^2 - 4 c^{-}\right) (y_{0}^{-})^2 
	+ 2\alpha \left[(y_{0}^{+})^4 - (y_{0}^{-})^4 \right]
	+ \left[(y_{0}^{+})^6 - (y_{0}^{-})^6 \right]. \label{Eq26}
\end{align}
\end{widetext}
The positions of the deformed minima are given by
\begin{equation}
	\left(\phi_{01}^{\pi}\right)^2
	= \left(y_{0}^{\pi}\right)^2
	= \frac{1}{3}\left[-2\alpha + \sqrt{\alpha^2 + 12c^{\pi}}\right],
	\ \pi=\pm ,\label{Eq27}
\end{equation}
These considerations allow one to construct a solvable sextic potential describing axially symmetric quadrupole-octupole-hexadecapole nuclei. Following Ref.~\cite{b22}, we adopt the ansatz
\begin{equation}
	\chi_L^{\;(M)}(y)
	\sim P^{(M)}\!\left(y^2\right)
	\, y^{2s-\frac{1}{2}}
	\exp\!\left(-\frac{y^4}{4}-\frac{\alpha y^2}{2}\right),\label{Eq28} 
\end{equation}

where $P^{(M)}(y^2)$ denotes a polynomial of degree $M$ in $y^2$.
Substituting Eq.~(\ref{Eq28}) into Eq.~(\ref{Eq11}), we obtain a quasi-exactly solvable differential equation :
\begin{align}
	\Bigg[
	-\left(\frac{d^2}{dy^2}
	+ \frac{4s-1}{y}\frac{d}{dy}\right)
	&+ 2\alpha y\frac{d}{dy}  
	+ 2y^2\left(y\frac{d}{dy}-2M\right) \nonumber
	\\ &\Bigg]
	P^{(M)}\!\left(y^2\right)
	= \lambda^{M} P^{(M)}\!\left(y^2\right).
	\label{Eq29}
\end{align}
The eigenvalue $\lambda^{M}$ is expressed as
\begin{widetext}
\begin{align}
	\lambda_{n_{\phi_1},L}^{M} ={}
	\lambda_{L,\phi_1}^{\pi}
	- 4\alpha s
	- u_0^{K}(\alpha)
	+ \frac{(4s-1)(4s-3)}{4\,\langle y^2\rangle_{ML}} 
	- \frac{s(4s-3)-\tfrac{1}{2}}
	{3(\sin^2\phi_{01}^{\pi}+1)+10\cot^2\phi_{02}} .\label{Eq30}
\end{align}
\end{widetext}
Here, $\langle y^2\rangle_{ML}$ denotes the expectation value of $y^2$ evaluated with the wave functions $\chi_L^{\;(M)}(y)$ given by Eq.~(\ref{Eq28}). 
The eigenvalues $\lambda^{M}_{n_{\phi_1}}$ are obtained by diagonalizing the matrix representation of Eq.~(\ref{Eq29}) for each value of $M$, following the analytical procedure described in the Appendix of Ref.~\cite{b25}. 
For a given $M$, there are $M+1$ solutions characterized by the quantum number $n_{\phi_1}$, with $n_{\phi_1}=0$ corresponding to the lowest eigenvalue and $n_{\phi_1}=M$ to the highest one. The $\phi_{1}$-dependent contribution to the separation constant is therefore written as
\begin{widetext}
\begin{align}
	\lambda_{L,\phi_1}^{\pi}
	\left(\phi_{01}^{\pi},\phi_{02}\right)
	={} \lambda_{n_{\phi_1},L}^{M}
	+ 4\alpha s
	+ u_0^{K}(\alpha)
	- \frac{(4s-1)(4s-3)}{4\,\langle y^2\rangle_{ML}} 
	+ \frac{s(4s-3)-\tfrac{1}{2}}
	{3(\sin^2\phi_{01}^{\pi}+1)+10\cot^2\phi_{02}} .\label{Eq31}
\end{align}
\end{widetext}
Finally, the normalized eigenfunctions take the form
\begin{equation}
	\chi_{n_{\phi_1}}^{\;(M_s)}(\phi_1)
	= N_{n_{\phi_1}}^{(M_s)}
	\, P^{(M_s)}\!\left(\phi_1^2\right)
	\, \phi_1^{2s-\frac{1}{2}}
	\exp\!\left(-\frac{\phi_1^4}{4}-\frac{\alpha \phi_1^2}{2}\right).\label{Eq32}
\end{equation}
\subsection{Radial equation and total energy}

For the radial variable $\tilde{\beta}$, we adopt the Davidson potential
\begin{equation}
	v(\tilde{\beta})=\tilde{\beta}^{2}+\frac{\beta_{0}^{4}}{\tilde{\beta}^{2}},\label{Eq33}
\end{equation}
which leads to exact solutions in terms of Laguerre polynomials. The energy eigenvalues are given by
\begin{equation}
	\epsilon_{n_{\tilde{\beta}},L}
	=2n_{\tilde{\beta}}+1
	+\sqrt{\frac{\lambda_{L,\phi_{1}}^{\pi}
			+\lambda_{\phi_{2}}^{\pi}}{\sin^{2}\phi_{02}}
		+\frac{1}{4}+\beta_{0}^{4}}.\label{Eq34}
\end{equation}

In the present work, we restrict our analysis to bands characterized by
$$
	(n_{\tilde{\beta}},n_{\phi_{1}},n_{\phi_{2}})=(n,0,0),
$$
and to the case $\beta_{0}=0$. In fact, calculation tests for several selected nuclei show that $\beta_0$ has a negligible effect on the results, and it is therefore set to zero, reducing the Davidson potential (\ref{Eq33}) to a harmonic oscillator, without loss of accuracy. The total energy then reads
\begin{equation}
	E(n,c_{2},\phi_{02},\phi_{01}^{\pi})
	=\frac{\hbar^{2}}{2B}
	\left[
	2n+1
	+\sqrt{\frac{\lambda_{L,\phi_{1}}^{\pi}
			+\lambda_{\phi_{2}}^{\pi}}{\sin^{2}\phi_{02}}
		+\frac{1}{4}}
	\right],\label{Eq35}
\end{equation}
and the complete collective wave function is finally written as
\begin{align}
	\Phi_{L,n}^{\pi}
	(\tilde{\beta},\phi_{1},\phi_{2},\theta)
	&=(\beta_{2}\beta_{3}\beta_{4})^{-3/2}
	\,N_{\tilde{\beta}}
	\,\tilde{\beta}^{-1/2}
	F_{L,n}^{\pi}(\tilde{\beta}) \notag \\
	&\quad \times \chi_{0}^{(M_{s})}(\phi_{1})
	\,\chi_{0}(\phi_{2})
	\,|LM_{L}0,\pm\rangle,\ \pi=\pm.\label{Eq36}
\end{align}
where
\begin{align}
	F^{\,\pi}_{L,\,n_{\tilde \beta}}(\tilde \beta) = \sqrt{2n_{\tilde \beta}! \over \Gamma( n_{\tilde \beta}+a^{\pi}_L+1)} \tilde\beta^{a^\pi_L} 
	L_{n_{\tilde \beta}}^{a^{\pi}_L}(\tilde \beta^2) e^{-\tilde \beta^2/2}, \label{Eq37} \\  with \qquad \quad a^{\pi}_L=\sqrt{\frac{\lambda^{\pi}_{L}}{\sin^2 \phi_{02}}+\frac{1}{4} +\beta_0^4}.\label{Eq38}
\end{align}
\subsection{Determination of $B(EL)$ transition rates}
The reduced electric transition probabilities $B(EL)$ serve as a sensitive indicator of nuclear collectivity and shape dynamics, reflecting the strength of electromagnetic transitions between nuclear states. They carry crucial information about deformation, vibrational motion, and angular momentum coupling. In collective models, $B(EL)$ values result from the interaction between intrinsic collective variables and rotational degrees of freedom, and their comparison with experimental measurements provides a stringent test of nuclear structure models. For axially symmetric nuclei, these transitions are computed using standard electric dipole, quadrupole, and octupole operators, with the latter two expressed as\cite{b20}:
\begin{align}
	&T^{(E2)}_\mu = t_2 \sqrt{\frac{B}{B_2}}\, \tilde{\beta}\, \sin \phi_{2} \cos \phi_{1}\,
	\mathcal{D}^{(2)}_{0,\mu}(\theta), \label{Eq39}\\
	&T^{(E3)}_\mu = t_3 \sqrt{\frac{B}{B_3}}\, \tilde{\beta}\, \sin \phi_{2} \sin \phi_{1}\,
	\mathcal{D}^{(3)}_{0,\mu}(\theta),\label{Eq40}
\end{align}
while the electric dipole operator takes the form \cite{b20}:
\begin{widetext}
\begin{align}
	T^{(E1)}_\mu = t_1 \Bigg[
	a_{2}\,\frac{B}{\sqrt{B_2 B_3}}\,\tilde{\beta}^2 \sin^{2}\phi_{2}\,
	\frac{\sin 2\phi_{1}}{2} 
 + a_{3}\,\frac{B}{\sqrt{B_3 B_4}}\,\tilde{\beta}^2 \sin \phi_{1}\,
	\frac{\sin 2\phi_{2}}{2}
	\Bigg]\mathcal{D}^{(1)}_{0,\mu}(\theta).\label{Eq41}
\end{align}
\end{widetext}
Here, $t_1$, $t_2$, and $t_3$ denote scaling parameters. For convenience, we introduce the constants
\begin{equation}
	F1 = t_1 a_{2}\frac{B}{\sqrt{B_2 B_3}}, \qquad
	F1^{\prime} = t_1 a_{3}\frac{B}{\sqrt{B_3 B_4}}, \label{Eq42}
\end{equation}
\begin{equation}
	F2 = t_2 \sqrt{\frac{B}{B_2}}, \qquad
	F3 = t_3 \sqrt{\frac{B}{B_3}}. \label{Eq43}
\end{equation}
The reduced electric transition probabilities are given by
\begin{equation}
	B(EL; L_i n_i \rightarrow L_f n_f)
	= \frac{|\langle L_f n_f || T^{(EL)} || L_i n_i \rangle|^2}{2L_i+1}.
	\label{Eq44}
\end{equation}
Using the Wigner–Eckart theorem, the reduced matrix element can be written as
\begin{widetext}
\begin{equation}
	\langle L_f n_f || T^{(EL)} || L_i n_i \rangle
	= \frac{\sqrt{2L_f+1}}{(L_i\, L\, L_f | \mu_i\, \mu\, \mu_f)}
	\langle L_f \mu_f n_f | T^{(EL)}_\mu | L_i \mu_i n_i \rangle .	\label{Eq45}
\end{equation}
\end{widetext}
The volume element associated with the collective coordinates reads \cite{b20}:
\begin{equation}
	d\tau = (\beta_2 \beta_3 \beta_4)^3\, d\beta_2\, d\beta_3\, d\beta_4\, d^3\theta, \label{Eq46}
\end{equation}

which, after introducing the reduced deformation parameters and applying the transformations of Eq. (\ref{Eq5}) and the corresponding Jacobian, becomes

\begin{equation}
	d\beta_2\, d\beta_3\, d\beta_4
	= \sqrt{\frac{B^3}{B_2 B_3 B_4}}\,
	\tilde{\beta}^2 d\tilde{\beta}\, \sin \phi_{02}\, d\phi_2\, d\phi_1 .\label{Eq47}
\end{equation}
As a consequence, the reduced matrix elements factorize into independent contributions,
\begin{equation}
	\langle L_f n_f || T^{(EL)} || L_i n_i \rangle
	\propto I_{\theta}^{(EL)}\, I_{\tilde{\beta}}^{(EL)}\,
	I_{\phi_1}^{(EL)}\, I_{\phi_2}^{(EL)} .\label{Eq48}
\end{equation}
\begin{itemize}
\item The angular integrals over the Euler angles yield
\begin{equation}
	I_{\theta}^{(EL)} = \sqrt{2L_i+1}\,(L_i\, L\, L_f | 0\, 0\, 0),\label{Eq49}
\end{equation}
\item The radial integrals $I_{\tilde{\beta}}^{(EL)}(n_i,n_f)$ are evaluated over
$\tilde{\beta}^2 d\tilde{\beta}$, leading to closed analytical expressions in terms
of Gamma functions (see Appendix \ref{app:B}). It is worth noting that
$I_{\tilde{\beta}}^{(E2)}(n_i,n_f)=I_{\tilde{\beta}}^{(E3)}(n_i,n_f)$, and the detailed
derivations follow those given in Appendix~C of Ref.~\cite{b15}.
\item The integrals $I_{\phi_1}^{(EL)}$ and $I_{\phi_2}^{(EL)}$ are performed over the
$\phi_1$ and $\phi_2$ variables, respectively, and depend on the trigonometric
structure of the transition operator. Their explicit forms are also given in
Appendix \ref{app:B}.
\end{itemize}
Finally, the reduced matrix elements for the $E2$, $E3$, and $E1$ transitions are
obtained as
\begin{widetext}
\begin{align}
	\langle L_f n_f || T^{E2} || L_i n_i \rangle &=
	F2\, \sqrt{2L_i+1}(L_i\,2\,L_f|0\,0\,0)\,
	I_{\tilde{\beta}}^{(E2)} I_{\phi_1}^{(E2)} I_{\phi_2}^{(E2)}, \label{me1} \\
	\langle L_f n_f || T^{E3} || L_i n_i \rangle &=
	F3\, \sqrt{2L_i+1}(L_i\,3\,L_f|0\,0\,0)\,
	I_{\tilde{\beta}}^{(E3)} I_{\phi_1}^{(E3)} I_{\phi_2}^{(E3)}, \label{me2} \\
	\langle L_f n_f || T^{E1} || L_i n_i \rangle &=
	\sqrt{2L_i+1}(L_i\,1\,L_f|0\,0\,0)\, I_{\tilde{\beta}}^{(E1)}\Big[
	F1\, I_{\phi_1}^{(E1)} I_{\phi_2}^{(E1)}
	+ F1^{\prime}\, I_{\phi_1}^{(E1)\prime} I_{\phi_2}^{(E1)\prime}
	\Big]. \label{me3}
\end{align}  
\end{widetext}
\section{NUMERICAL RESULTS AND DISCUSSION}
\subsection{ Fits to specific nuclei}

\begin{figure}[!h]
	\begin{center}
		\includegraphics[width=.48\textwidth]{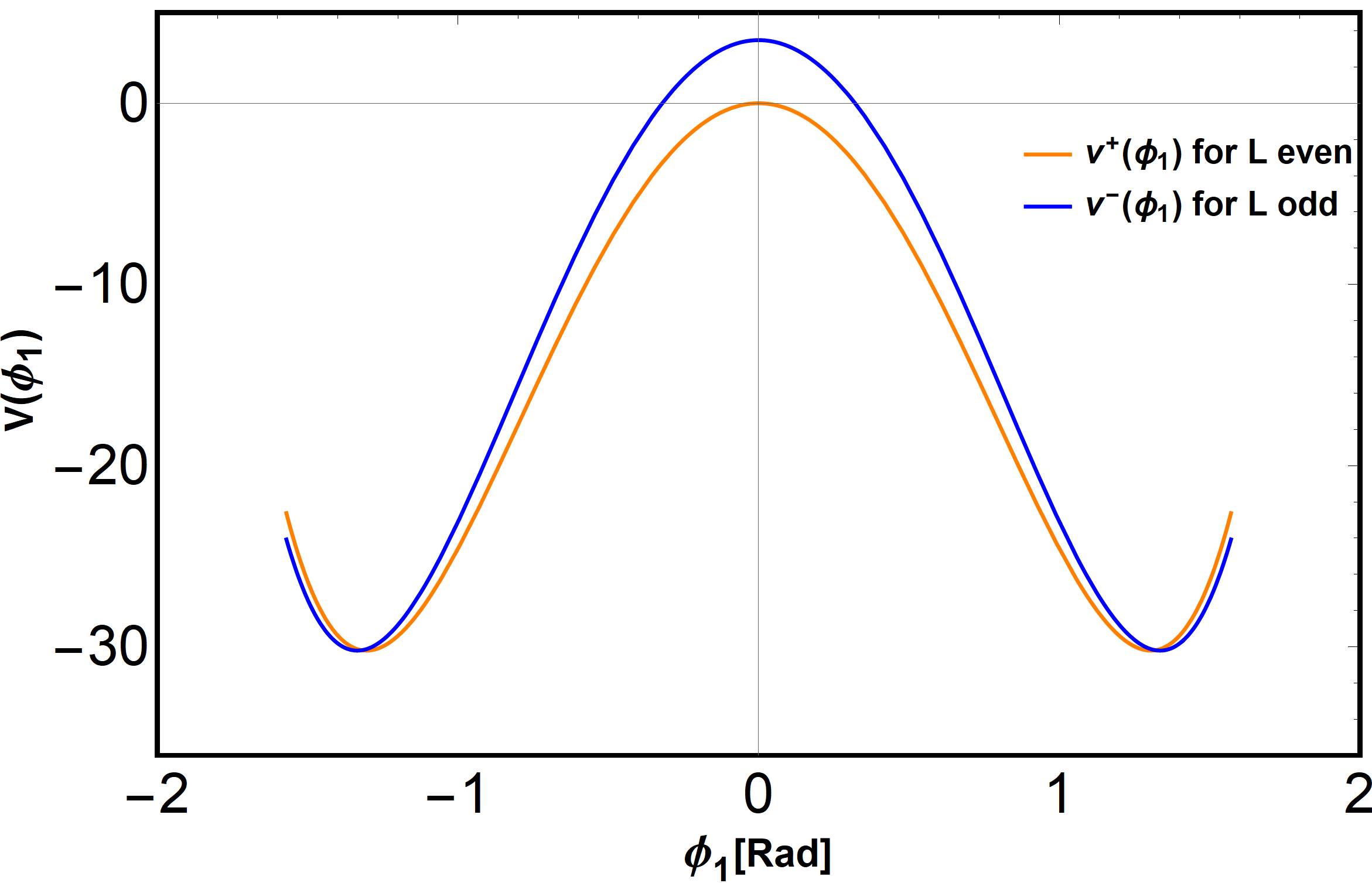}
	\end{center}
	\caption{Evolution of the potential $v(\phi_{1})$ Eq.(\ref{Eq24}) as a function of $\phi_{1}$ ; the blue line corresponds to odd states and the orange line corresponds to even states.} \label{vv11}
\end{figure}

In this work, we expand the analysis to a wider range of octupole deformed nuclei. To do so, we adjusted the parameters entering the total energy, namely $c_{2}$, which represents the stiffness constant, $\phi_{02}$, which specifies the hexadecapole deformation, and $\alpha$, which defines the sextic potential parameter. These parameters were determined by minimizing the squared differences between the calculated and experimental normalized energies of the selected bands. The optimization was carried out for even–even nuclei found in regions characterized by stable octupole deformation or octupole vibrational behavior, using the standard deviation:
\begin{equation}
	\sigma=\sqrt{\frac{\sum_{i=1}^{n}\left(E_{i}(\exp )-E_{i}(\mathrm{th})\right)^{2}}{(n-1) E\left(2_{g.s}^{+}\right)^{2}}},\label{Eq53}
\end{equation}
As a result, we found that a considerable number of nuclei, specifically 54 nuclei, as shown in Table (\ref{Tab1}), can be successfully described by our model, demonstrating its effectiveness. It is worth noting that :
\begin{itemize}
	\item For most nuclei, the fit quality measure is less than unity, indicating good agreement between the QOH theoretical predictions and the experimental data.
	\item The proposed model successfully describes two distinct structural behaviours in octupole deformed nuclei : nuclei exhibiting an octupole vibrational negative parity band, offset from the ground state band, and those with stable octupole deformation, where positive and negative parity bands merge to form a single unified band. This ability to capture both structural behaviours makes the model an effective tool for studying the evolution of octupole correlations across different isotopic regions.  
	\item The stiffness constant $c_{2}$ and the oscillator minimum $\phi_{02}$, exhibited in Table (\ref{Tab1}), are indicative of the presence of hexadecapole deformation within the isotopes under consideration. 
	\item As shown in Fig.(\ref{vv11}), the potential $v(\phi_{1})$ is a finite barrier with two symmetric minima, which allows for an improved reproduction of the negative parity states $1^{-}$ and $3^{-}$ for most nuclei, compared to the results in Refs\cite{b15,b19}, where these states are undervalued, as presented in Table (\ref{Tab3}).
\end{itemize}
\begin{table*}
	\caption{Parameters $c_{2}$, $\phi_{02}$ and $\alpha$ of the QOH Model determined from RMS fits to experimental spectra, taken from Refs\cite{b27,b28,b29,b30} of ${ }^{218-230} \mathrm{Ra}$, ${ }^{222-234} \mathrm{Th}$, ${ }^{148-154} \mathrm{Sm}$, ${ }^{120} \mathrm{Ba}$, ${ }^{140-148} \mathrm{Ba}$, ${ }^{230-240} \mathrm{U}$, ${ }^{154-160} \mathrm{Gd}$, ${ }^{146-152} \mathrm{Nd}$,${ }^{218-226} \mathrm{Rn}$, ${ }^{236-244} \mathrm{Pu}$, ${ }^{112-114} \mathrm{Xe}$, ${ }^{140-142} \mathrm{Xe}$ and ${ }^{144-146} \mathrm{Ce}$. The Minima $\phi_{01}^{+}$ and $\phi_{01}^{-}$, which represent the interplay between octupole and quadrupole deformations  for even L and odd L, are calculated  using Eq.(\ref{Eq27}). The angular momenta of the highest levels of the ground state and negative parity bands considered in the RMS fits are denoted by $L_{g}$ and $L_{o}$, respectively, while n represents the total number of levels included in the fit and $\sigma_{QOH}$ is the quality metric of Eq.(\ref{Eq53}).}
	$$\label{Tab1}
	\begin{array}{|lcccccccccc|}               
		\hline \text { \textbf{Nucleus} } & R_{4 / 2} & c_{2} & \phi_{02} & \alpha & \phi_{01}^{+} & \phi_{01}^{-}& L_g  & L_o & n & \sigma_{\textbf{QOH}}\\
		\hline \hline
		& &  && & &  &  &  &  &     \\
		{ }^{218} \mathrm{Ra}& 1.905 &19.88 &1.205  & 3.26 & 1.358 & 1.387 & 20 &   21&20 & 0.502 \\
		{ }^{220} \mathrm{Ra} &2.297& 22.99 & 1.301 &3.74  & 1.255 &1.287 & 20 &   21&20 & 0.722 \\
		{ }^{222} \mathrm{Ra} & 2.715& 5.67 & 1.749 & 3.73 & 1.257 & 1.289 &20 &   19 & 19 &0.716 \\
		{ }^{224} \mathrm{Ra} & 2.970 & 8.37 & 1.698 &3.68  & 1.267 & 1.298 &  20 &   21 & 20 & 0.554 \\
		{ }^{226} \mathrm{Ra} & 3.127 & 13.42 & 1.453 & 3.42 & 1.323 &1.354  &  20 &   21 & 20 & 0.597  \\
		{ }^{228} \mathrm{Ra} & 3.207 & 6.02 & 1.458 & 3.41 & 1.386 &1.415 & 22 &  19 & 20 & 0.901 \\
		{ }^{230} \mathrm{Ra} & 3.251 & 19.05 & 1.678 & 3.20 &  1.429 & 1.456 &  18 &  5 &11  & 0.196 \\
		{ }^{222} \mathrm{Th}& 2.399 &7.47 & 1.832 & 3.40 &  1.327& 1.358 & 20 &   21&19 &  0.828 \\
		{ }^{224} \mathrm{Th} & 2.896 & 4.207 & 1.416 & 3.68  &1.267& 1.298 & 18 &  17 & 17 & 0.426 \\
		{ }^{226} \mathrm{Th} & 3.136 & 9.07 & 1.695 & 3.59 &  1.347 & 1.376 & 20 &  19 & 19 & 0.561 \\
		{ }^{228} \mathrm{Th}& 3.234 & 3.65 & 1.483 & 3.50 & 1.367 & 1.395 & 22 &  19 &20 & 0.699 \\
		{ }^{230} \mathrm{Th}& 3.271 & 8.18 & 1.656 & 3.31 & 1.407 & 1.434 & 22 &   23 & 22 &  1.10 \\
		{ }^{232} \mathrm{Th}& 3.283 & 3.70 & 1.484 & 3.18 &1.433 &1.460 &  22 &  23 & 22 & 0.802 \\
		{ }^{234} \mathrm{Th}& 3.290 & 1.57 & 1.501 & 3.24 & 1.422 & 1.449 & 22 &  23 & 19 &1.022 \\
		{ }^{148} \mathrm{Sm}& 2.14 & 3.47 & 1.832 & 3.74 & 1.316 & 1.345 & 22 & 15 & 17 & 0.347 \\
		{ }^{150} \mathrm{Sm}&2.315&14.94 & 1.295& 3.54 & 1.299 & 1.330&  20 &   21& 19& 0.487  \\
		{ }^{152} \mathrm{Sm}& 3.009 & 1.16 & 1.403& 3.35 & 1.399& 1.427 & 20 & 17 & 18 & 0.638 \\
		{ }^{154} \mathrm{Sm}& 3.254 &0.10 & 1.484 & 3.24 & 1.420& 1.448 & 22 &  13 &17 &  0.531     \\
		{ }^{120} \mathrm{Ba}& 2.924 &2.70 &1.370 & 3.38 & 1.393& 1.421 & 12 &17 & 12& 0.413 \\
		{ }^{140} \mathrm{Ba} & 1.876 & 1.46 & 1.301 & 3.60 & 1.345& 1.374 & 12 & 13 & 11  &0.255 \\
		{ }^{142} \mathrm{Ba}& 2.321 & 1.95 & 1.832 & 3.57 & 1.353& 1.381 &  10 &  11& 9 & 0.370 \\
		{ }^{144} \mathrm{Ba}& 2.659 & 1.75 & 1.356 & 3.60 & 1.346& 1.375 & 16 &  19& 17 & 0.728 \\
		{ }^{146} \mathrm{Ba}& 2.837 & 13.79 & 1.359 & 3.62 & 1.342& 1.370 & 16 & 13& 14 &0.594  \\
		{ }^{148} \mathrm{Ba}& 2.983 & 3.71 & 1.373 & 3.49 & 1.368 & 1.397 & 12  & 11 &  11 &0.558 \\
		{ }^{230} \mathrm{U} &3.273 & 1.51 & 1.499 & 3.37  & 1.395 & 1.423 &22  &  17& 19  &0.459 \\
		{ }^{232} \mathrm{U}& 3.291 & 6.69 &1.637  & 3.22 & 1.426&1.453 &  20&  15& 16  &0.287 \\
		{ }^{234} \mathrm{U} & 3.295 & 3.89 & 1.639 & 3.12 & 1.445 & 1.472 & 22 & 11 & 16  &0.243 \\
		{ }^{236} \mathrm{U}& 3.303 & 2.66 & 1.630 & 3.18 & 1.434 & 1.461 &22  &19 &20   & 0.671 \\
		{ }^{238} \mathrm{U}& 3.303 & 0.84 & 1.504 & 3.19 & 1.431& 1.459 & 22 &  23 &  22 & 1.228 \\
		{ }^{240} \mathrm{U}& 3.346 & 3.57 & 1.528 &3.15 & 1.440 & 1.467 &  12&  9&9   &0.618 \\
		{ }^{154} \mathrm{Gd} & 3.014 & 2.64 & 1.385 & 3.01 & 1.410 & 1.439 & 20 &  21 & 20 &0.770 \\
		{ }^{156} \mathrm{Gd}& 3.239 & 5.55 & 1.692 &2.87& 1.439 & 1.468 & 20 &5 & 12  &0.308 \\
		{ }^{158} \mathrm{Gd}&3.288&  3.50 & 1.613 & 2.88 &1.437 &1.466 & 12& 5& 8 &0.193 \\
		{ }^{160} \mathrm{Gd}&3.301&  3.57 &1.518 & 2.90 &1.433 & 1.462 &16 & 11& 13  &1.047\\
		{ }^{146} \mathrm{Nd}& 2.298 &1.84 & 1.832 & 3.60 & 1.285 & 1.317 & 16 & 15 &  14 & 0.325 \\
		{ }^{148} \mathrm{Nd}& 2.493 & 5.20 & 1.301 &3.40& 1.329 & 1.360 & 12&  13& 12  &0.490 \\
		{ }^{150} \mathrm{Nd}& 2.926 &6.11 & 1.780 & 3.09 & 1.393 & 1.423 & 16 & 7 & 11  &0.418 \\
		{ }^{152} \mathrm{Nd}& 3.267 & 3.52 & 1.643 & 2.88 & 1.437 & 1.466 & 20 &  5 &  12 &0.119 \\
		{ }^{218} \mathrm{Rn} & 2.014  & 19.85& 1.230  & 3.30 & 1.349 & 1.379 &20  &19 & 18 & 0.386 \\
		{ }^{220} \mathrm{Rn} & 2.214 &9.6 & 1.287 & 3.42 & 1.324 & 1.354 & 20  & 21 & 20  &0.487  \\
		{ }^{222} \mathrm{Rn} & 2.408 &9.58 & 1.306  & 3.43  & 1.323 & 1.353 & 16 & 21 & 18  &0.461 \\
		{ }^{224} \mathrm{Rn}& 2.637 &9.29 & 1.317 & 3.29 &1.353 & 1.383& 14 & 11 & 11 & 0.369 \\
		{ }^{226} \mathrm{Rn}& 2.906 & 9.36& 1.343 & 3.10 & 1.391 & 1.421 & 14 & 11 & 11 &0.394\\
		{ }^{236} \mathrm{Pu} & 3.303 & 23.81 &1.527 &3.16& 1.438 & 1.465 & 16  &5 & 10  &0.226 \\
		{ }^{238} \mathrm{Pu}&3.311 & 0.70 & 1.520 & 3.21 & 1.428 & 1.455 & 20& 21 & 20 &1.234  \\
		{ }^{240} \mathrm{Pu}&3.308 & 3.10 & 1.512 & 3.22 & 1.427 & 1.454 &20& 21 & 20 & 1.408 \\
		{ }^{242} \mathrm{Pu}&3.309 & 2.22 & 1.506 & 3.16 & 1.438 & 1.466 &20 & 21 & 20 & 1.544\\
		{ }^{244} \mathrm{Pu}& 3.391 & 4.11 & 1.516 & 3.13 & 1.444 & 1.472 &20 & 21 & 19 & 1.608 \\
		{ }^{112} \mathrm{Xe}& 2.407 &1.88 & 1.311   & 3.6 & 1.347 & 1.376 & 12&13 & 11& 0.371 \\
		{ }^{114} \mathrm{Xe}& 2.375 & 7.78 &   1.832 & 3.65 & 1.335 & 1.364 & 22 &23 &21 &  0.446 \\
		{ }^{140} \mathrm{Xe}& 2.214 &4.00 & 1.832   & 3.54 & 1.298 & 1.329 & 18 & 17 &  16 &0.534 \\
		{ }^{142} \mathrm{Xe}& 2.404 &3.45 &1.301  & 3.29 & 1.353 & 1.383 & 16 & 13 & 13  &0.449 \\
		{ }^{144} \mathrm{Ce}& 2.361 & 1.99 & 1.832 & 3.51 &1.304  & 1.335 & 8&  11&  8 &0.260  \\
		{ }^{146} \mathrm{Ce}& 2.586  &6.59 & 1.304  & 3.33 & 1.343 & 1.373 & 10 & 15 & 12  & 0.487 \\
		&  & &    &  &  &  & & & &\\
		\hline 
	\end{array}
	$$
\end{table*} 

\begin{table*}
	\centering
	\caption{Comparison between the theoretical predictions of the QOH solution for the ratios $R_{1/2}=E(1^{-})/E(2^{+})$ and $R_{3/2}=E(3^{-})/E(2^{+})$ and the corresponding available experimental data taken from Refs\cite{b27,b28,b29,b30}.}\label{Tab3}
	$$ 
	\begin{array}{|lcccccccccccccc|}
		\hline\hline  
		&   &  &   &  &  &  & &   &  &   &  &    & &\\
		& { }^{218} \mathrm{Ra}  & { }^{220} \mathrm{Ra} &  { }^{222} \mathrm{Ra} & { }^{224} \mathrm{Ra} & { }^{226} \mathrm{Ra} & { }^{228} \mathrm{Ra} & { }^{230} \mathrm{Ra}& { }^{222} \mathrm{Th}  & { }^{224} \mathrm{Th}  &  { }^{226} \mathrm{Th} & { }^{228} \mathrm{Th} & { }^{230} \mathrm{Th} & { }^{232} \mathrm{Th} &{ }^{234} \mathrm{Th} \\
		&   &  &   &  &  &  & &   &  &   &  &    & &\\
		R_{1 / 2}\, \text{exp} & 2.193  & 2.314 &  2.18 & 2.56 & 3.75 & 7.429 & 12.385 & - & 2.56 & 3.19  & 5.67 & 9.546  & 14.471 & -\\
		R_{1 / 2}\, \text {th} & 2.352  & 1.280 &  1.292 & 1.386  & 2.361  & 5.347 &12.009&-  & 1.389 & 3.027 & 3.978 & 7.740 & 13.331 & -\\
		R_{3 / 2}\, \text {exp}& 2.039  & 2.656 &  2.85 &  3.44 & 4.75 & 8.421  & 13.389 & 2.547  & 3.11  &   4.26 & 6.85 & 10.741  & 15.686 & -\\
		R_{3 / 2}\, \text {th}& 2.807  & 2.399 & 2.614 & 2.783 & 3.783 & 6.758 & 13.371 & 3.241 & 2.751 & 4.44  & 5.44 & 9.185 & 14.738 & -\\
		&   &  &   &  &  &  & &   &  &   &  &    & &\\
		& { }^{148} \mathrm{Sm} & { }^{150} \mathrm{Sm} & { }^{152} \mathrm{Sm}  & { }^{154} \mathrm{Sm} & { }^{120} \mathrm{Ba}  & { }^{140} \mathrm{Ba} & { }^{142} \mathrm{Ba}  & { }^{144} \mathrm{Ba} &  { }^{146} \mathrm{Ba} & { }^{148} \mathrm{Ba} &   { }^{154} \mathrm{Gd}   & { }^{156} \mathrm{Gd} & { }^{158} \mathrm{Gd}& { }^{160} \mathrm{Gd}\\
		&   &  &   &  &  &  & &   &  &   &  &    & &\\
		R_{1 / 2}\, \text {exp} & -  & - & 7.91  &11.238  & -  & - &  -& 3.807  & 4.08 & 4.846  & 10.085 & 15.359 & 15.890 & 16.266\\
		R_{1 / 2}\, \text {th} &  - & - &  6.404 &10.109  & - & -  & - & 2.824  & 2.726 &3.86   & 8.136 & 16.018  &16.168 & 14.436\\
		R_{3 / 2}\, \text {exp} & 2.11  & 3.20 &  8.549 & 12.349  & - & 2.993 & 3.593 & 4.206  & 4.535 & 5.465  & 10.170 & 17.296 & 17.643 & 17.136\\
		R_{3 / 2}\, \text {th} & 2.485  & 2.710 & 7.650  & 11.536 & - & 2.708 & 3.420 & 3.944  & 3.912 & 5.051  & 9.326 & 17.304 &17.666 & 15.917\\
		&   &  &   &  &  &  & &   &  &   &  &    & &\\
		&  { }^{230} \mathrm{U} & { }^{232} \mathrm{U} & { }^{234} \mathrm{U} & { }^{236} \mathrm{U} &  { }^{238} \mathrm{U}& { }^{240} \mathrm{U} & { }^{146} \mathrm{Nd} &   { }^{148} \mathrm{Nd}& { }^{150} \mathrm{Nd} & { }^{152} \mathrm{Nd} & { }^{112} \mathrm{Xe} &{ }^{114} \mathrm{Xe} & { }^{140} \mathrm{Xe}& { }^{142} \mathrm{Xe}\\
		&   &  &   &  &  &  & &   &  &   &  &    & &\\
		R_{1 / 2}\, \text {exp} & 7.088 &11.838  & 18.076  & 15.197  &15.141  & - & - & 3.391  & 6.550 & 15.866 & - &  -  &- &-\\
		R_{1 / 2}\, \text {th} & 6.241  & 11.533 & 18.307 & 14.016 & 13.15 & - & - & 2.251  & 5.929 & 15.685  & - &  -  &- &-\\
		R_{3 / 2}\, \text {exp}  &  8.413 & 13.221 & 19.524 & 16.448 & 16.295 & 18.822 & 2.621 & 3.312 & 7.177 & 17.113 & 3.539 & 3.607 &4.016 &4.381\\
		R_{3 / 2}\, \text {th} & 7.721  & 12.997 & 19.737  & 15.487 & 14.608 & 17.948 & 2.464 & 3.137  & 7.083 & 17.108  & 3.258 &   3.109 &2.659 &3.846\\
		&   &  &   &  &  &  & &   &  &   &  &    & &\\
		&  {}^{236} \mathrm{Pu} & { }^{238} \mathrm{Pu} &  { }^{240} \mathrm{Pu} & { }^{242} \mathrm{Pu} & { }^{244} \mathrm{Pu} & { }^{218} \mathrm{Rn} & { }^{220} \mathrm{Rn}& { }^{222} \mathrm{Rn} & { }^{224} \mathrm{Rn} & { }^{226} \mathrm{Rn} & { }^{144} \mathrm{Ce}
		&  { }^{146} \mathrm{Ce} & &\\  
		
		&   &  &   &  &  &  & &   &  &   &  &    & &\\
		R_{1 / 2}\, \text {exp}    & 15.646  & 13.733 & 13.948  & 17.52 & - & - & 2.678 &  3.226 & - & -  & - & 3.577  & &\\
		R_{1 / 2}\, \text {th}  & 15.464  &  12.119& 11.764 & 15.544 & - & - & 2.084 &  2.177 & - & -  & - & 2.662   & &\\
		R_{3 / 2}\, \text {exp} & 16.984  & 15.010 &  15.151 & 18.684 & 21.651 & 2.590 &2.751 & 3.413 & 4.797 & 7.5   & 3.125 &  3.717 & &\\
		R_{3 / 2}\, \text {th} & 16.967  & 13.616 & 13.247 & 16.997 & 19.752 & 2.954 &2.914 & 3.148 & 4.155 & 6.79 & 2.667 &  3.595  & &\\
		&   &  &   &  &  &  & &   &  &   &  &    & &\\
		\hline \hline
	\end{array}
	$$
\end{table*}
\subsection{Fine structure of alternating-parity bands}
\subsubsection{The odd-even staggering }
\par
\begin{figure*}
	\includegraphics[width=.3\textwidth]{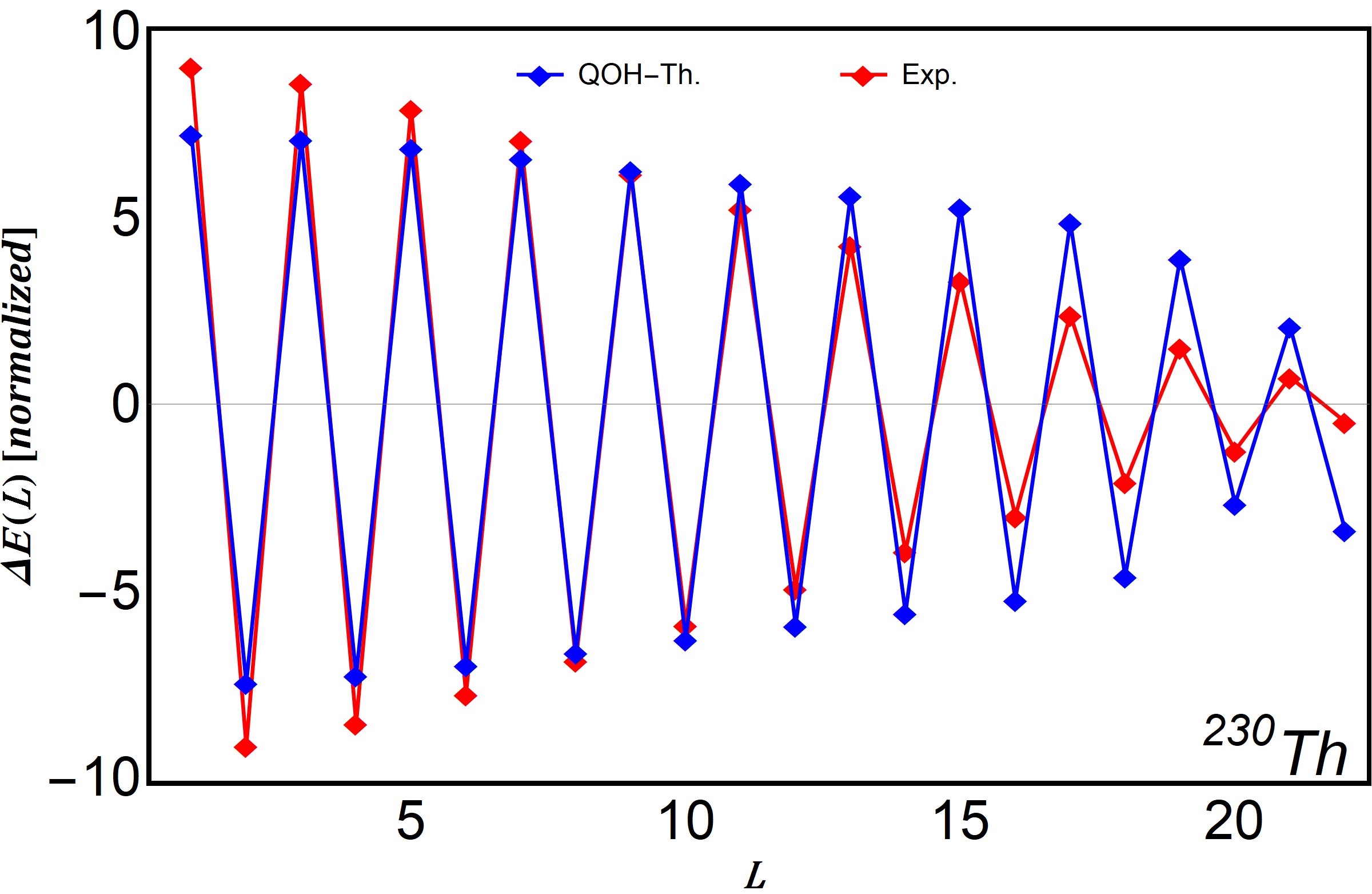}
	\includegraphics[width=.3\textwidth]{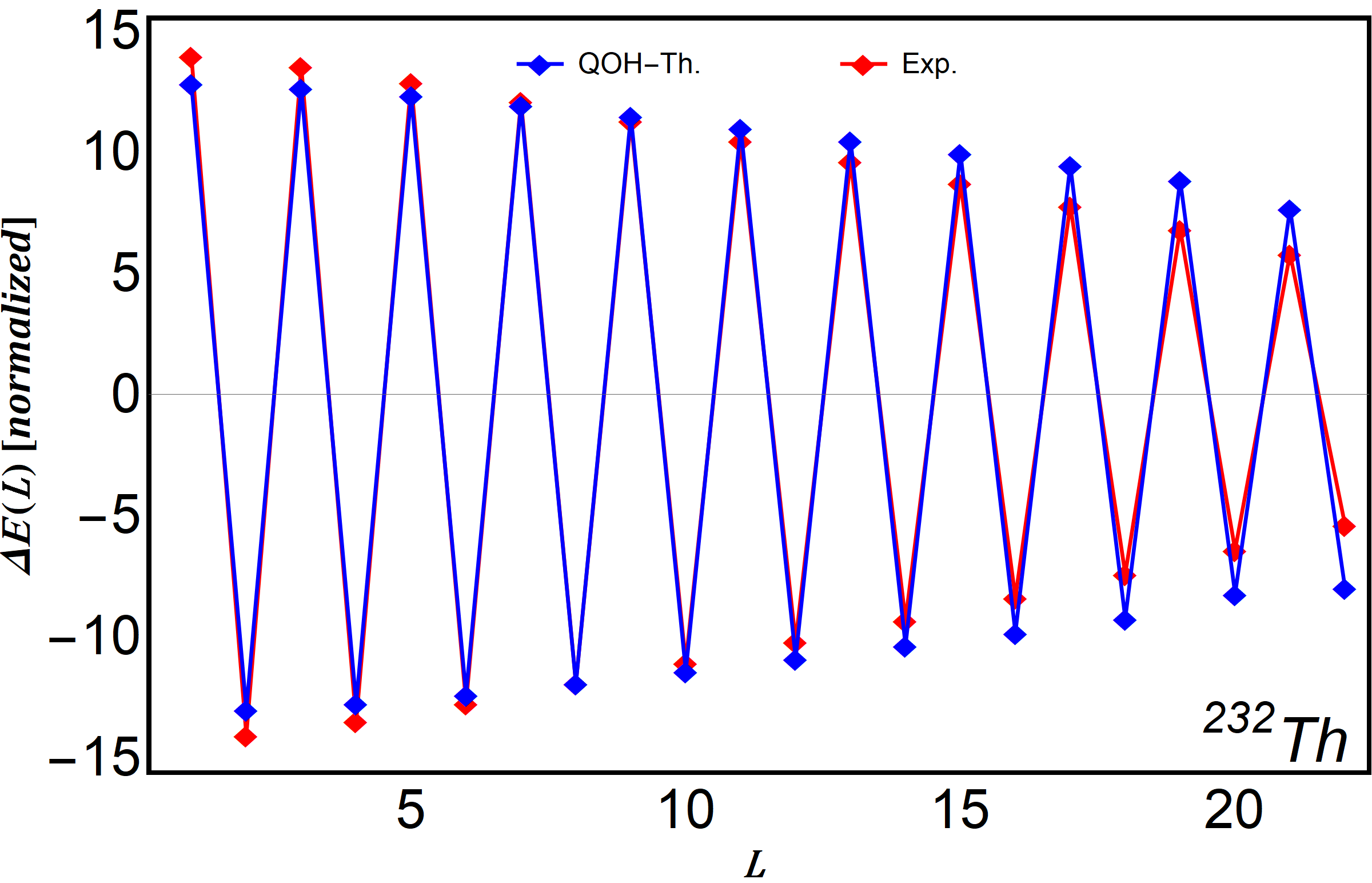}
	\includegraphics[width=.3\textwidth]{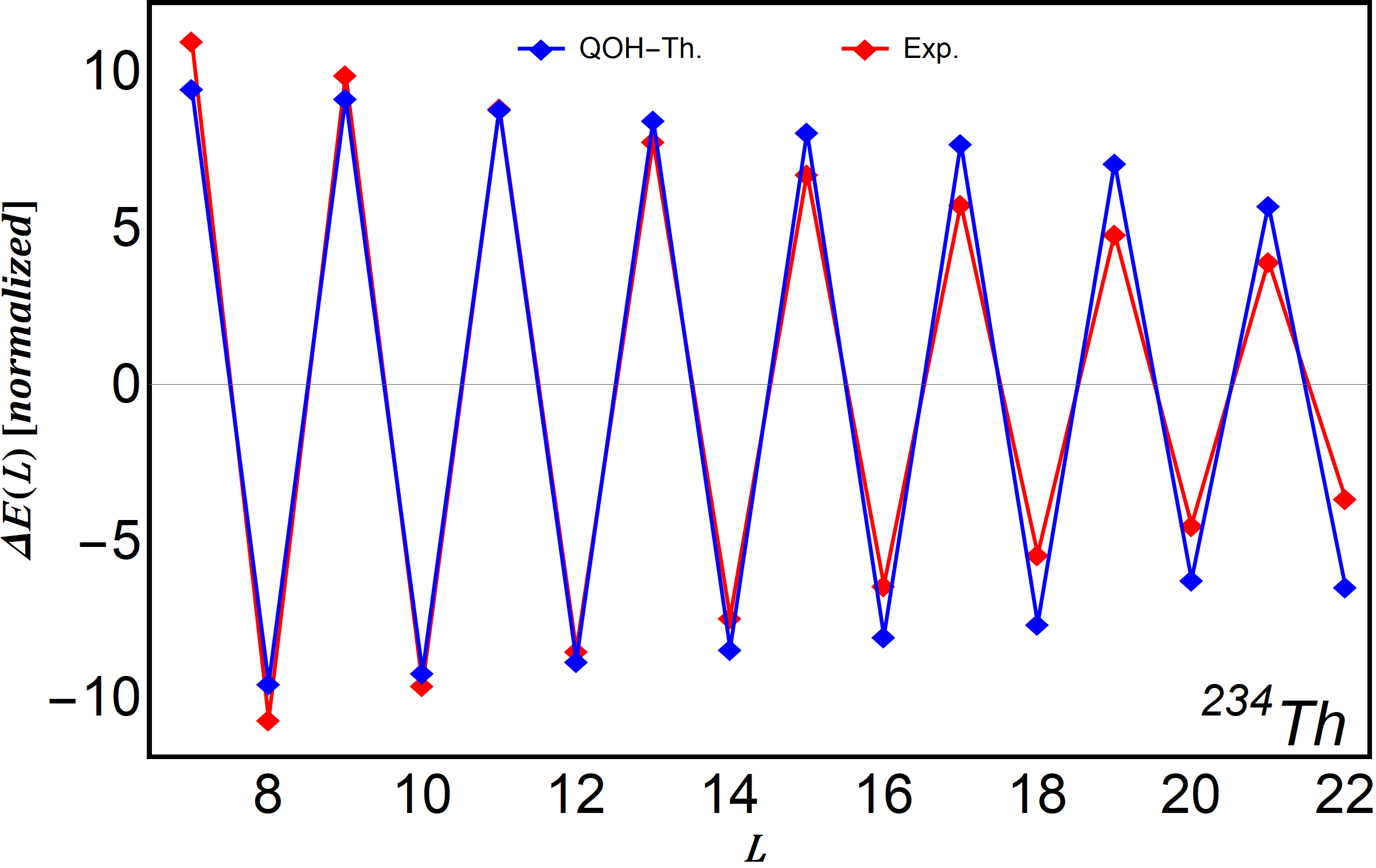}
	\includegraphics[width=.3\textwidth]{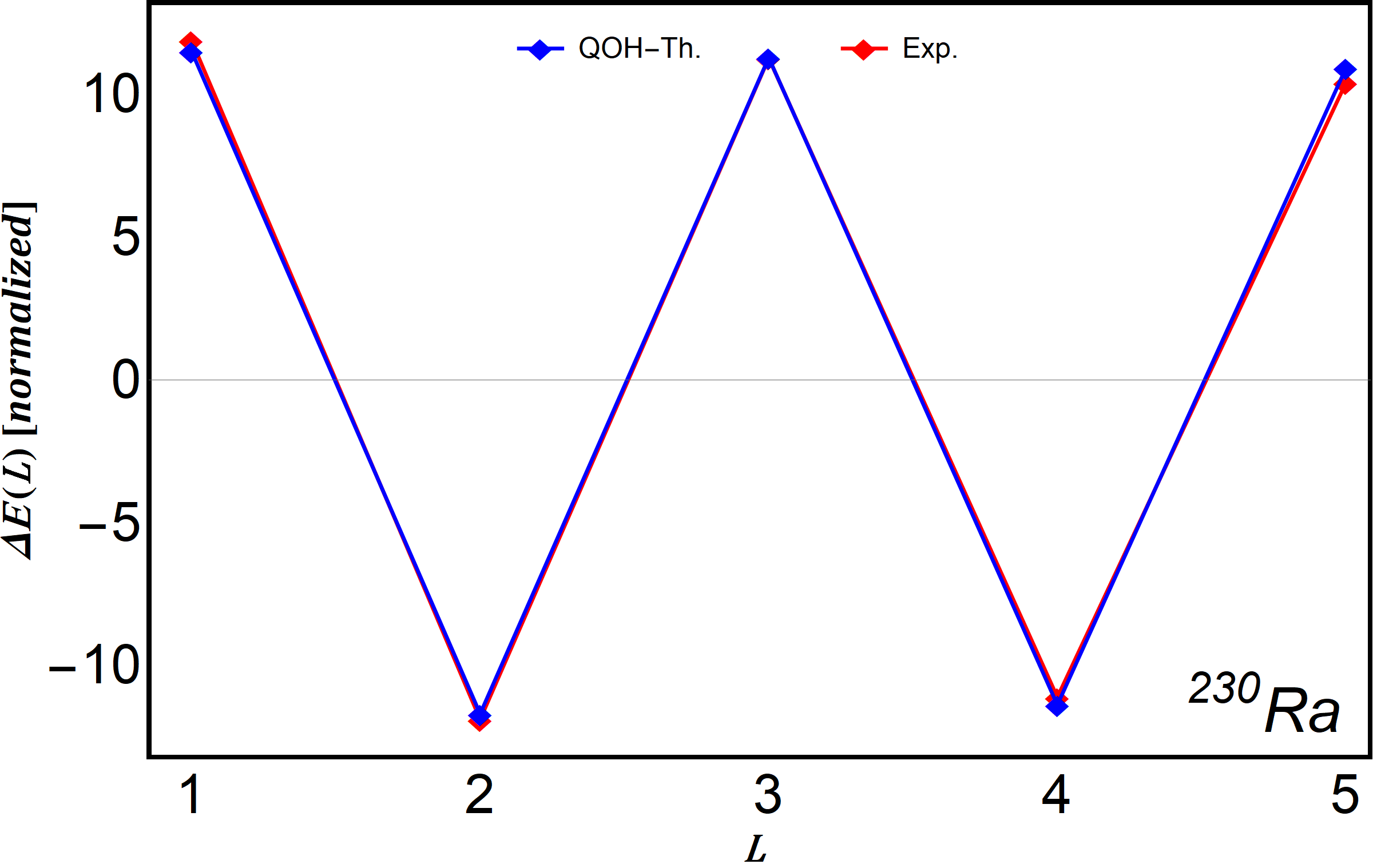}
	\includegraphics[width=.3\textwidth]{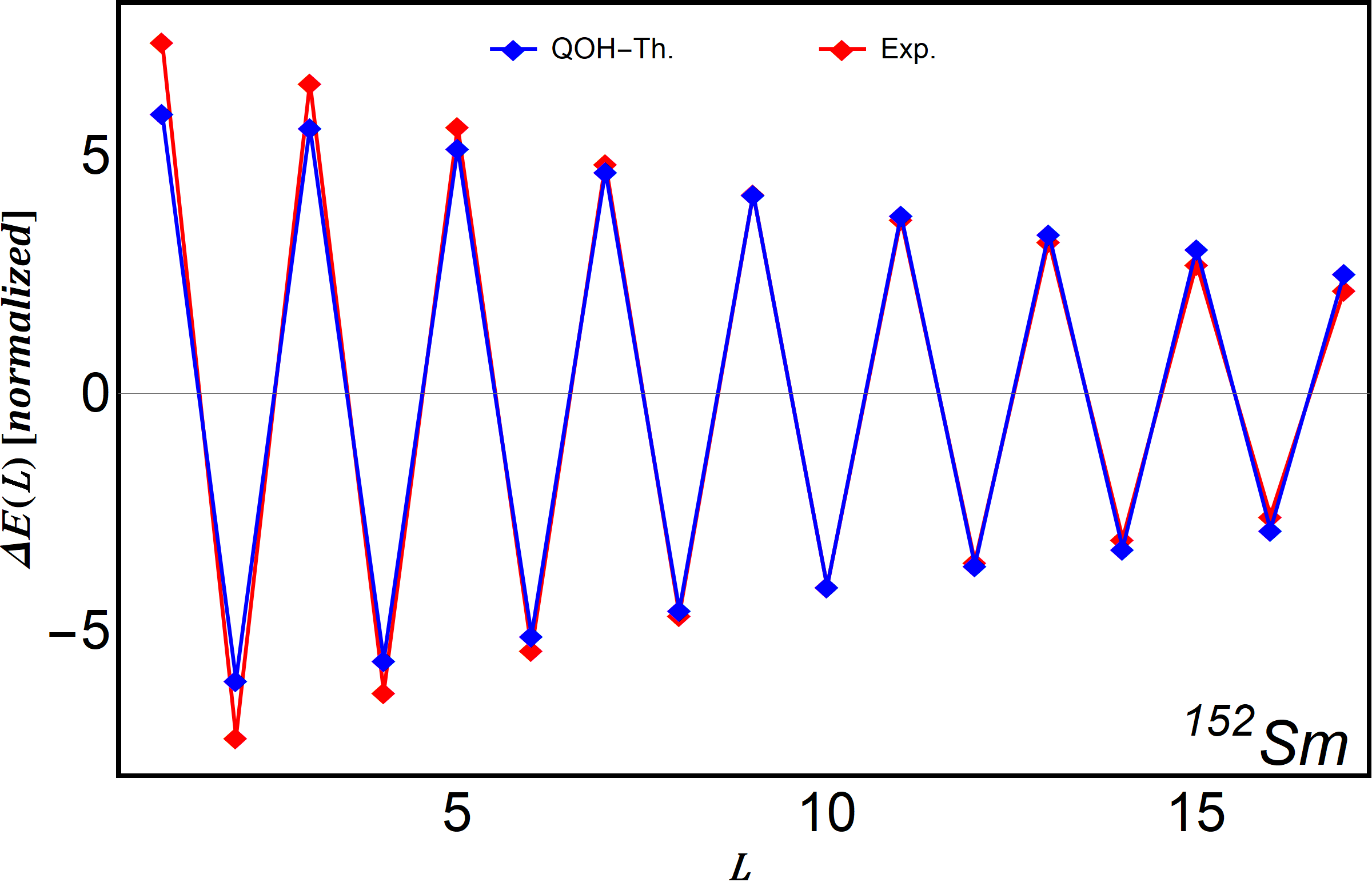}
	\includegraphics[width=.3\textwidth]{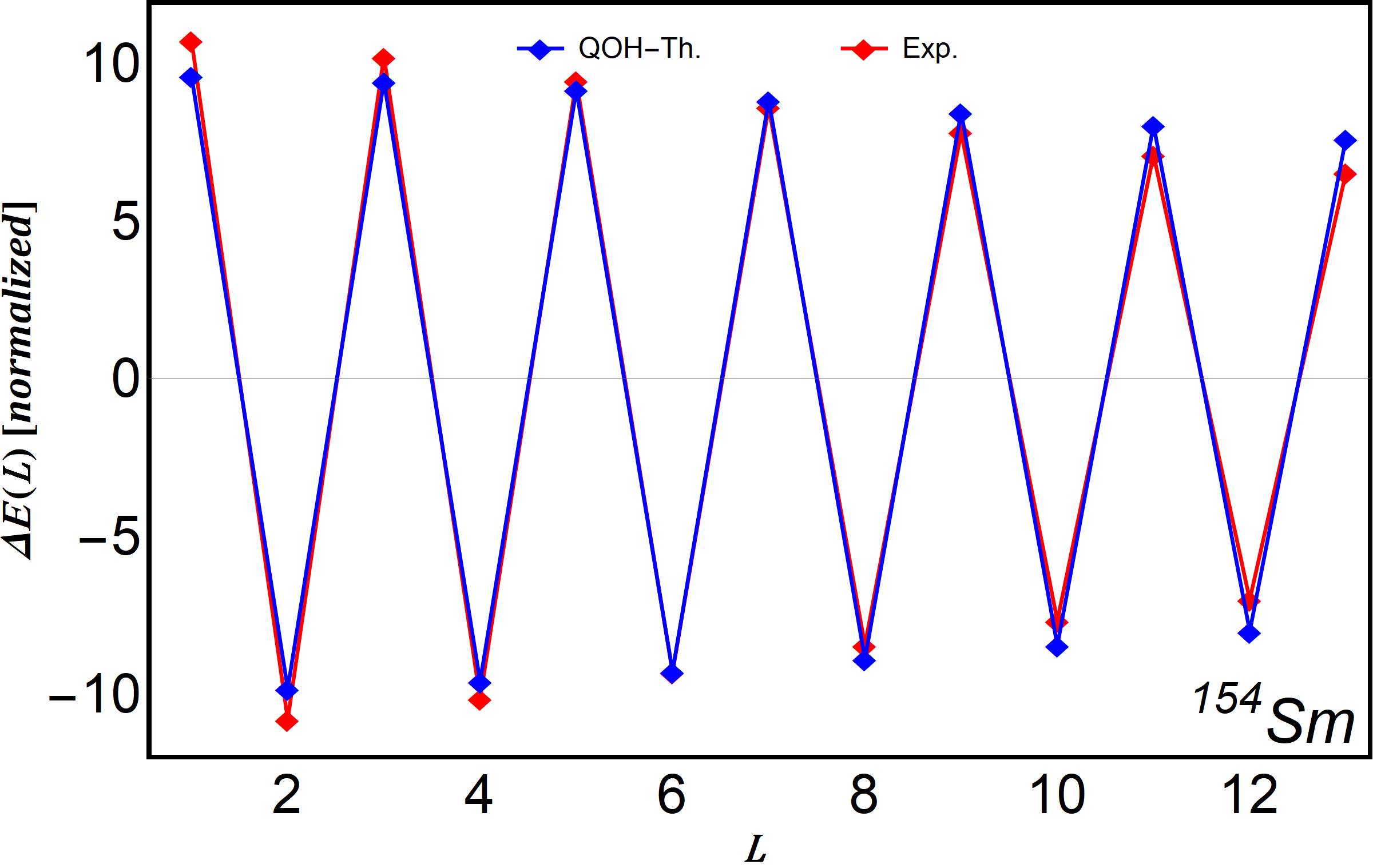}
	\includegraphics[width=.3\textwidth]{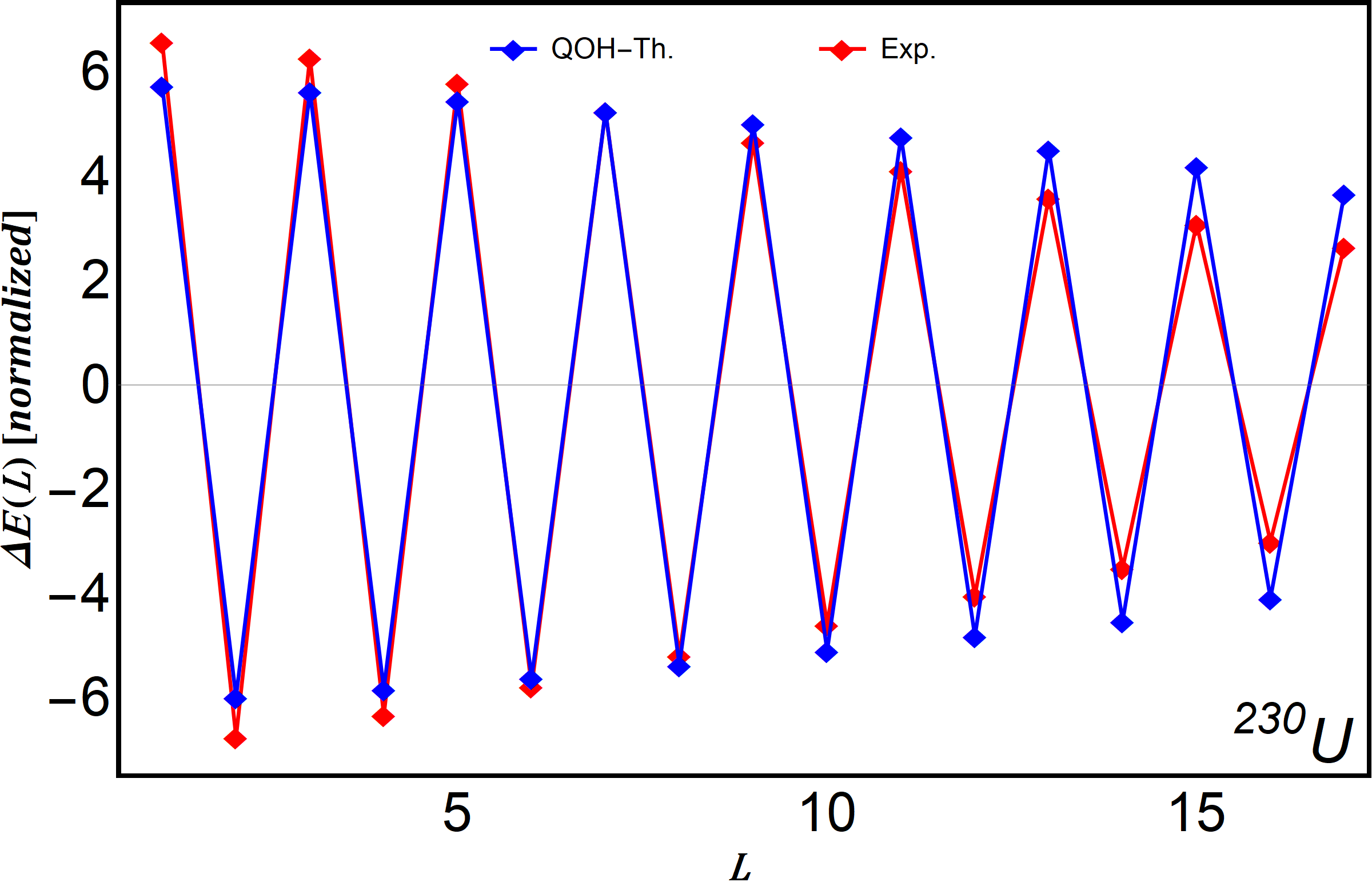}
	\includegraphics[width=.3\textwidth]{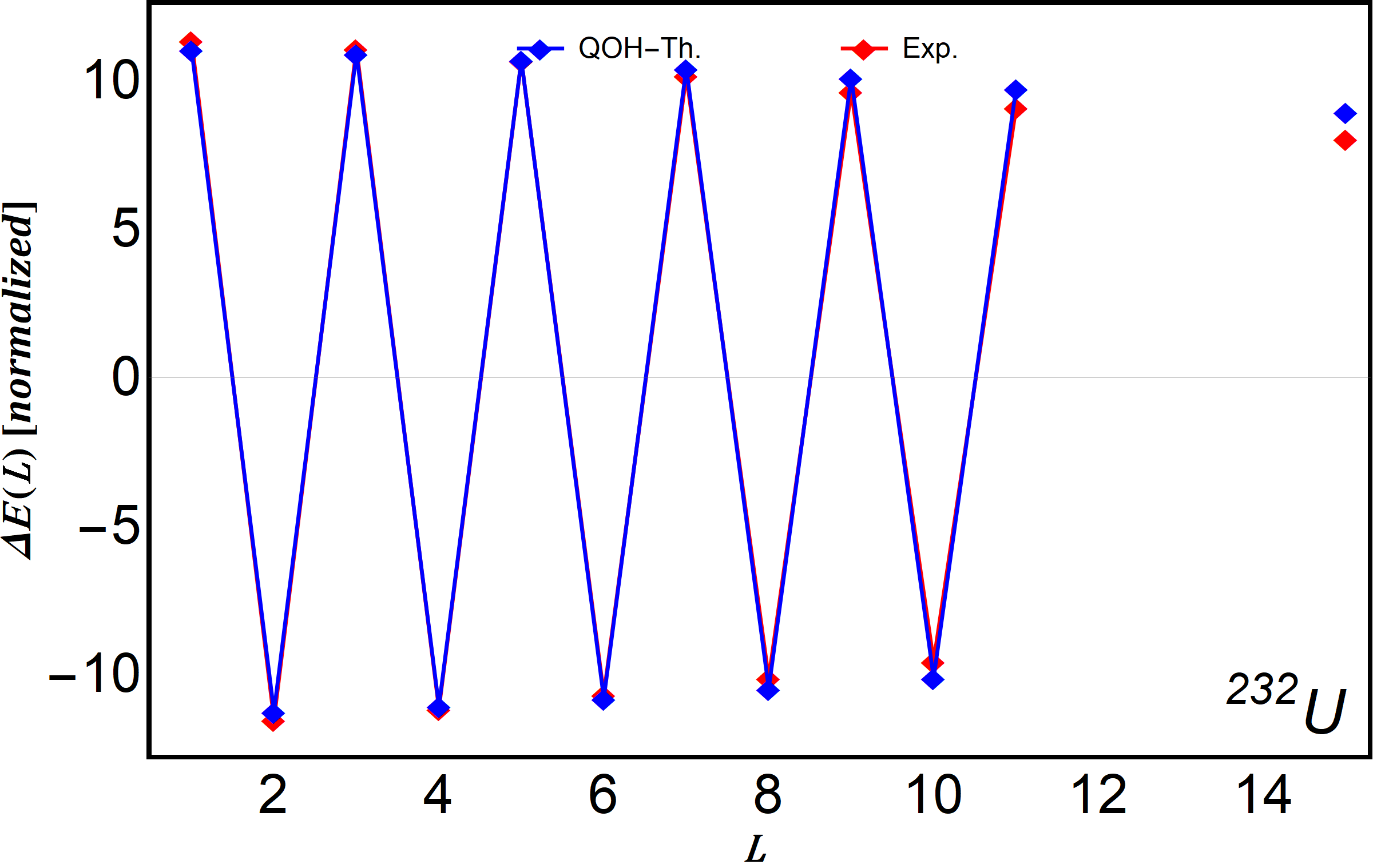}
	\includegraphics[width=.3\textwidth]{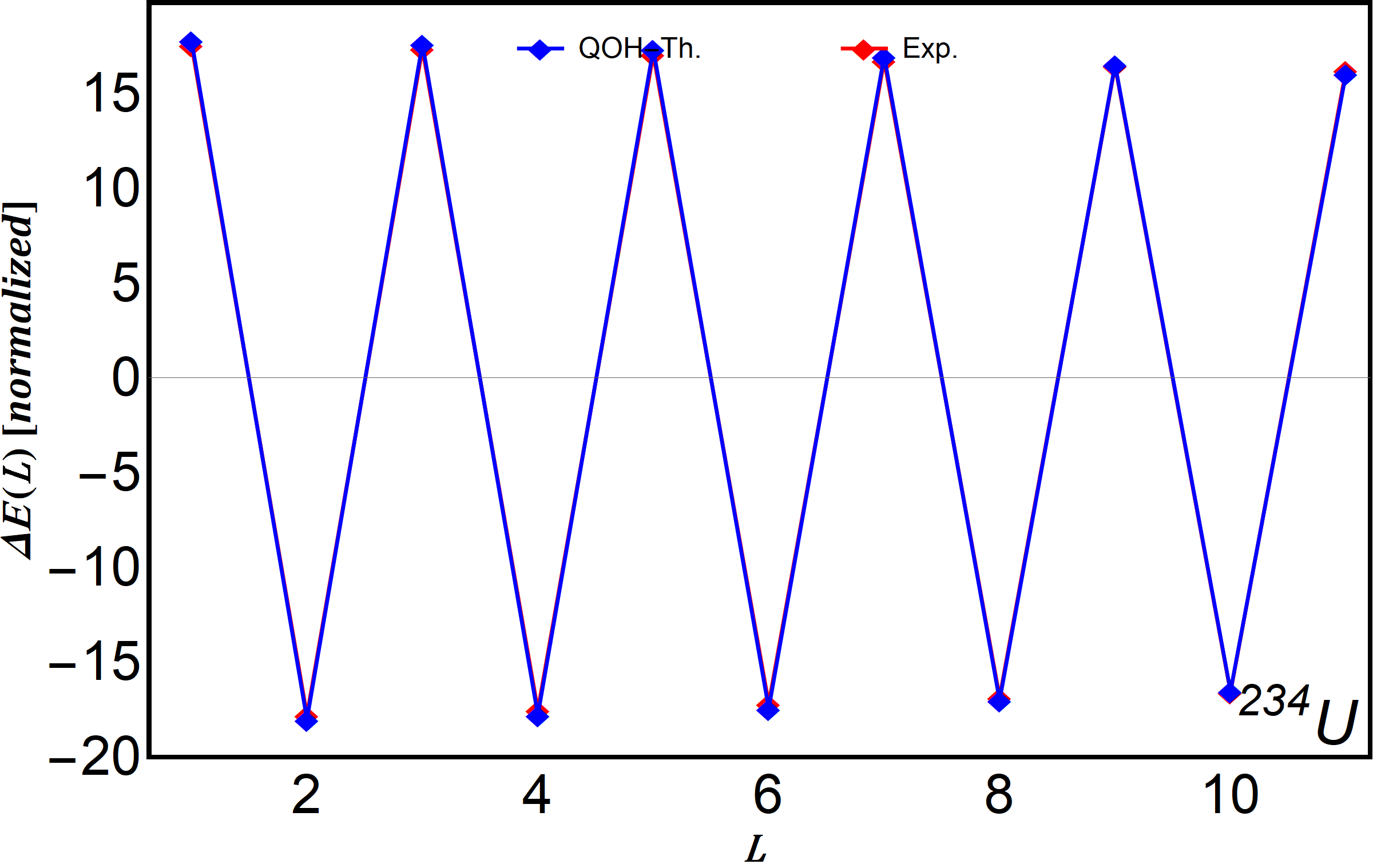}
	\includegraphics[width=.3\textwidth]{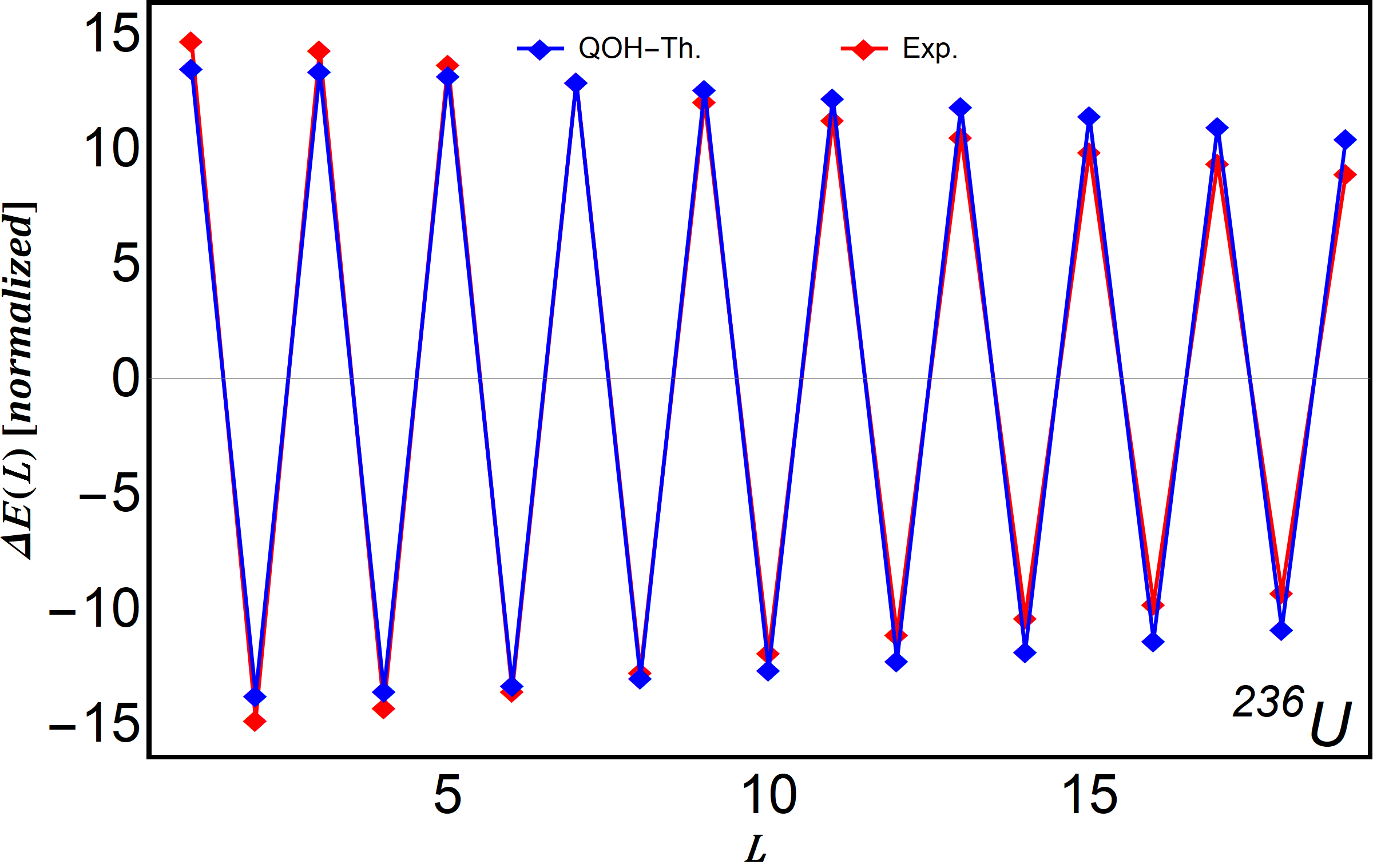}
	\includegraphics[width=.3\textwidth]{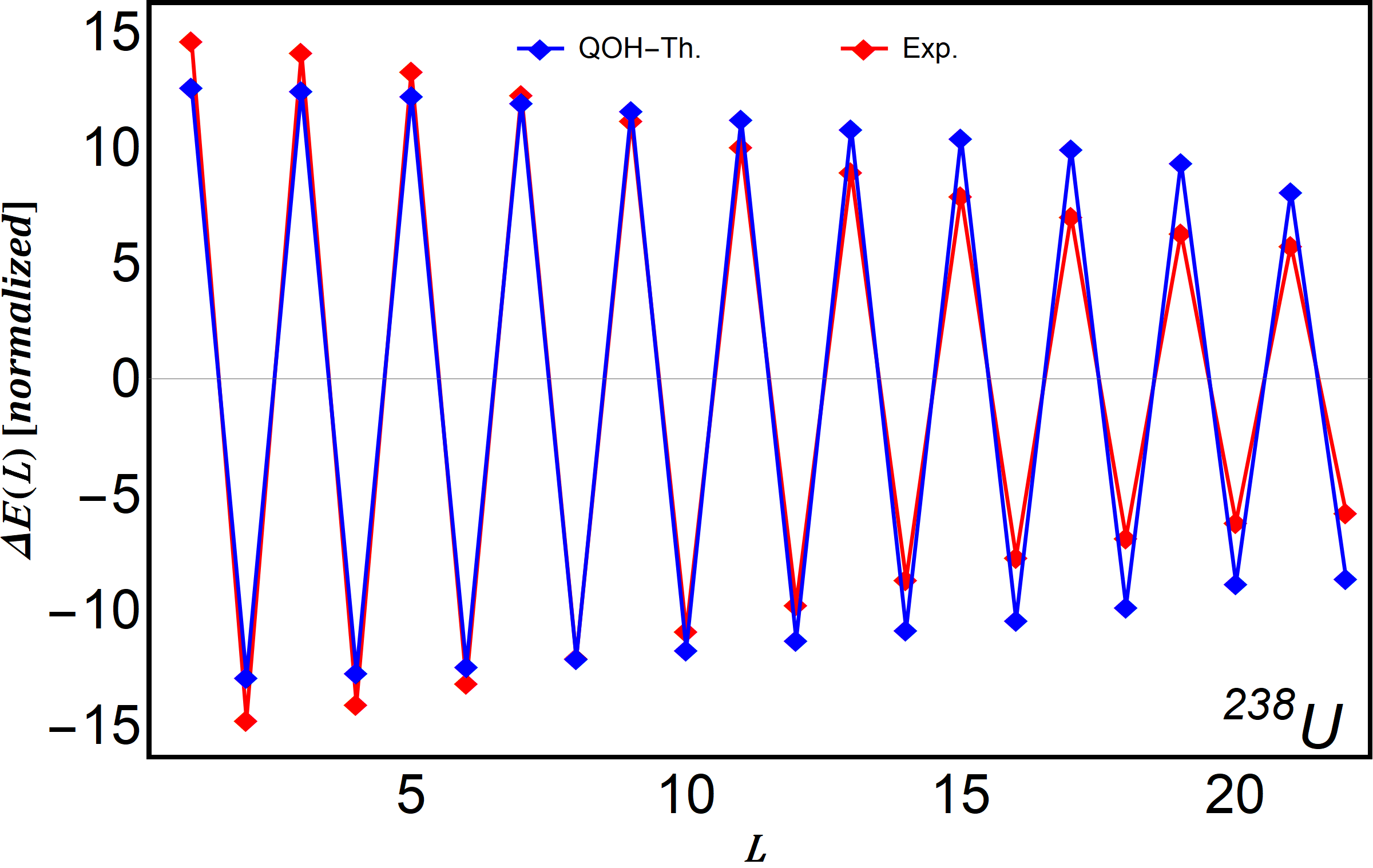}
	\includegraphics[width=.3\textwidth]{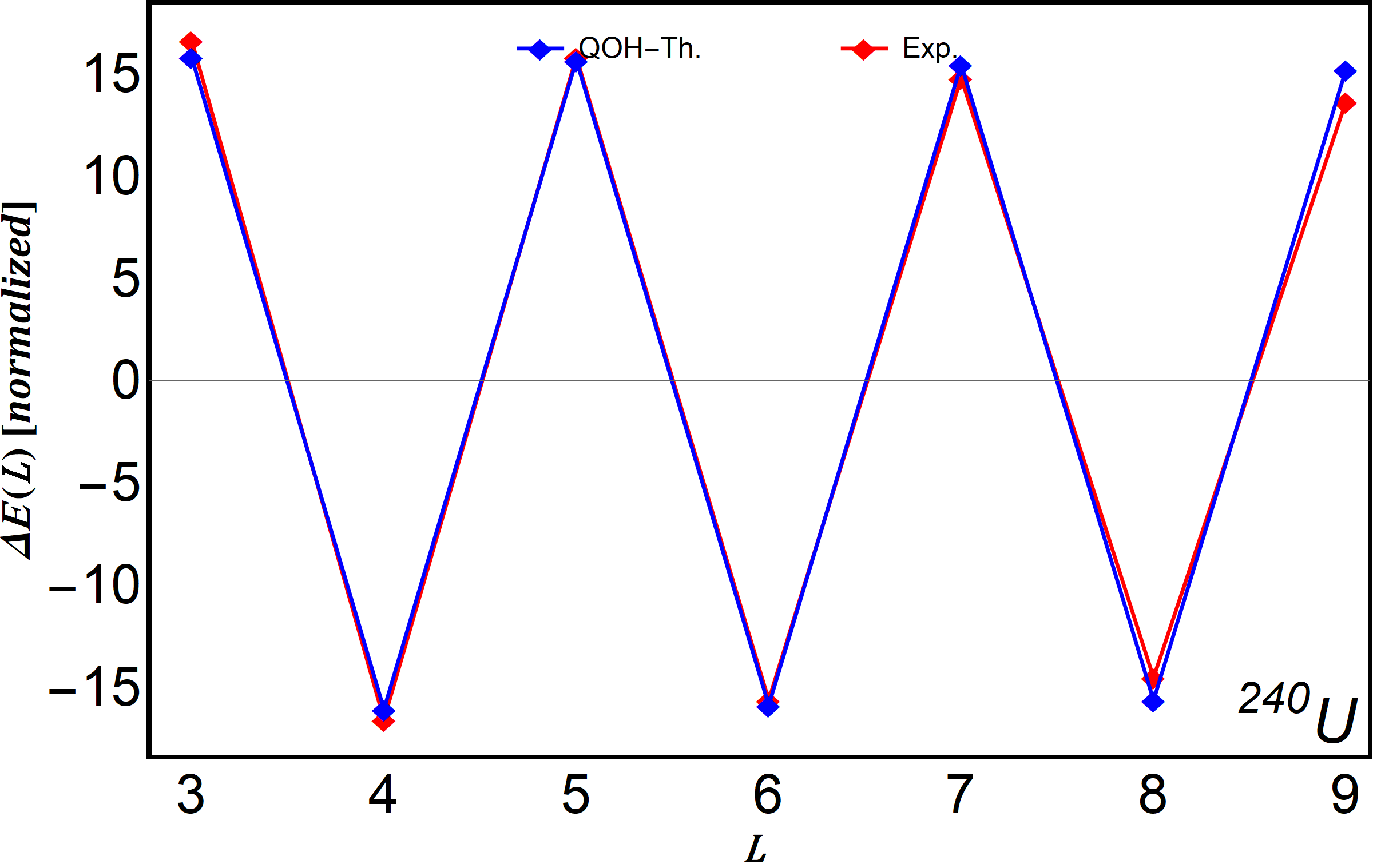}
	\includegraphics[width=.3\textwidth]{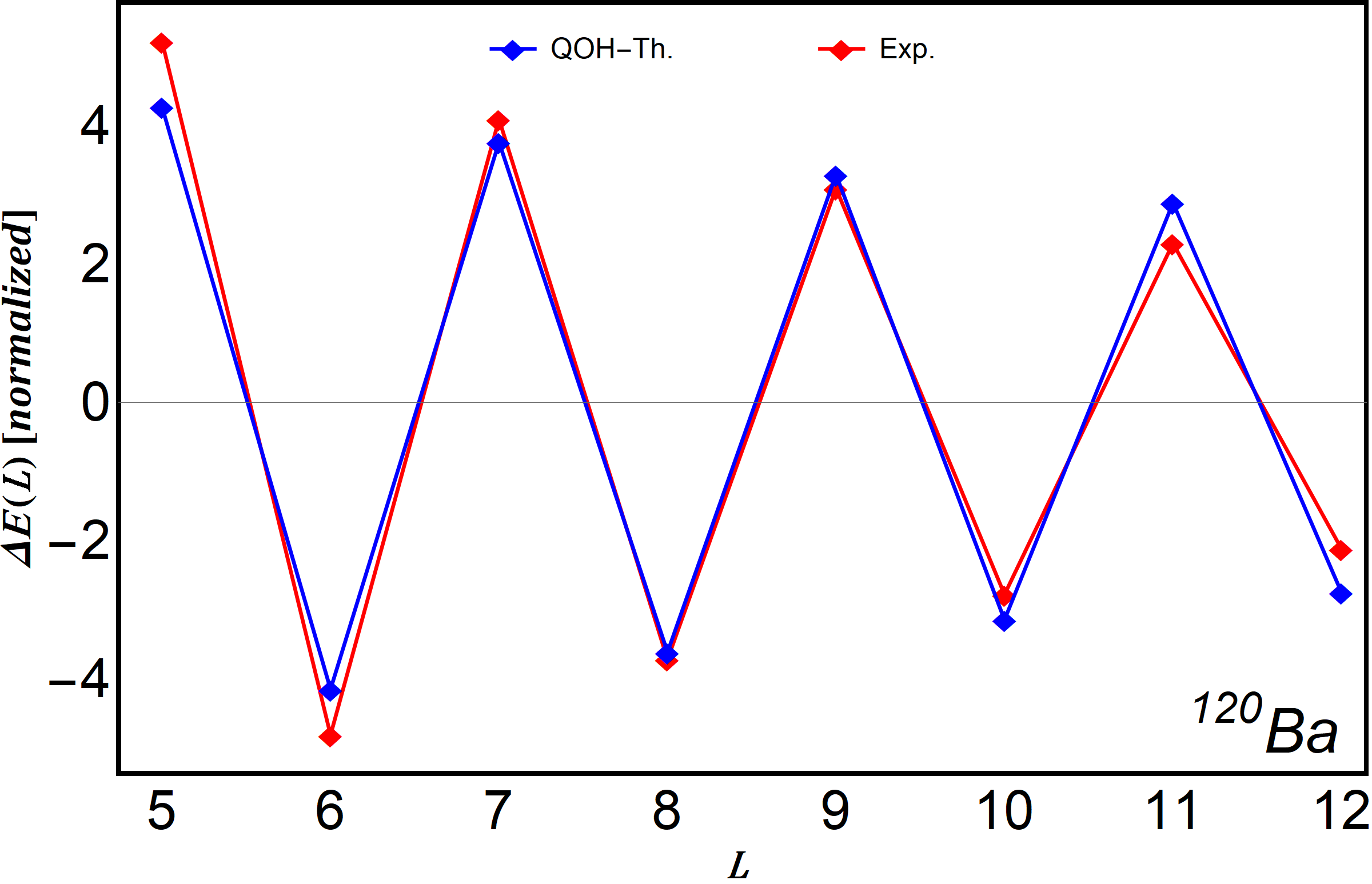}	
	\includegraphics[width=.3\textwidth]{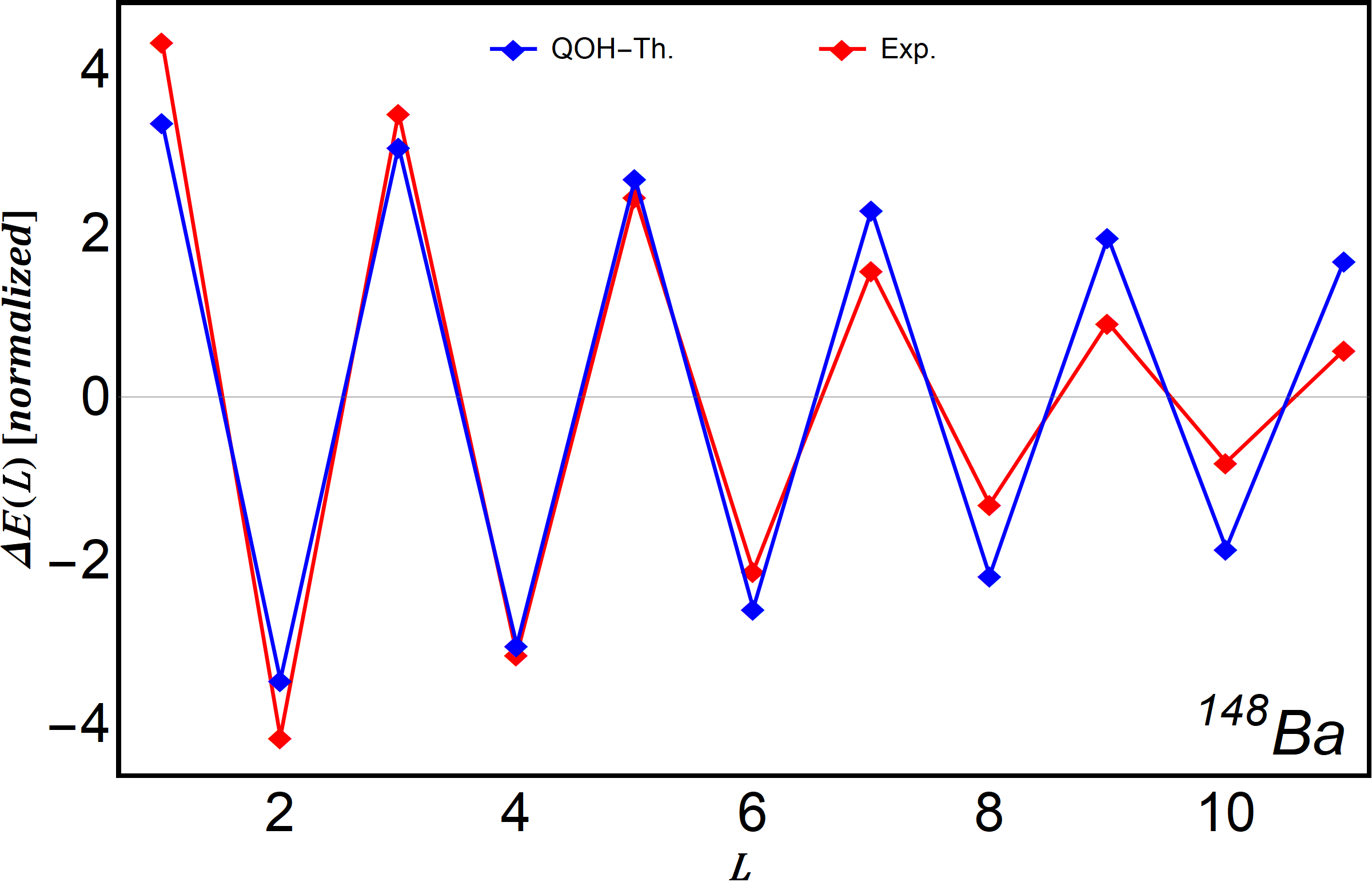}
	\caption{Comparison between Experimental and Theoretical Staggering $\Delta E(L)$ Eq. (\ref{stgg}) for strongly vibrational nuclei ${ }^{230-234} \mathrm{Th}$, ${ }^{230} \mathrm{Ra}$,${ }^{152-154} \mathrm{Sm}$, ${ }^{230-240} \mathrm{U}$, ${ }^{120} \mathrm{Ba}$ and ${ }^{148} \mathrm{Ba}$.}	\label{stgS1}
\end{figure*}

\begin{figure*}
	\includegraphics[width=.3\textwidth]{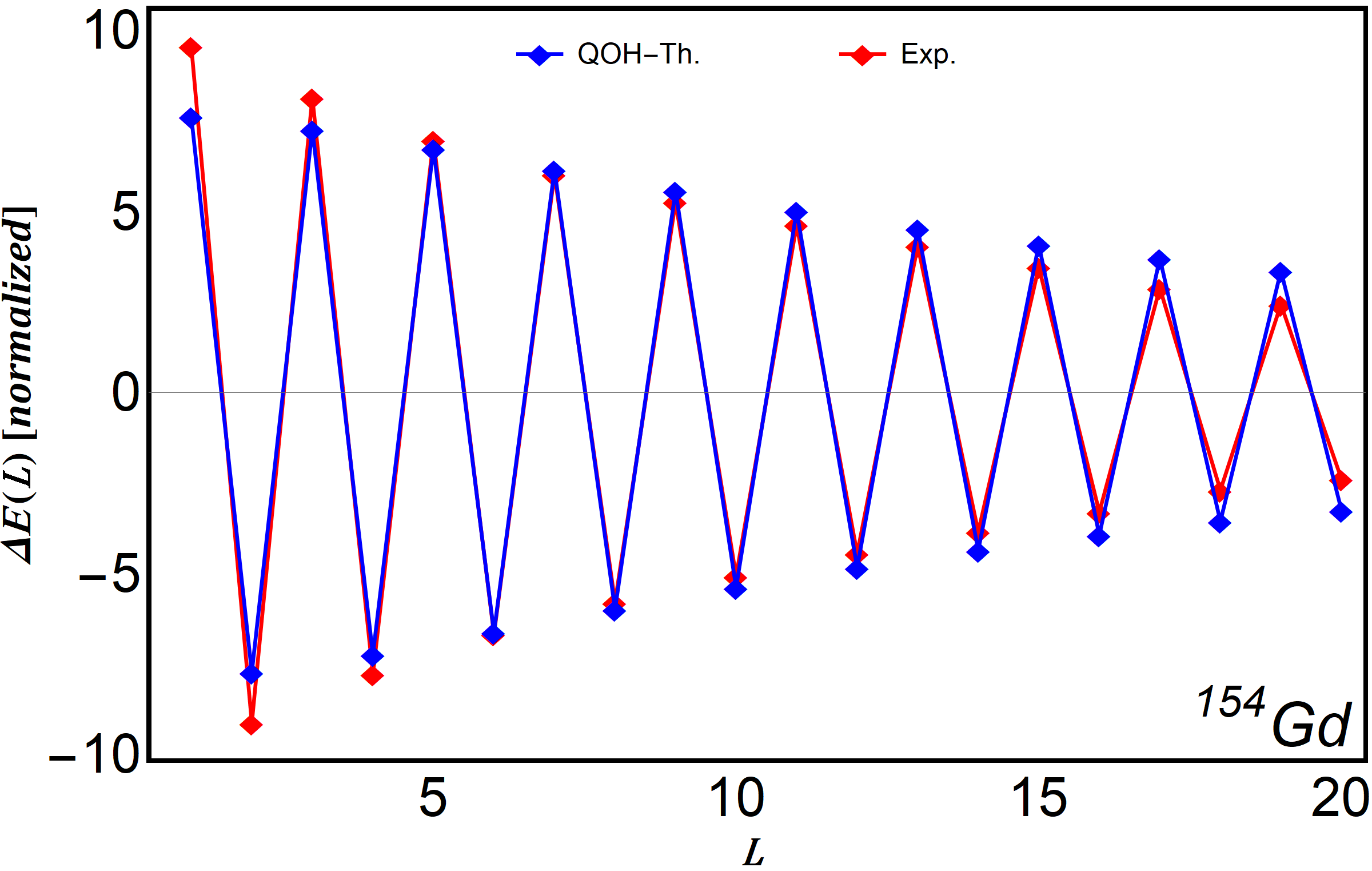}
	\includegraphics[width=.3\textwidth]{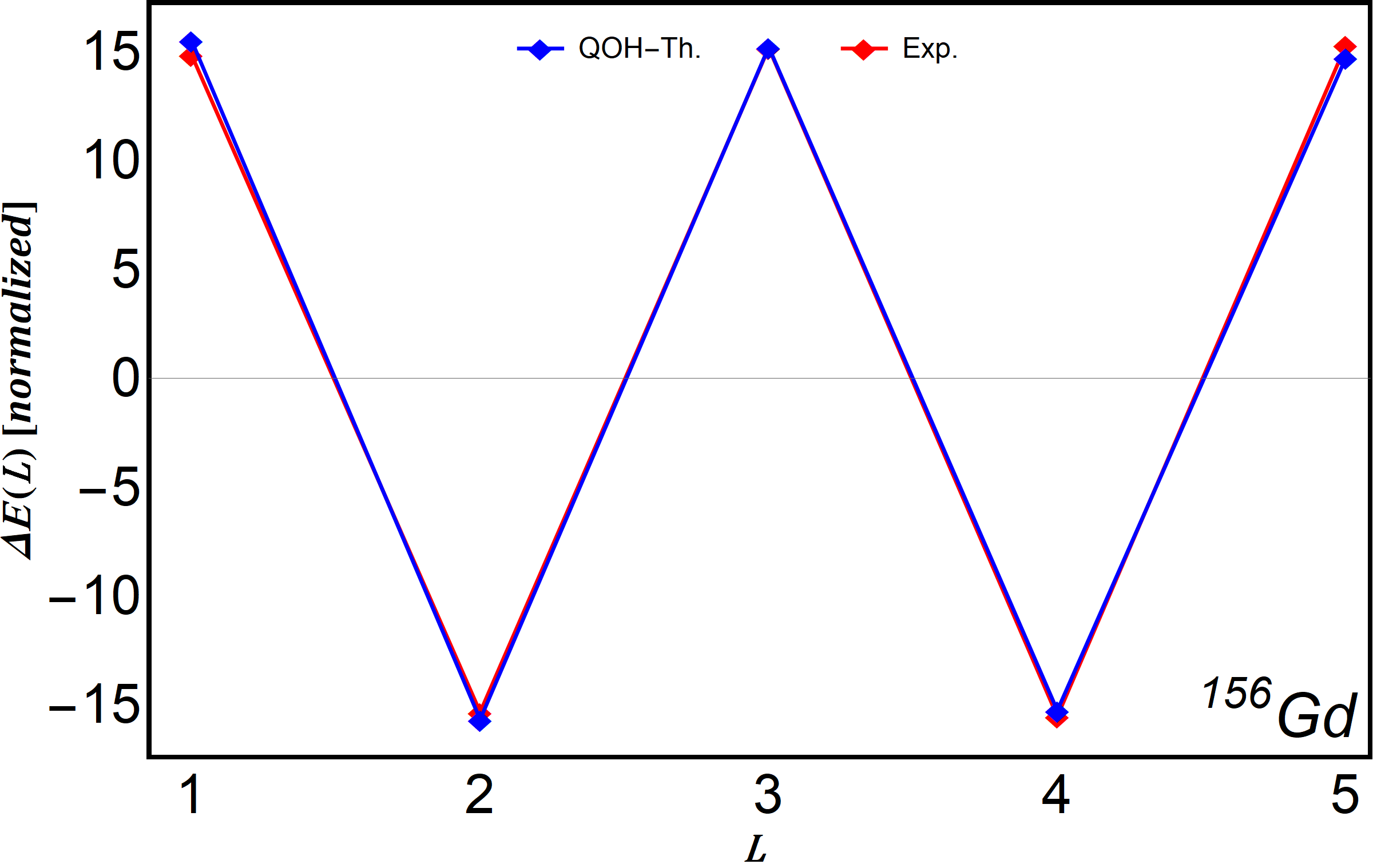}
	\includegraphics[width=.3\textwidth]{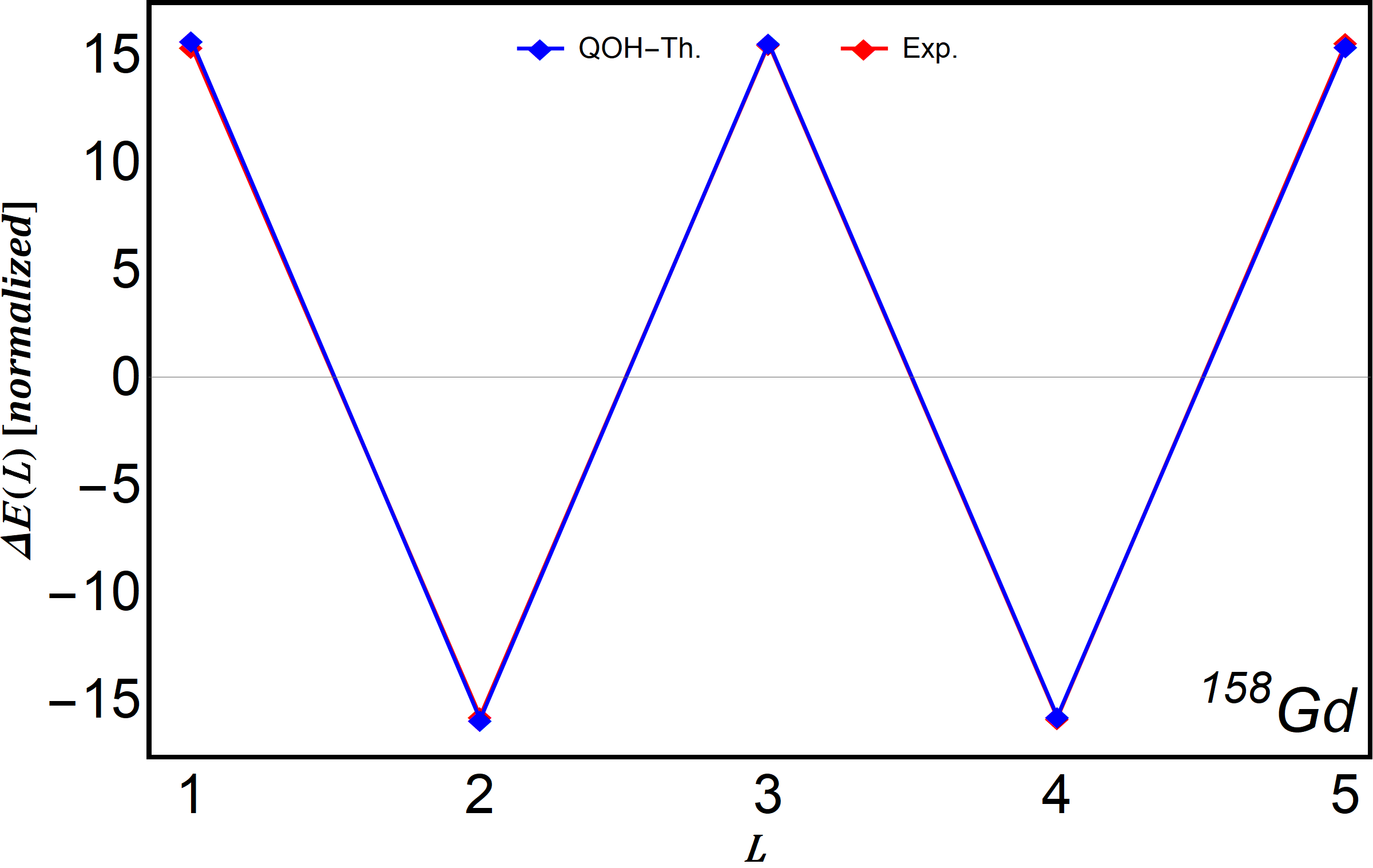}
	
	\includegraphics[width=.3\textwidth]{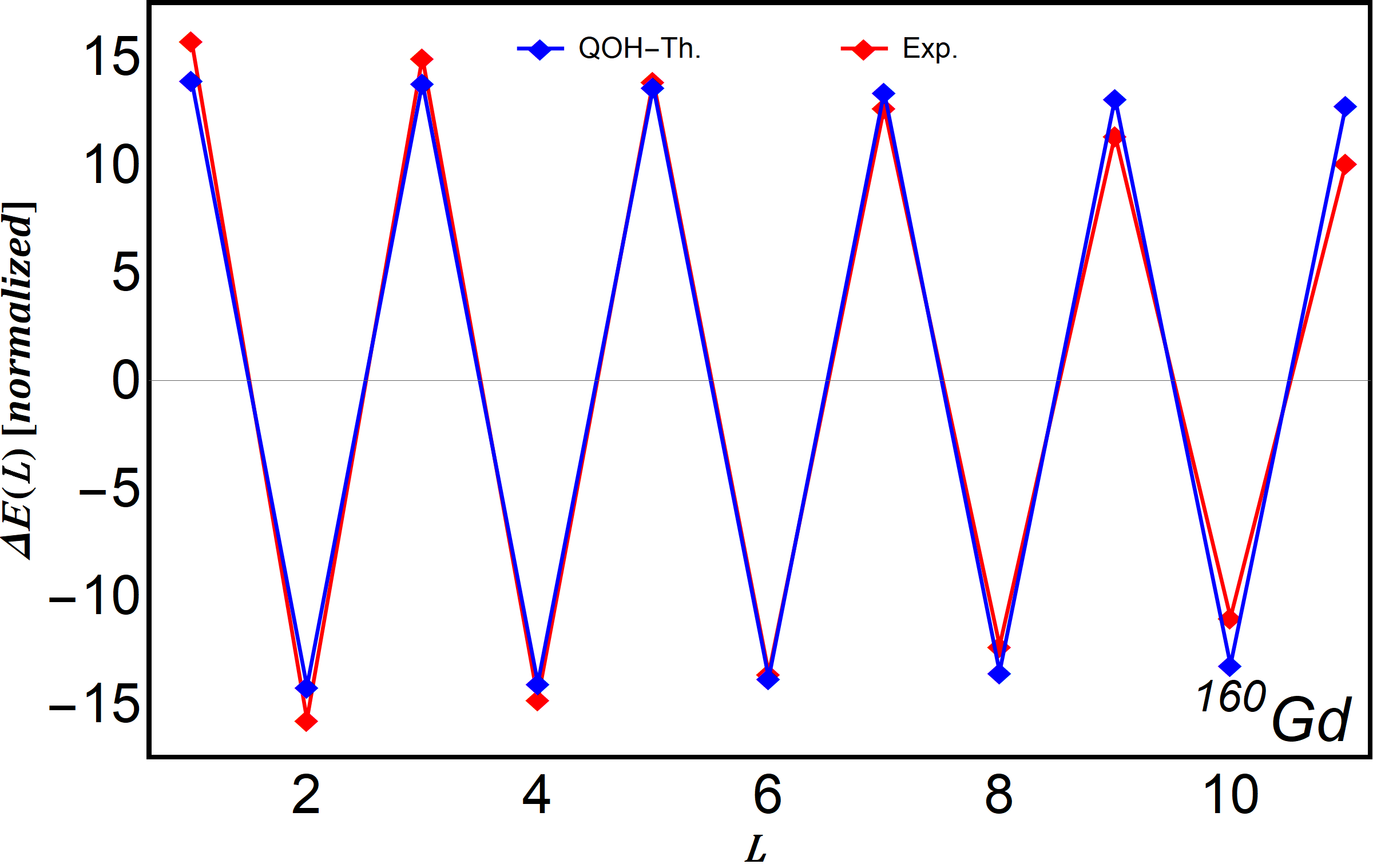}
	\includegraphics[width=.3\textwidth]{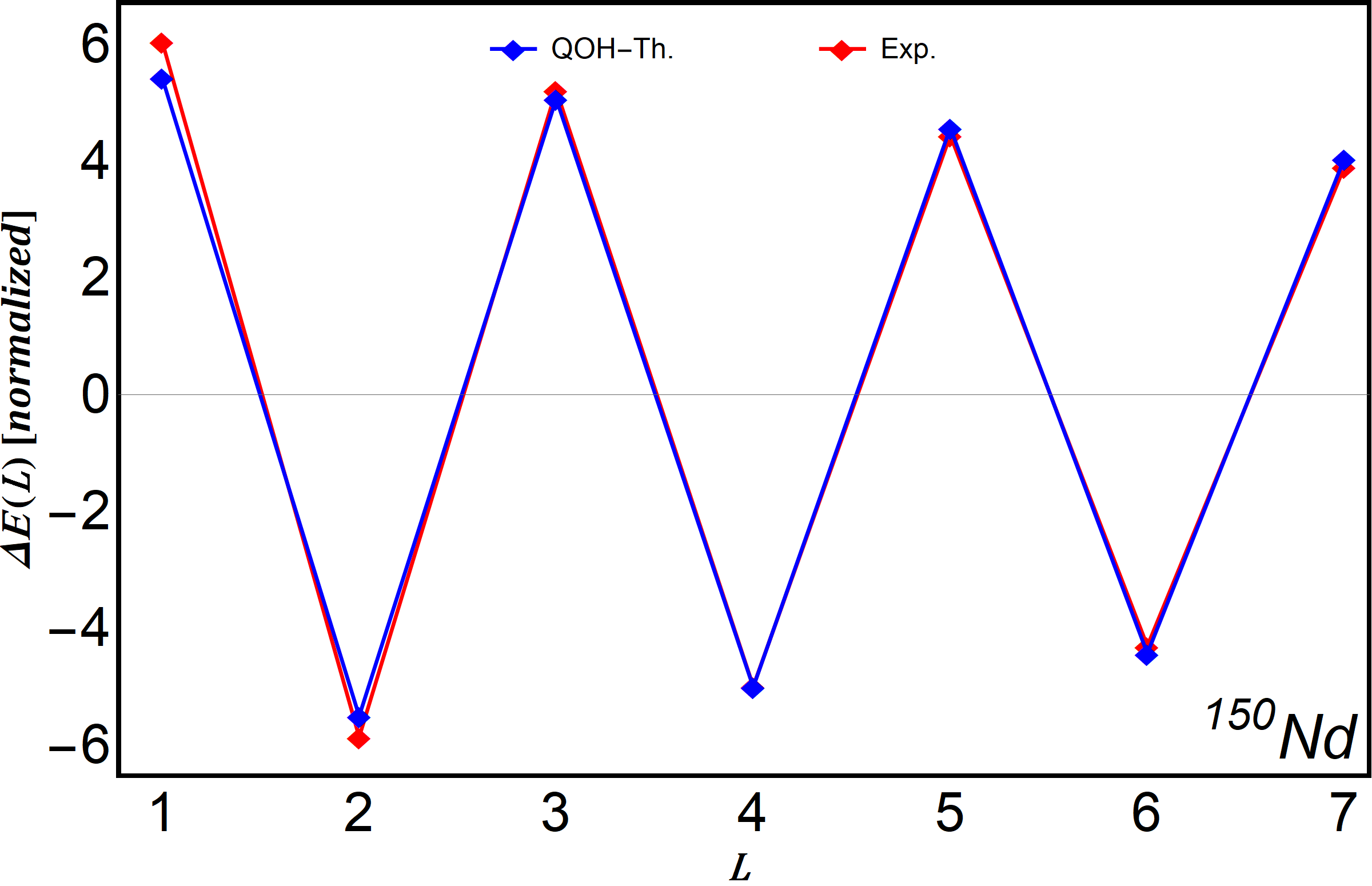}
	\includegraphics[width=.3\textwidth]{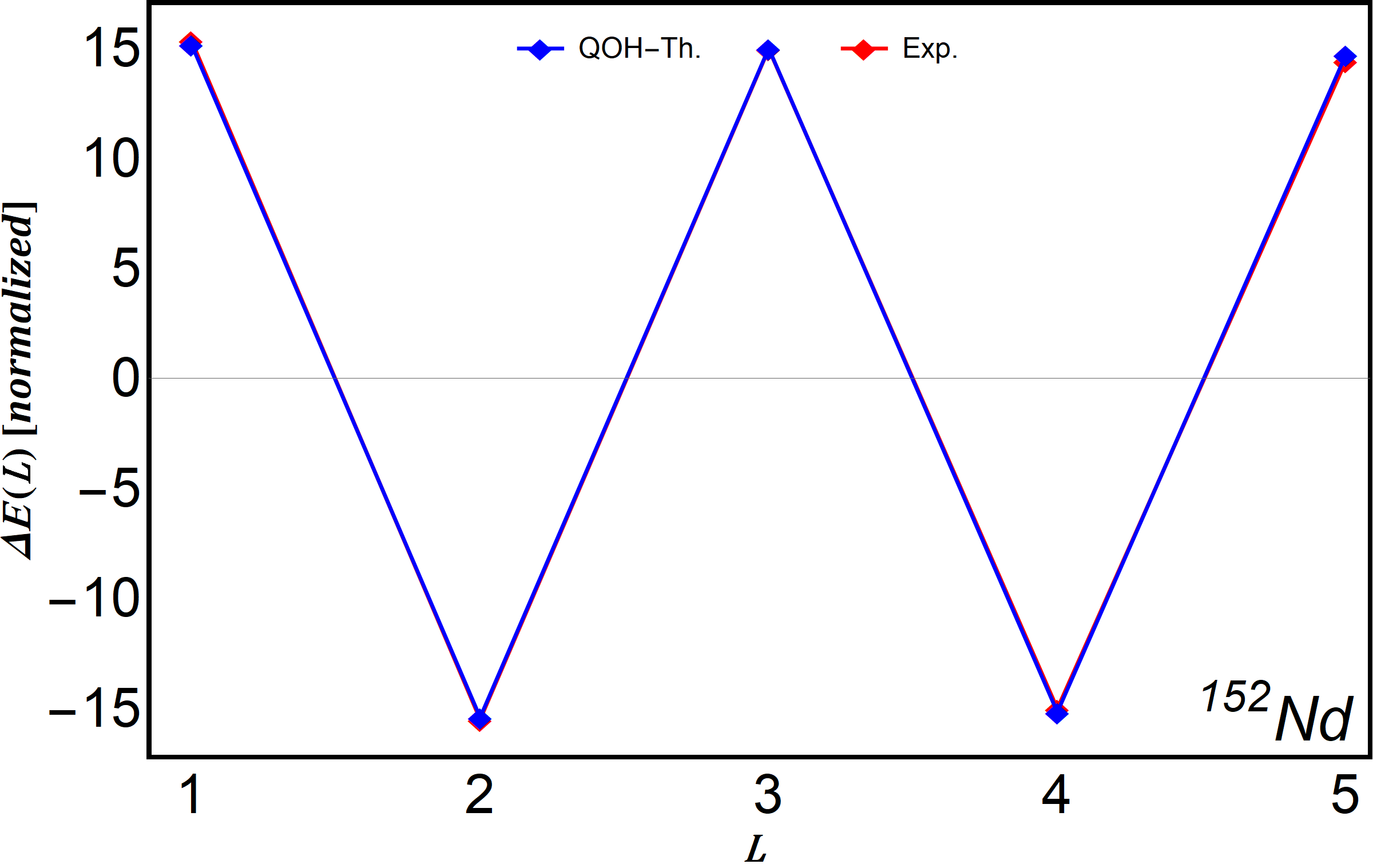}
	
	\includegraphics[width=.3\textwidth]{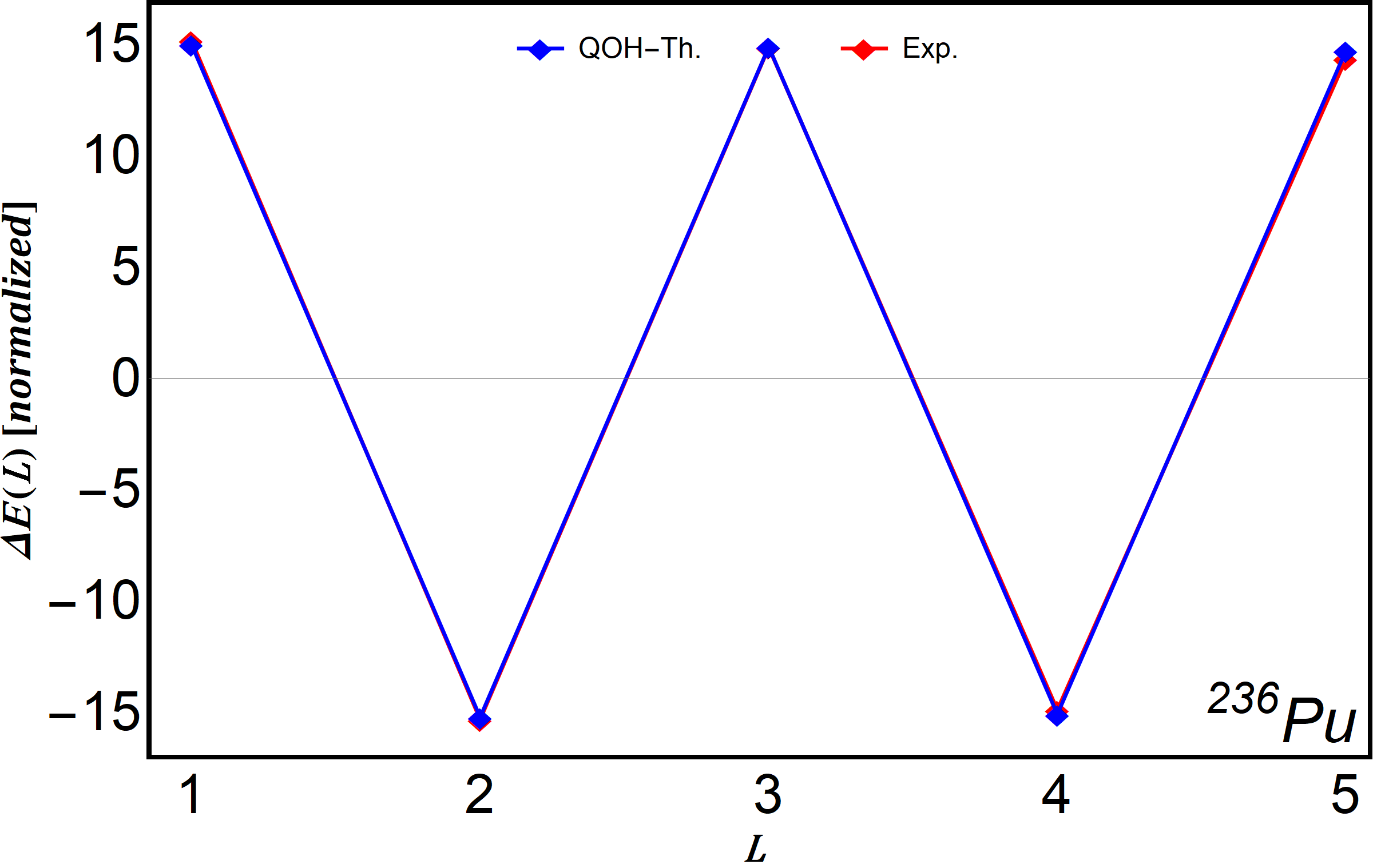}
	\includegraphics[width=.3\textwidth]{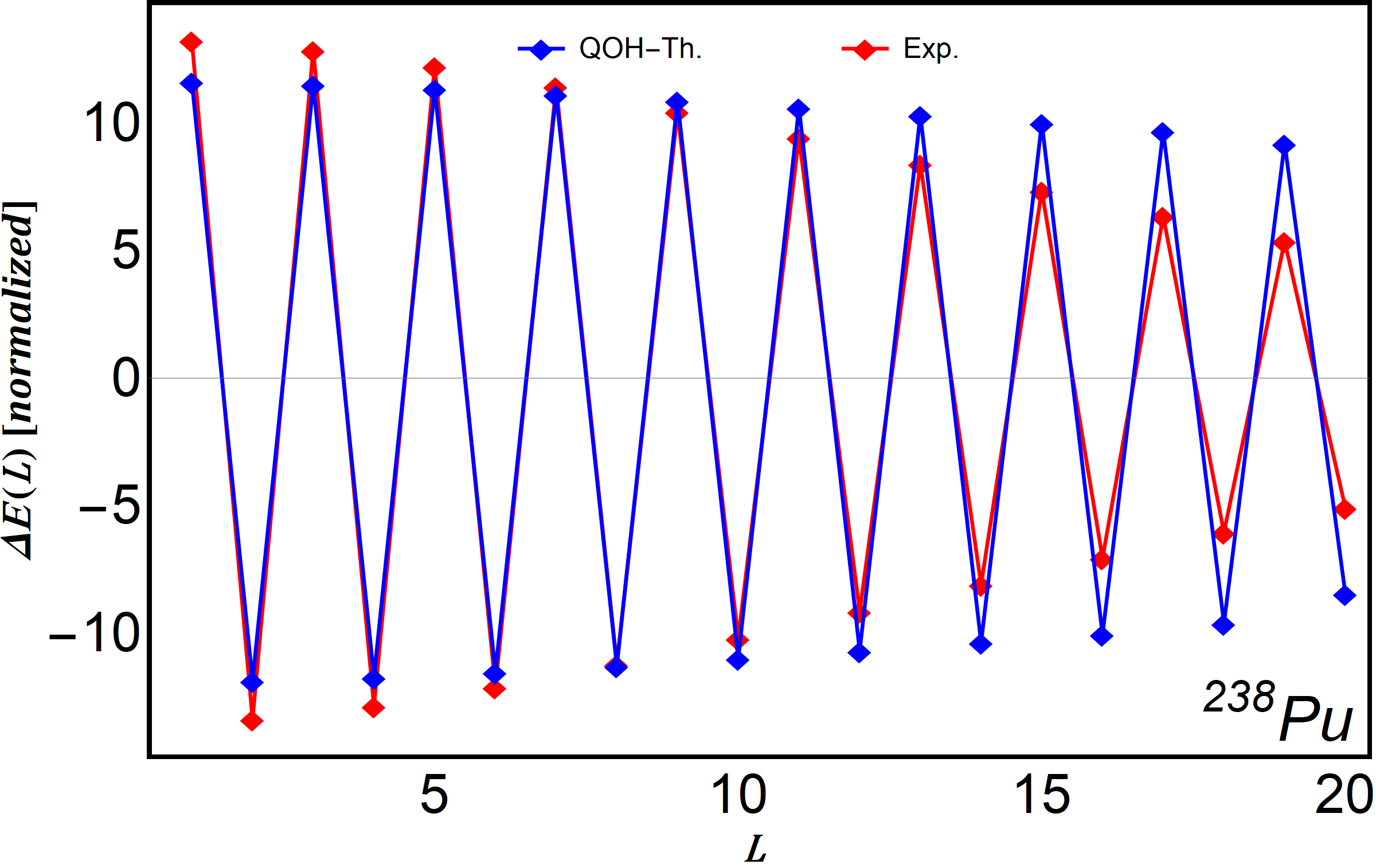}
	\includegraphics[width=.3\textwidth]{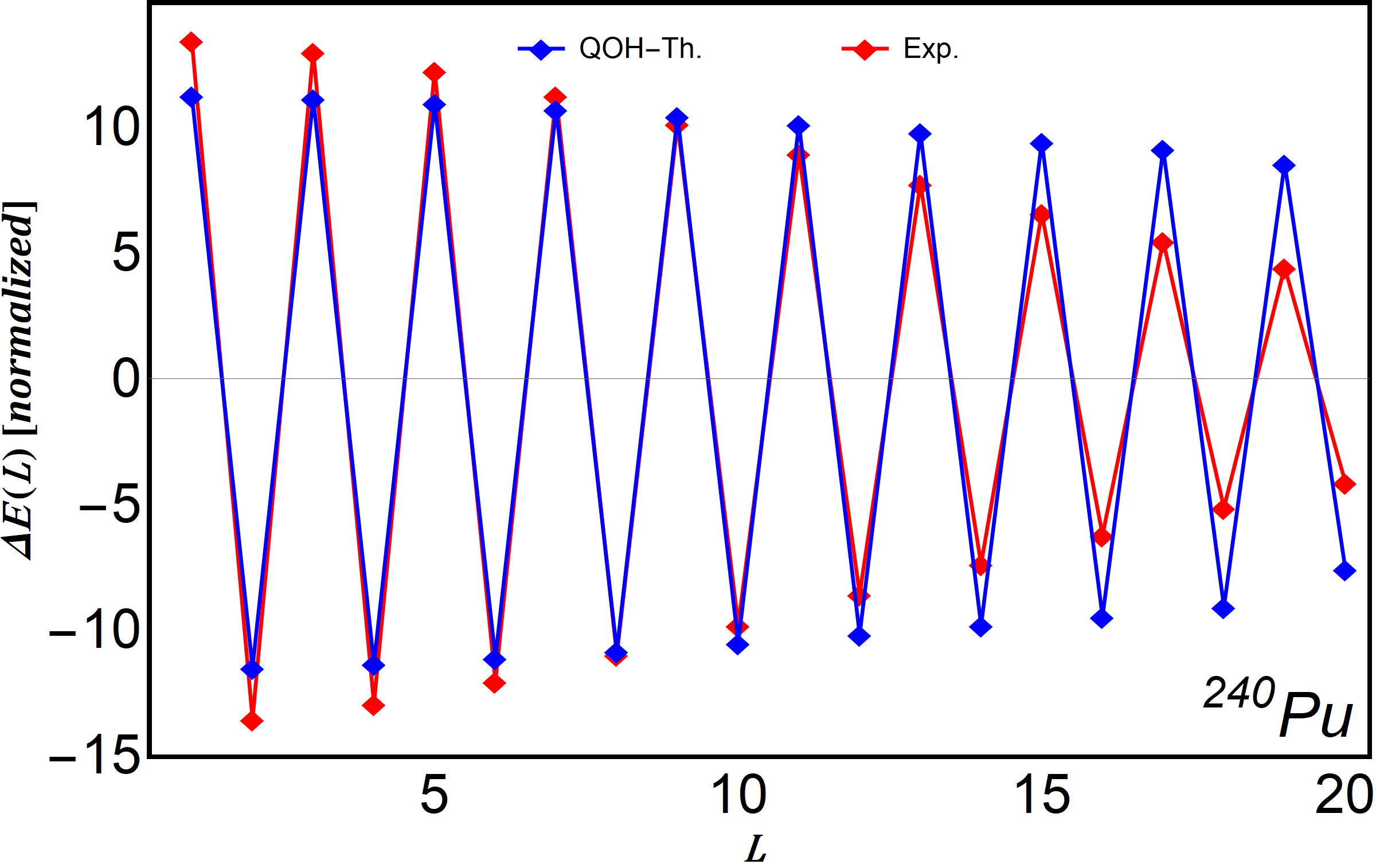}
	
	\includegraphics[width=.3\textwidth]{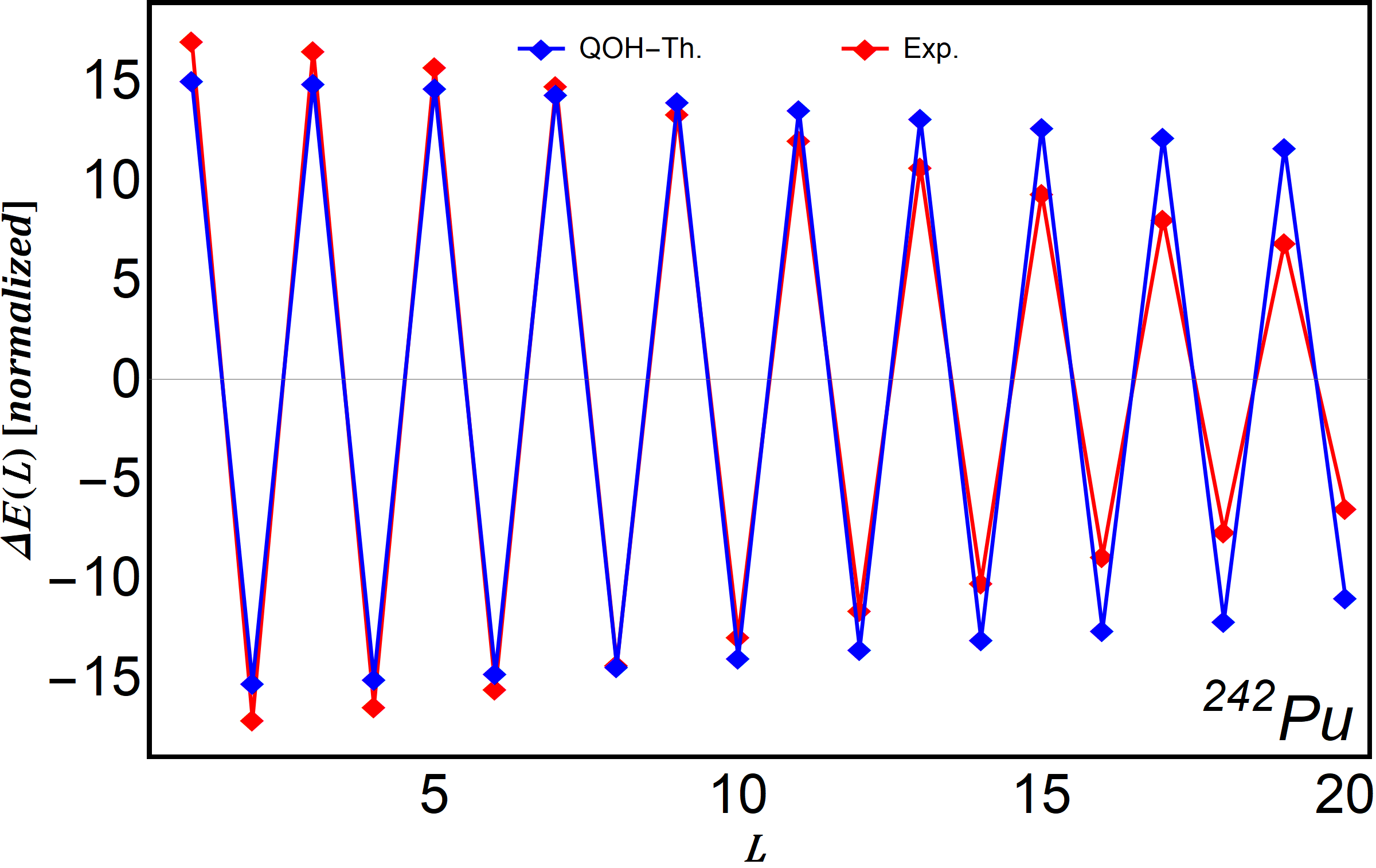}
	\includegraphics[width=.3\textwidth]{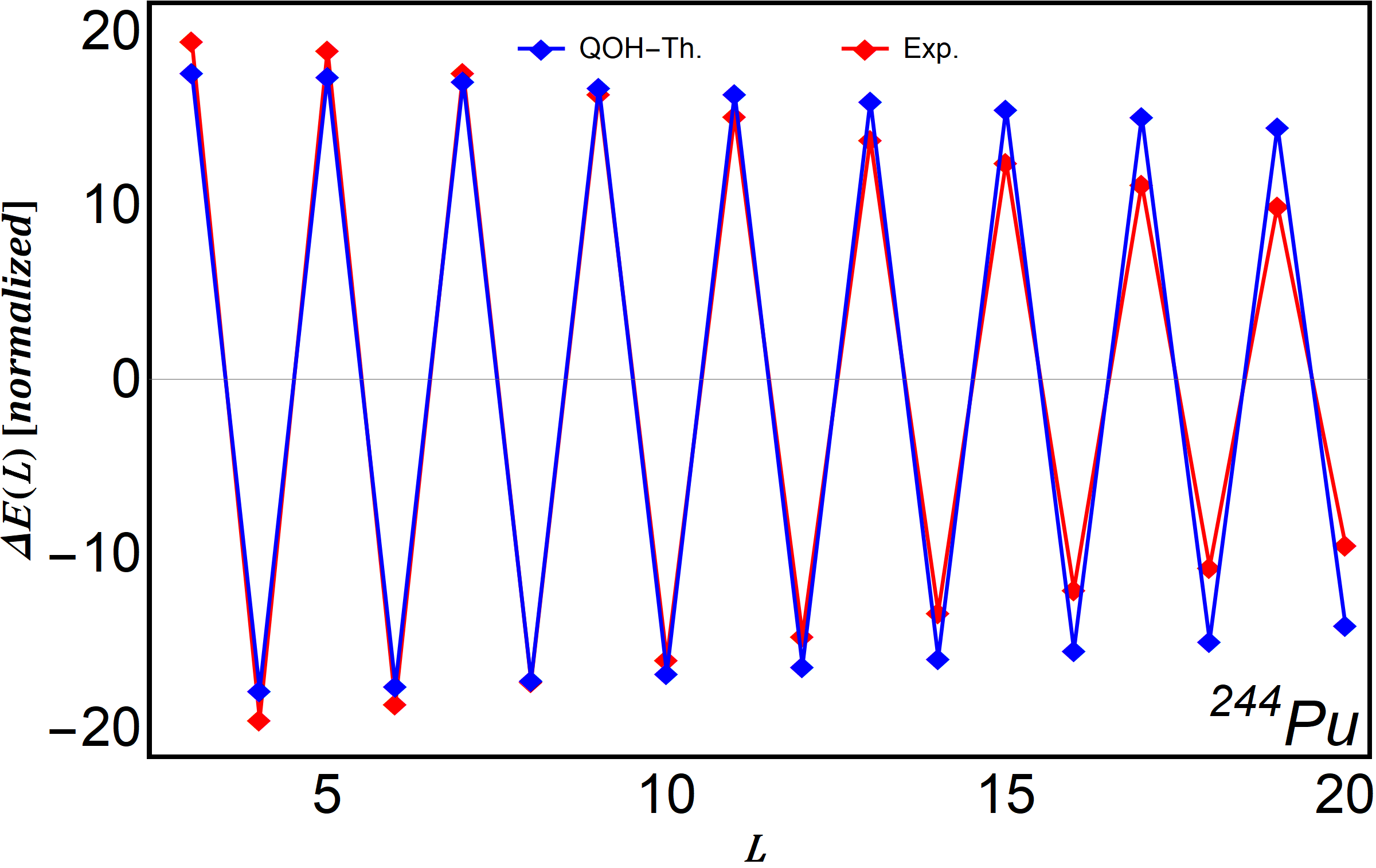}
	
	\includegraphics[width=.3\textwidth]{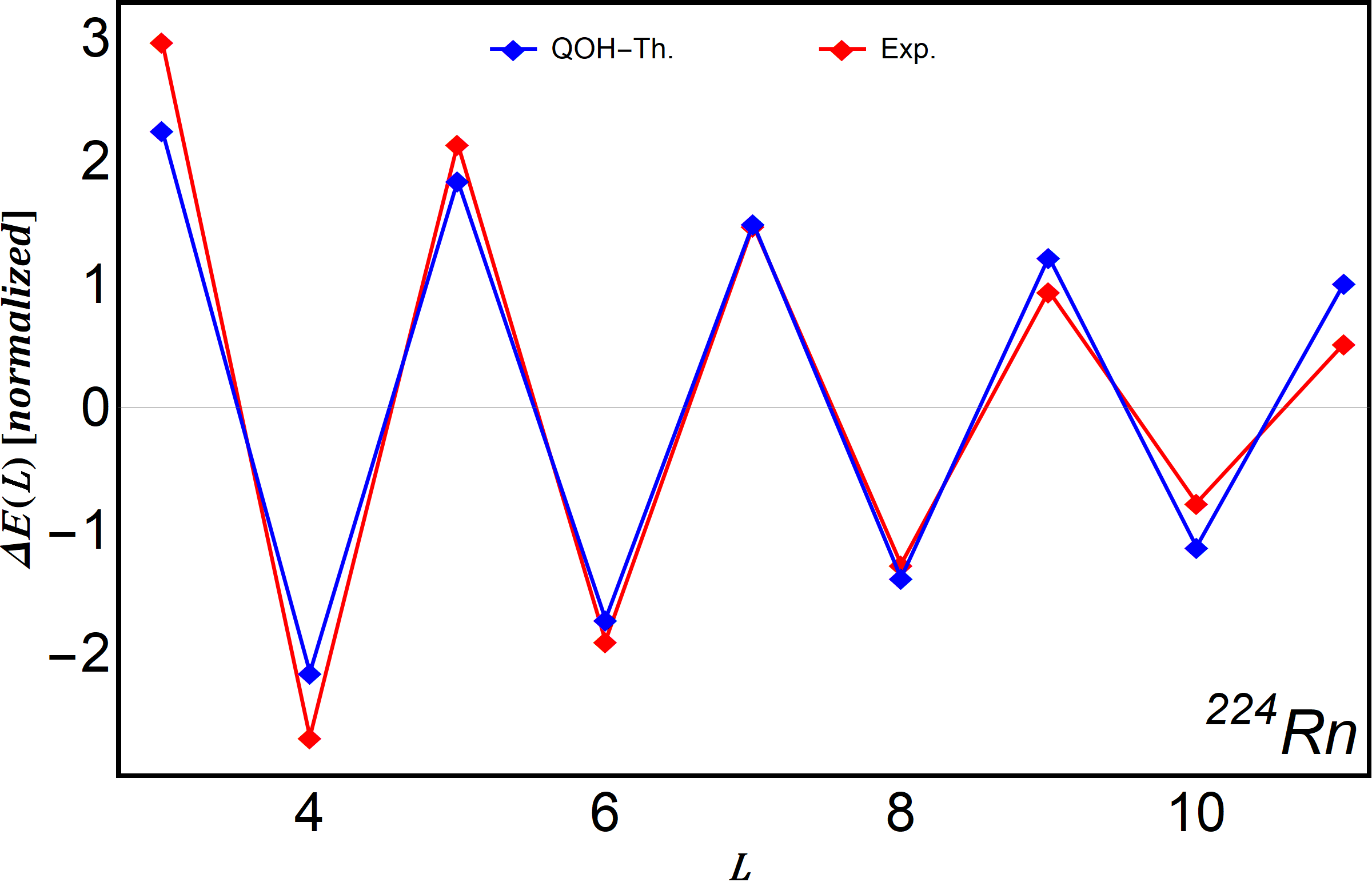}
	\includegraphics[width=.3\textwidth]{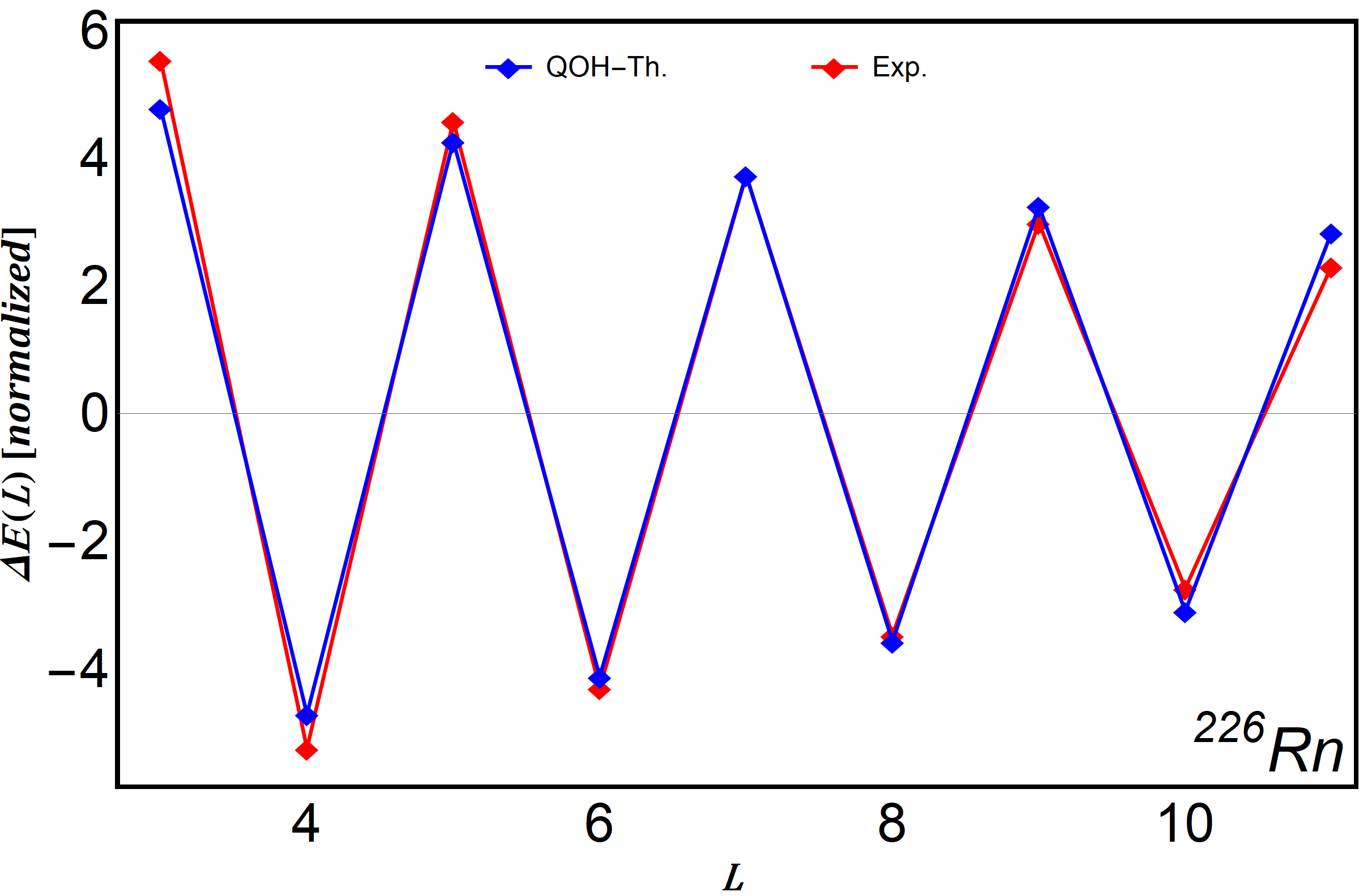}
	\includegraphics[width=.3\textwidth]{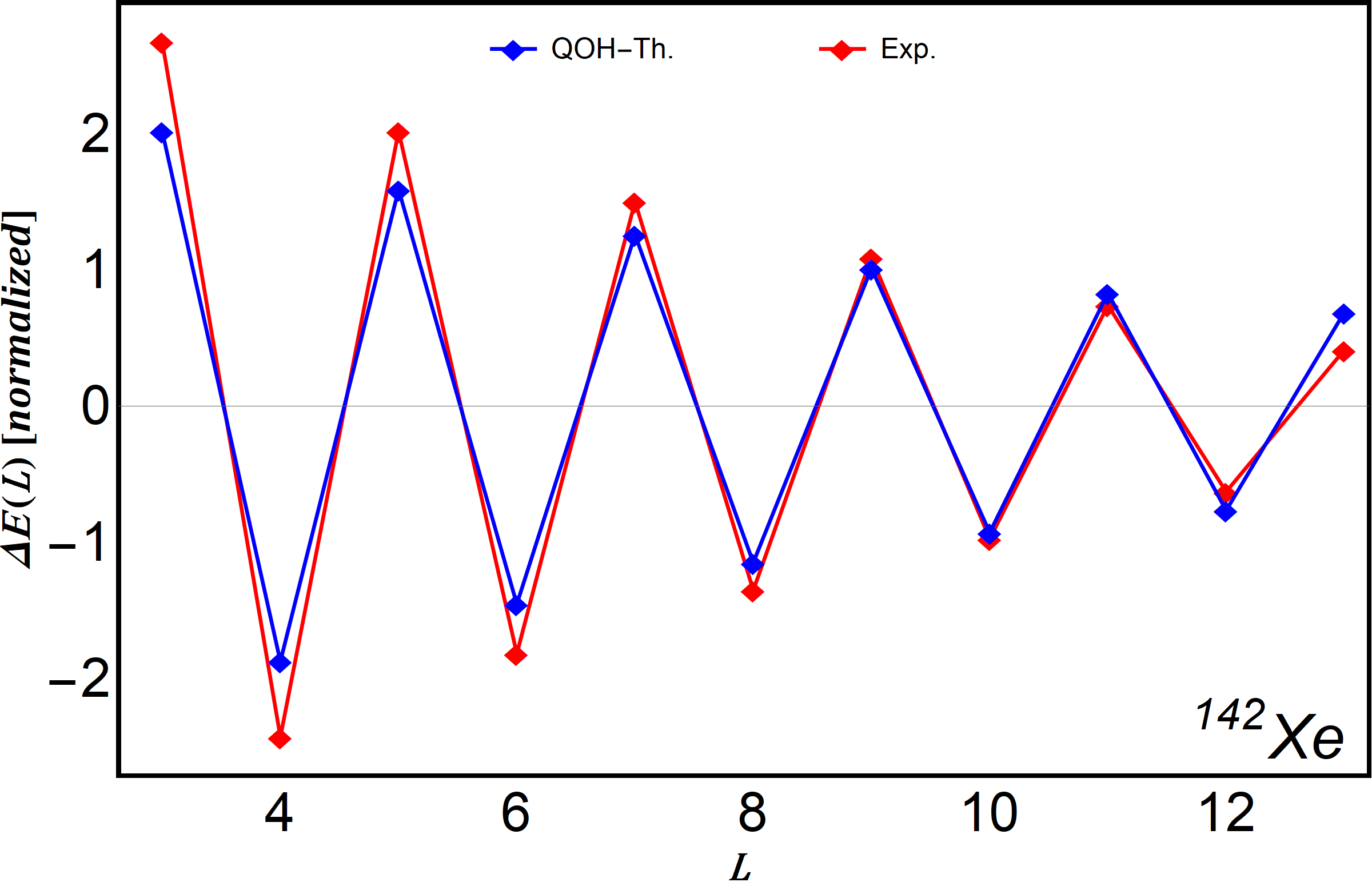}
	\caption{Same as in Fig. (\ref{stgS1}), but for ${ }^{154-160} \mathrm{Gd}$, ${ }^{150-152} \mathrm{Nd}$, ${ }^{236-244} \mathrm{Pu}$, ${ }^{224-226} \mathrm{Rn}$ and ${ }^{142} \mathrm{Xe}$.}\label{stgS2}
\end{figure*}

 In nuclear physics, the odd-even staggering serves as an important tool for exploring the fine structure of alternating parity bands in even-even nuclei. To evaluate the capacity of our model to reproduce its pattern in the nuclei under study, we use the following formula \cite{b38}:
\begin{equation}
	\Delta E(L)=E(L)-\frac{E(L-1)+E(L+1)}{2}\label{stgg} 
\end{equation}
The comparison between experimental and theoretical patterns reveals that our model successfully reproduces the staggering oscillations observed in the strongly vibrational nuclei considered in this study, including ${}^{230}$Ra, ${}^{230-234}$Th, ${}^{152-154}$Sm, ${}^{120}$Ba, ${}^{148}$Ba, ${}^{230-240}$U, ${}^{154-160}$Gd, ${}^{150-152}$Nd, ${}^{224-226}$Rn, ${}^{236-244}$Pu, and ${}^{142}$Xe. These nuclei exhibit a large staggering amplitude, as illustrated in Figs. (\ref{stgS1}) and (\ref{stgS2}), which reflects the pronounced energy gap between the ground state and negative parity bands.
\par
\begin{figure*}
	\includegraphics[width=.3\textwidth]{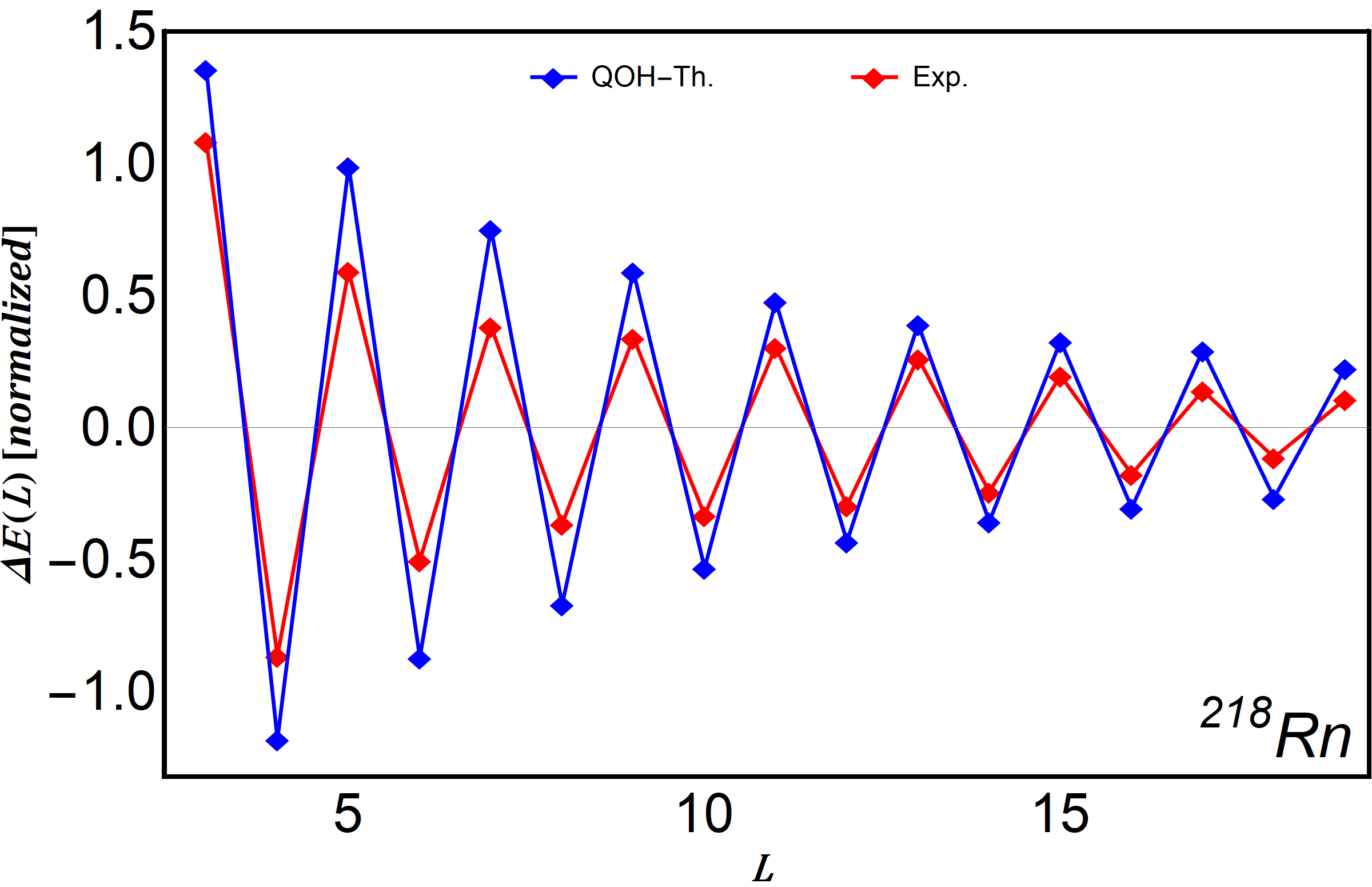}
	\includegraphics[width=.3\textwidth]{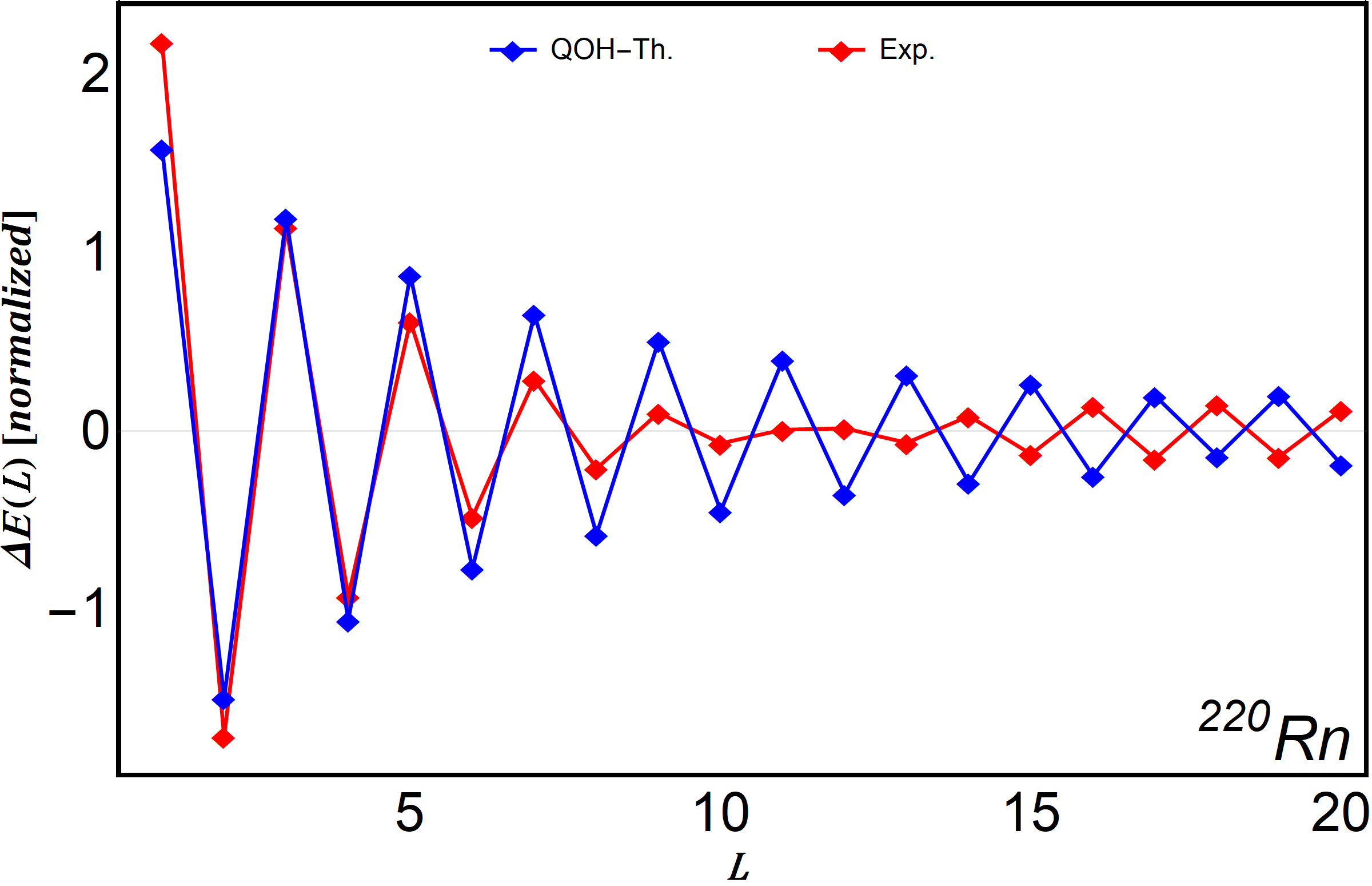}
	\includegraphics[width=.3\textwidth]{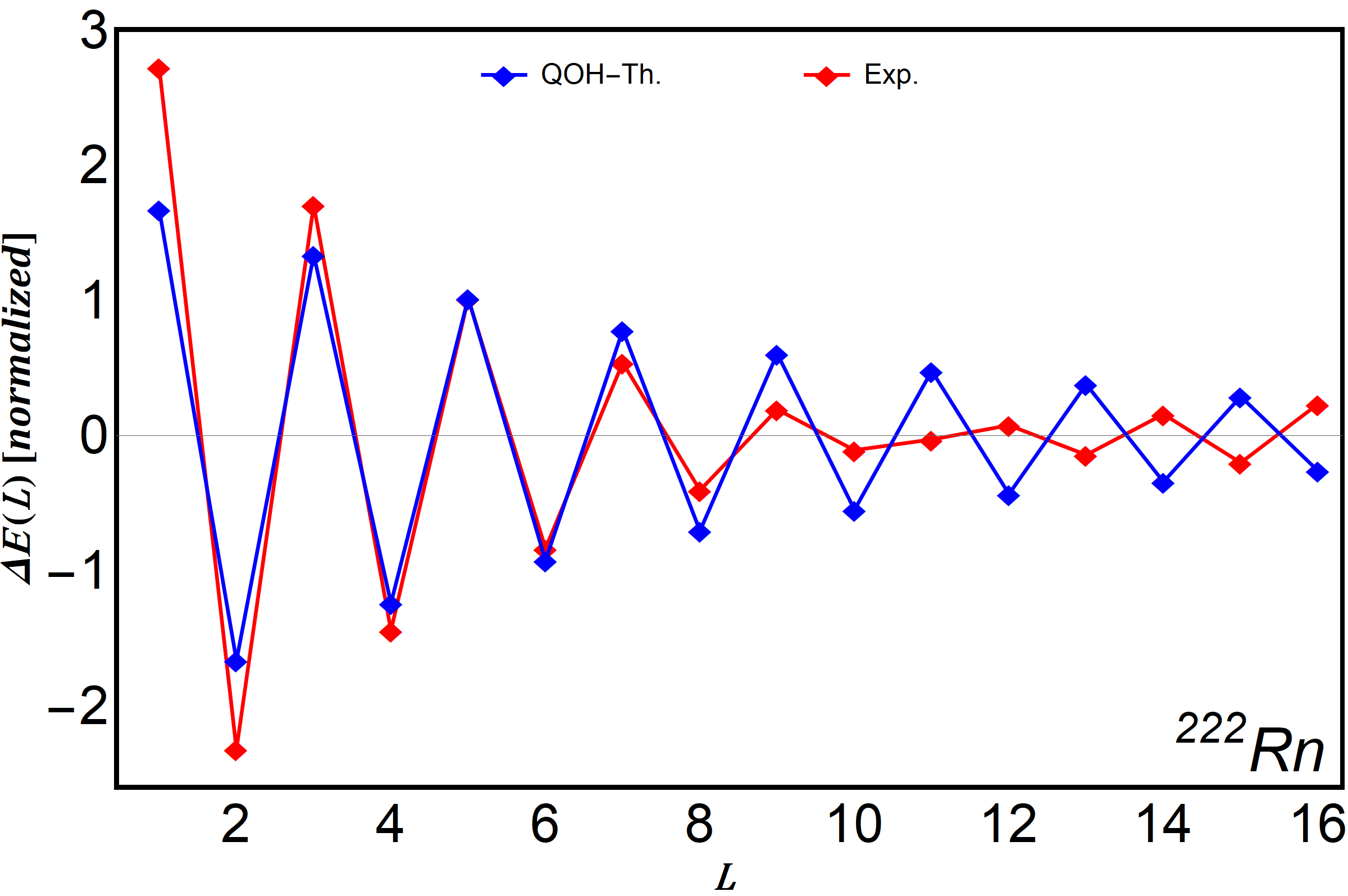}
	
	\includegraphics[width=.3\textwidth]{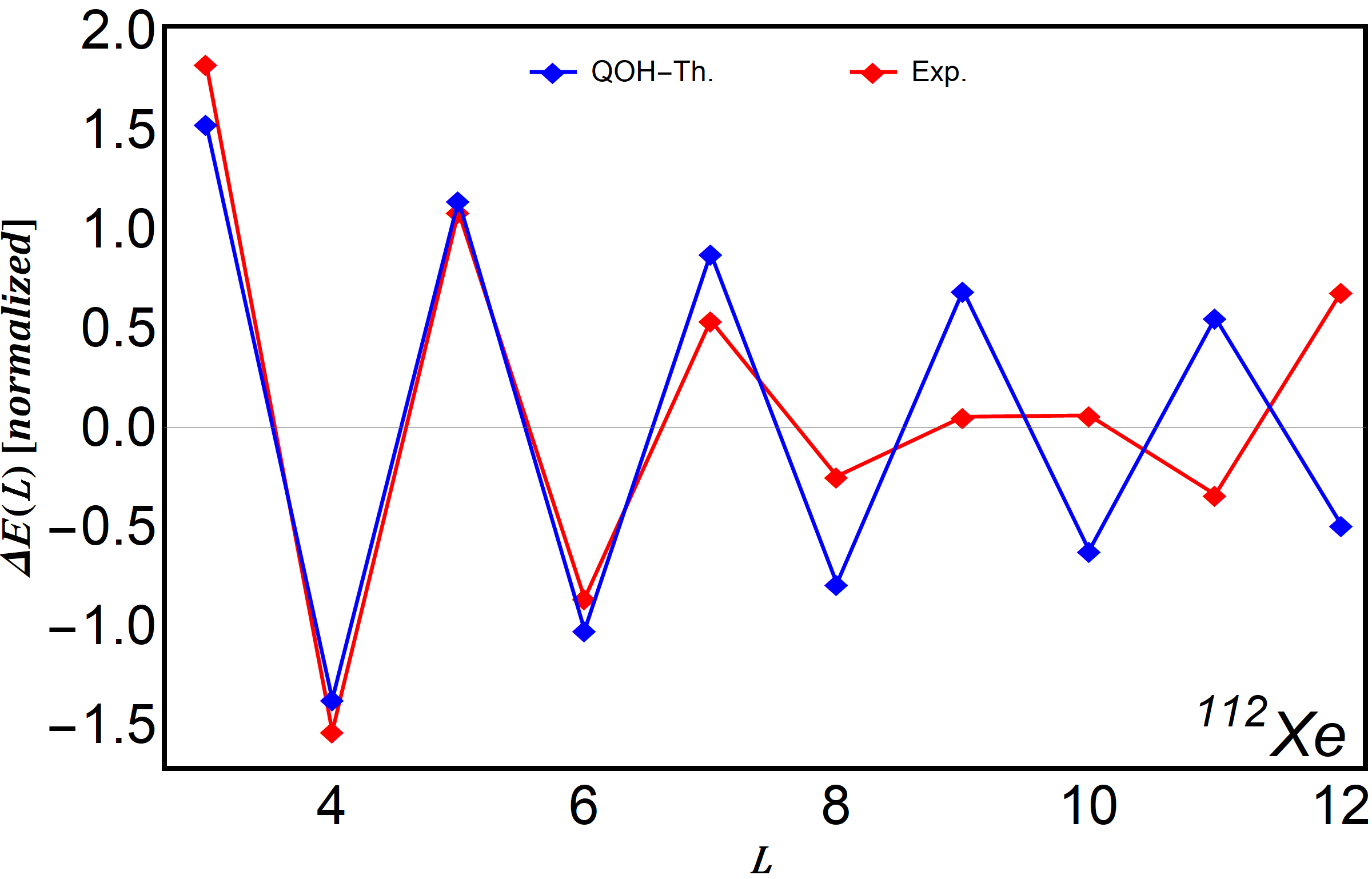}
	\includegraphics[width=.3\textwidth]{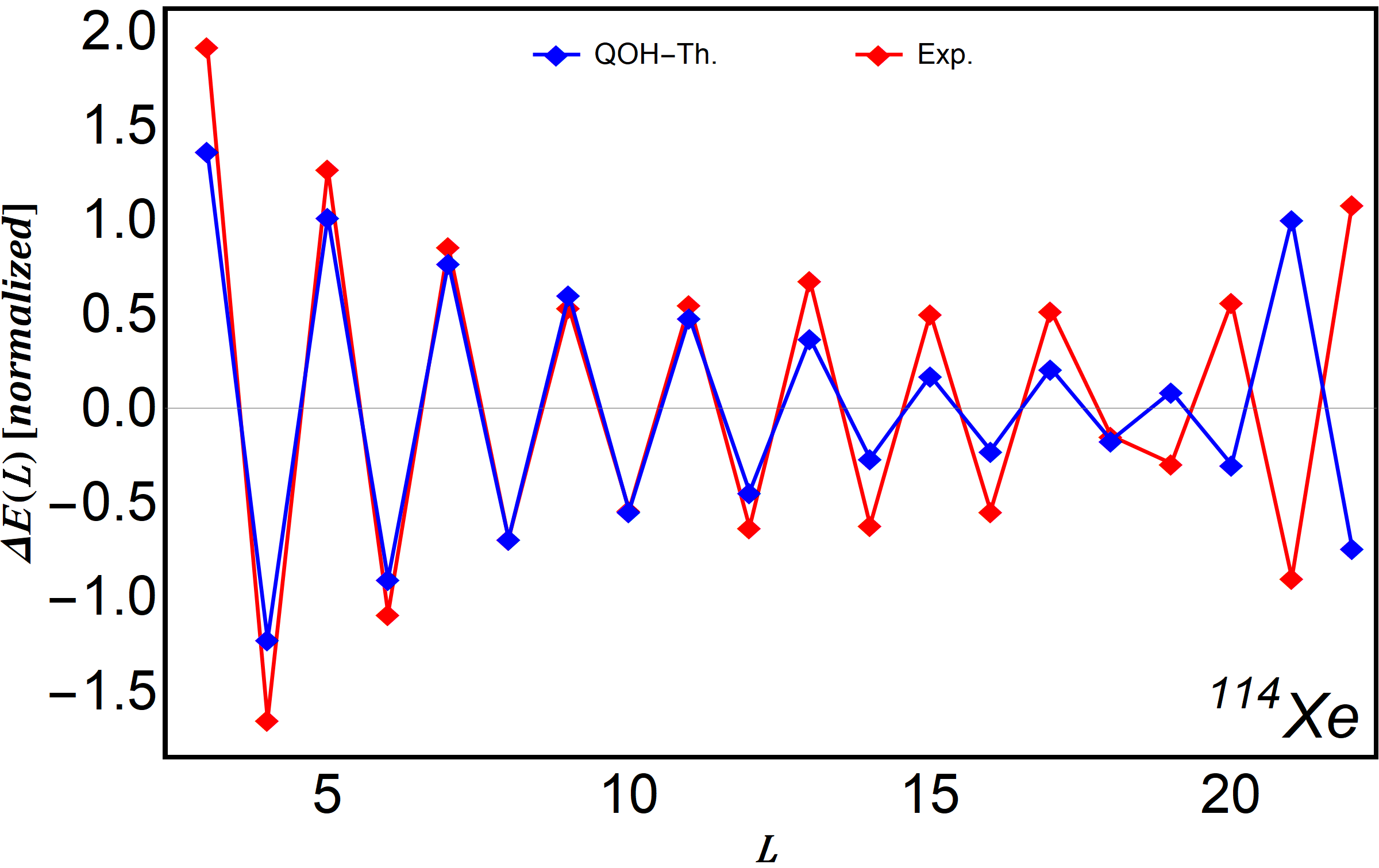}
	\includegraphics[width=.3\textwidth]{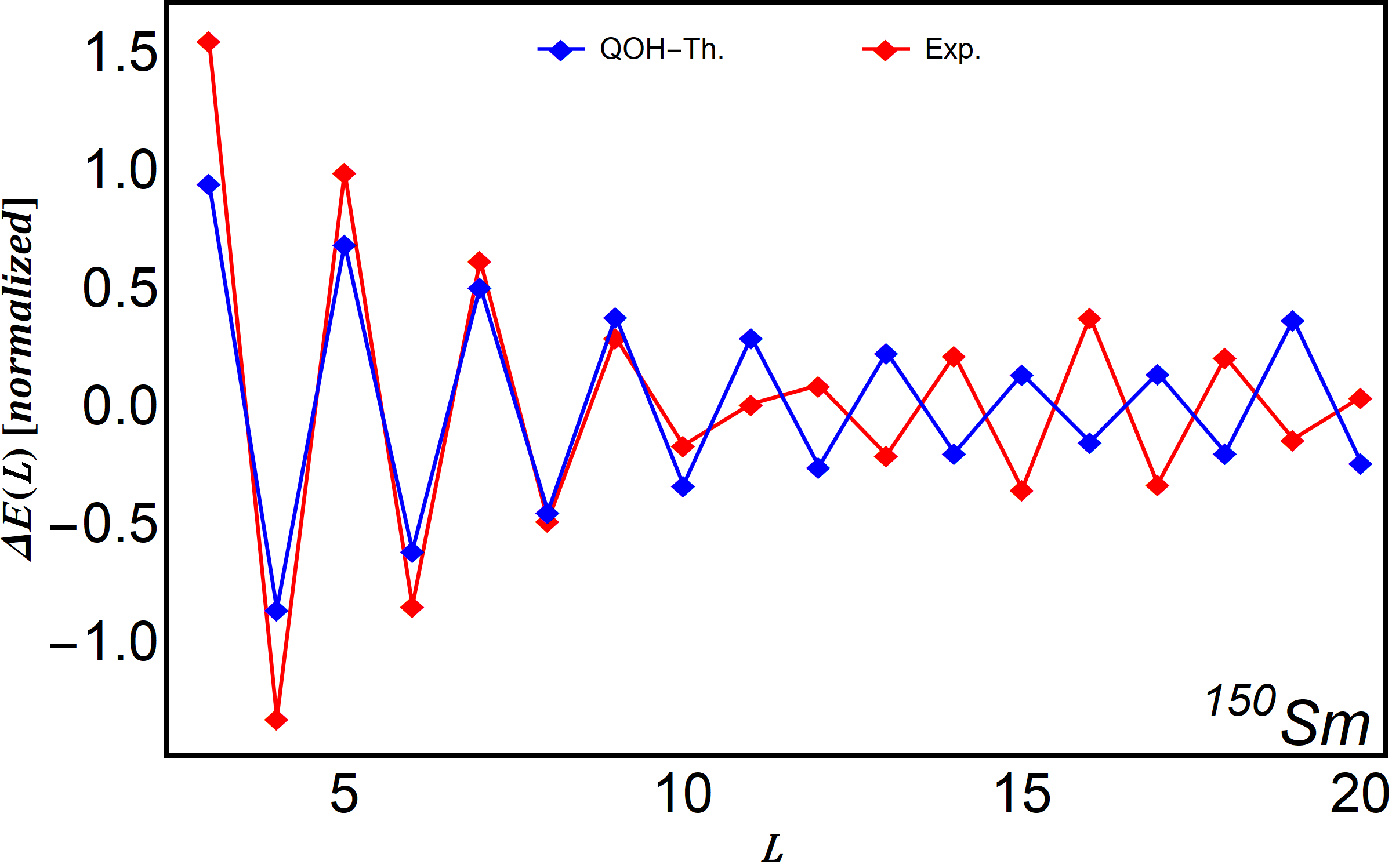}
	
	\includegraphics[width=.3\textwidth]{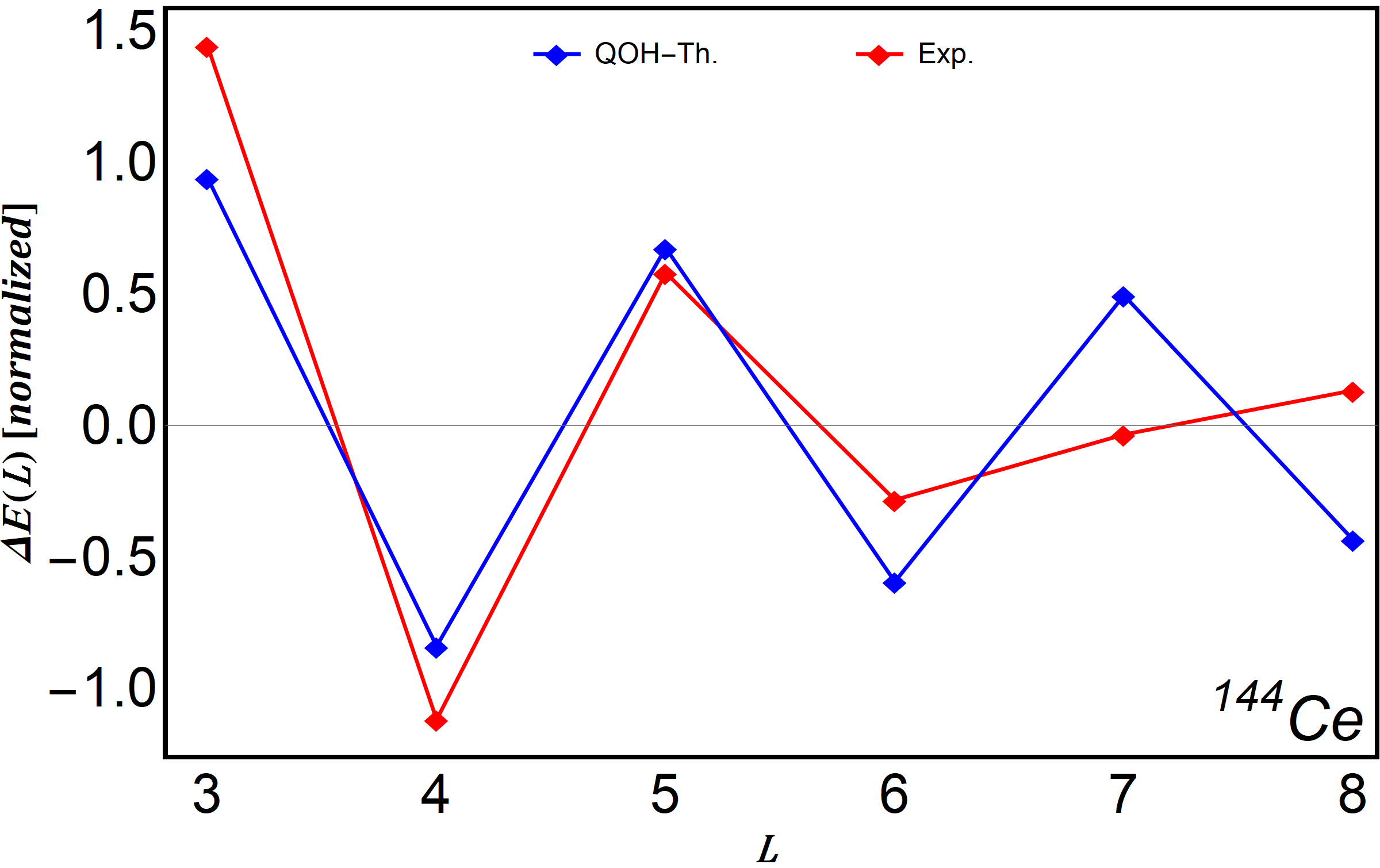}
	\includegraphics[width=.3\textwidth]{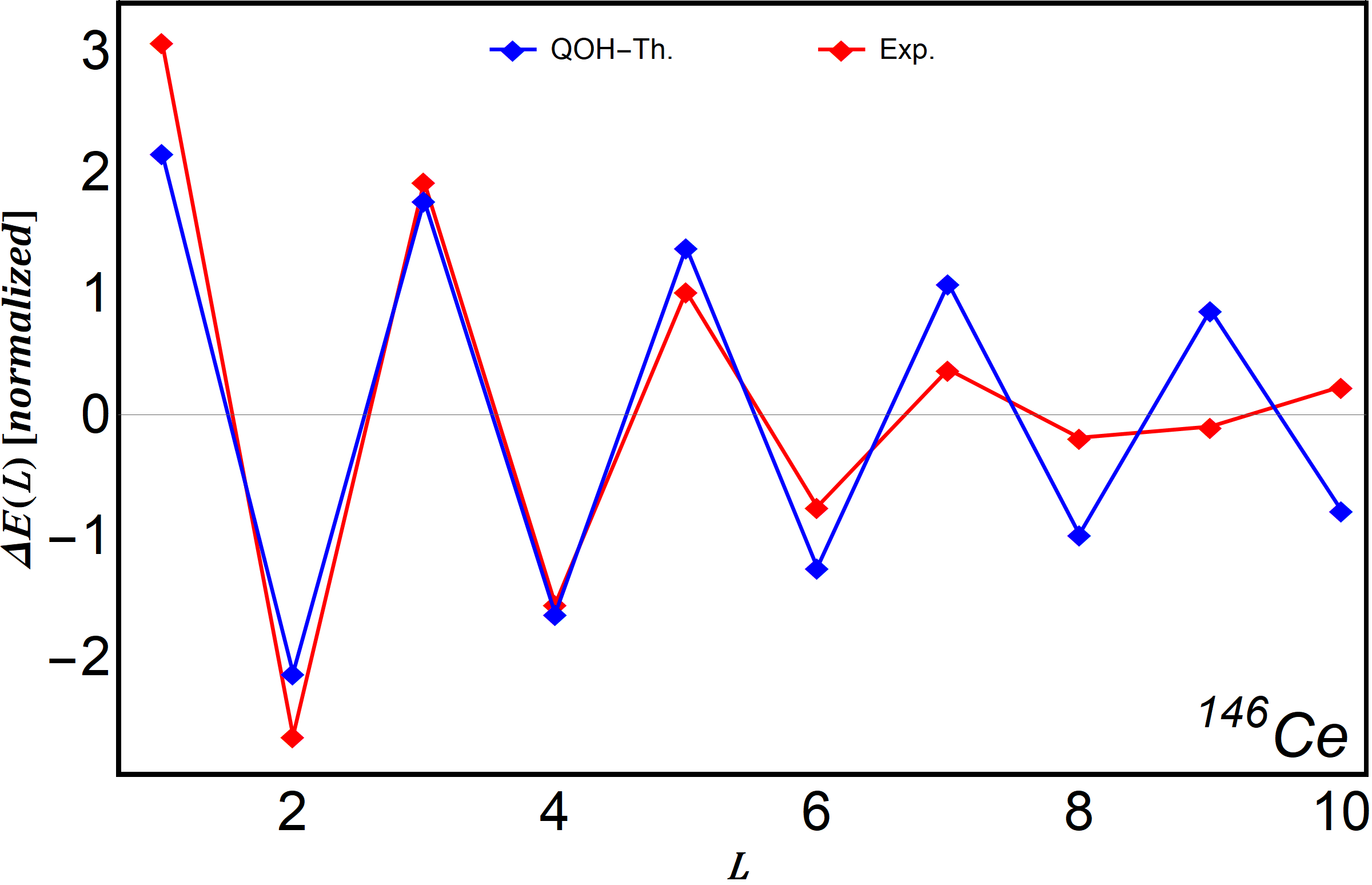}
	\includegraphics[width=.3\textwidth]{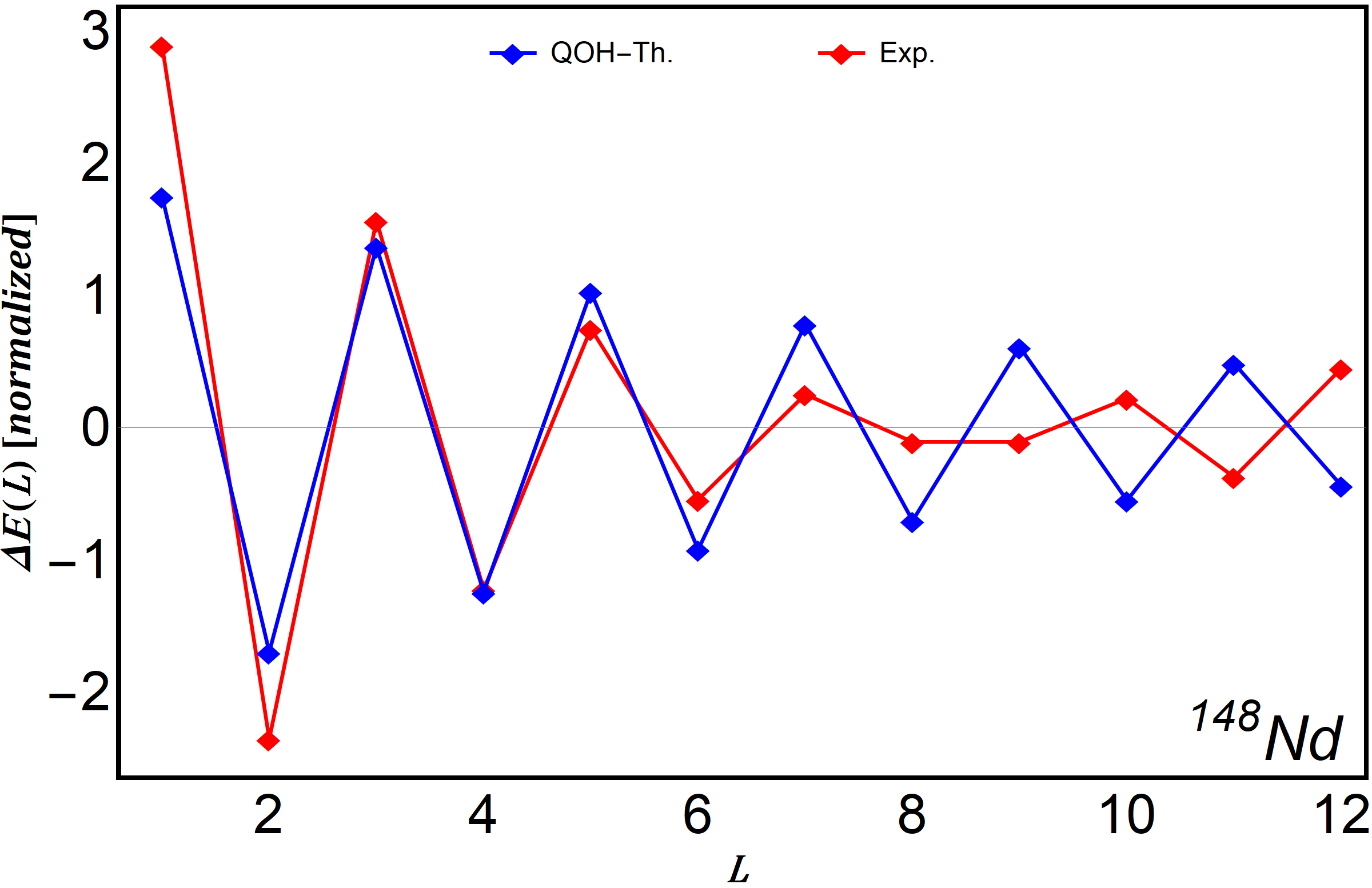}
	
	\includegraphics[width=.3\textwidth]{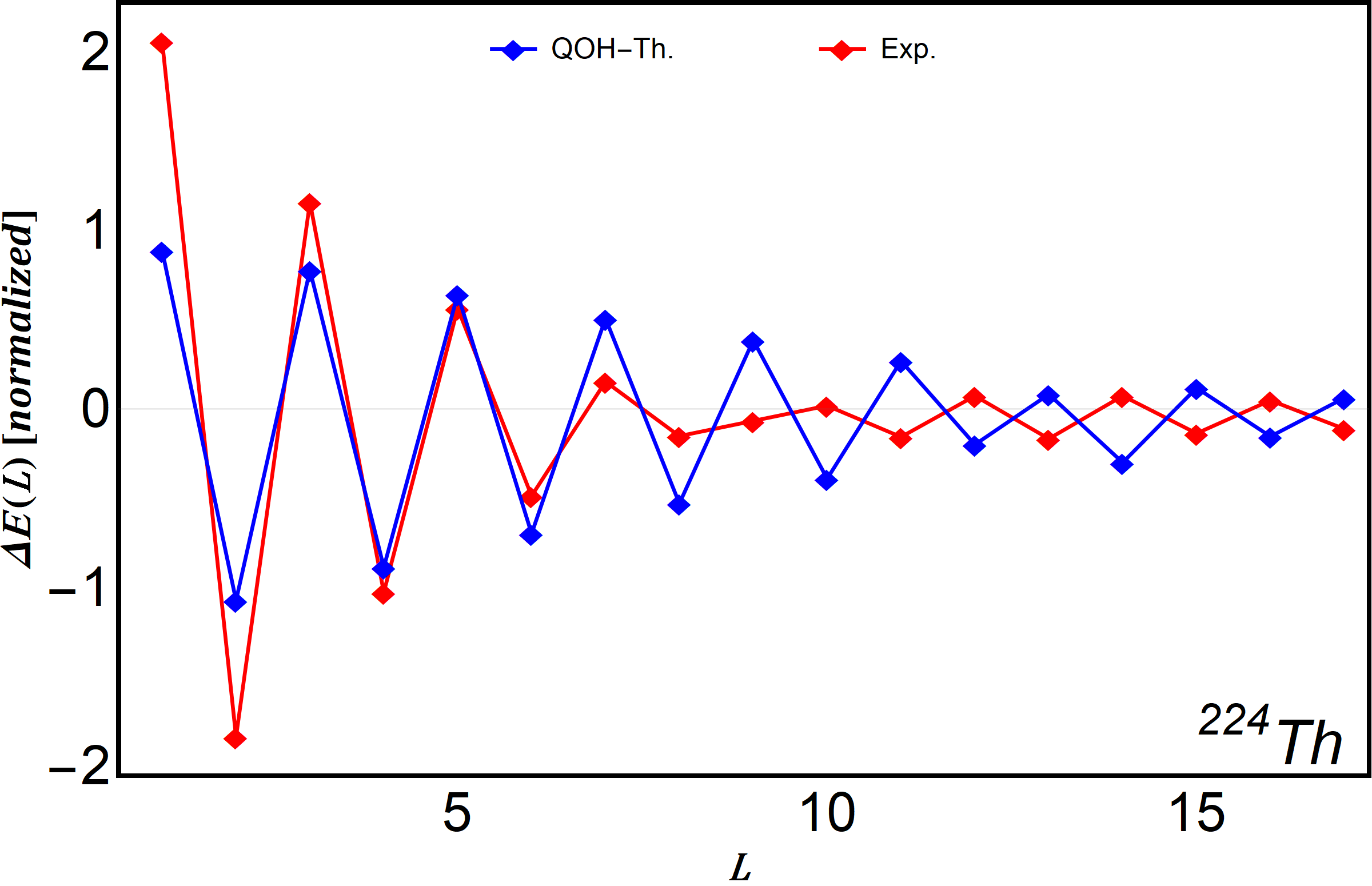}
	\includegraphics[width=.3\textwidth]{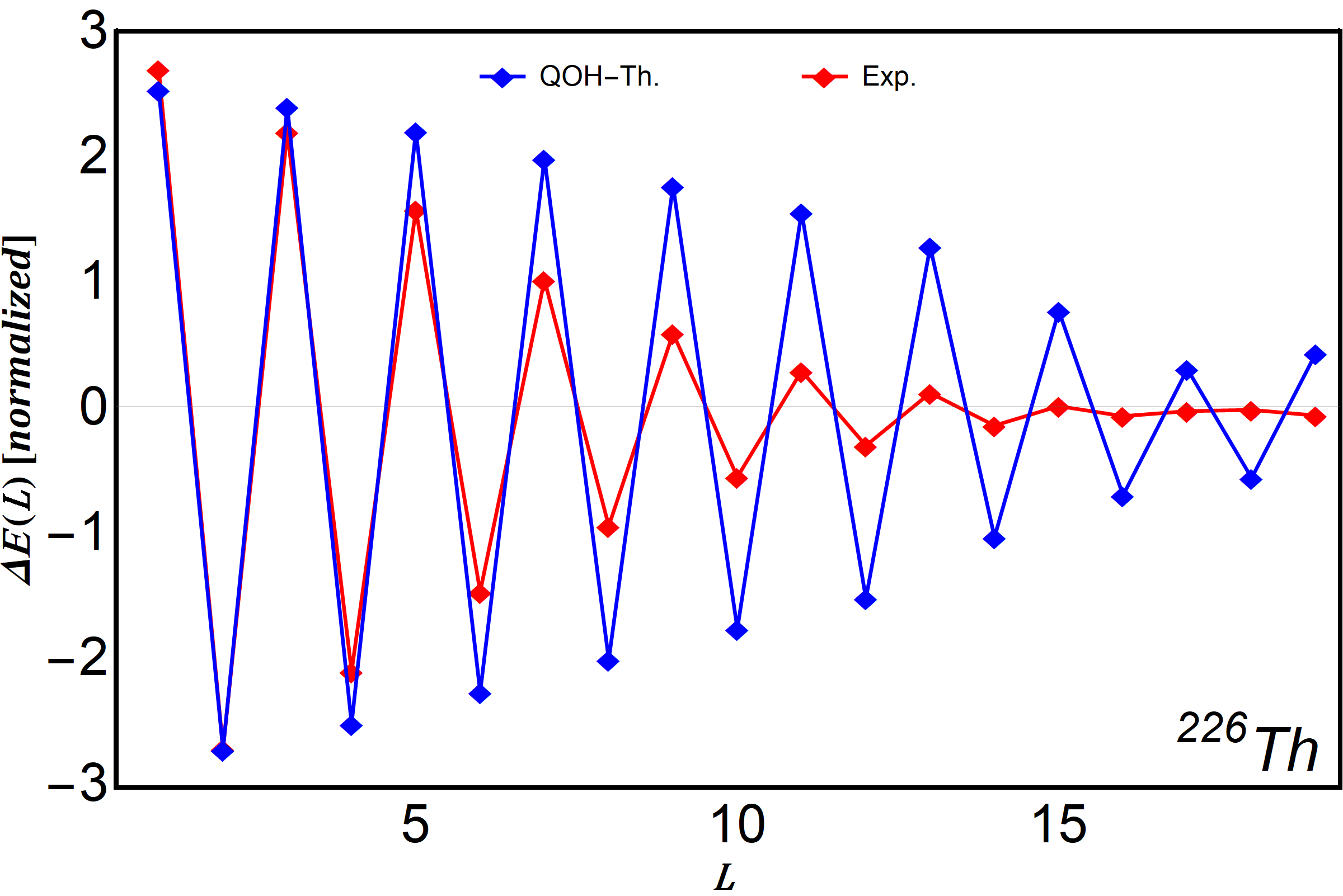}
	\includegraphics[width=.3\textwidth]{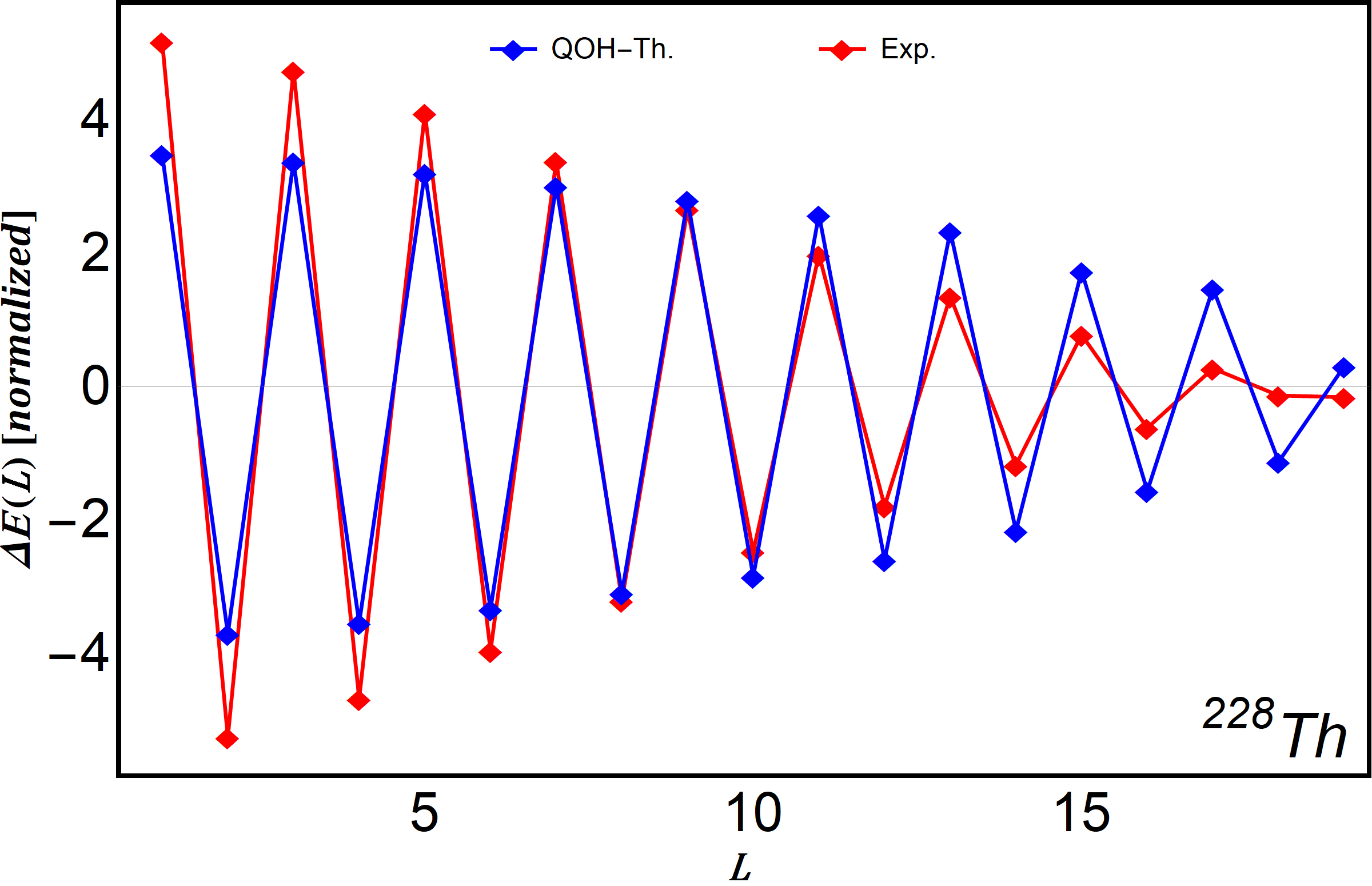}
	
	\includegraphics[width=.3\textwidth]{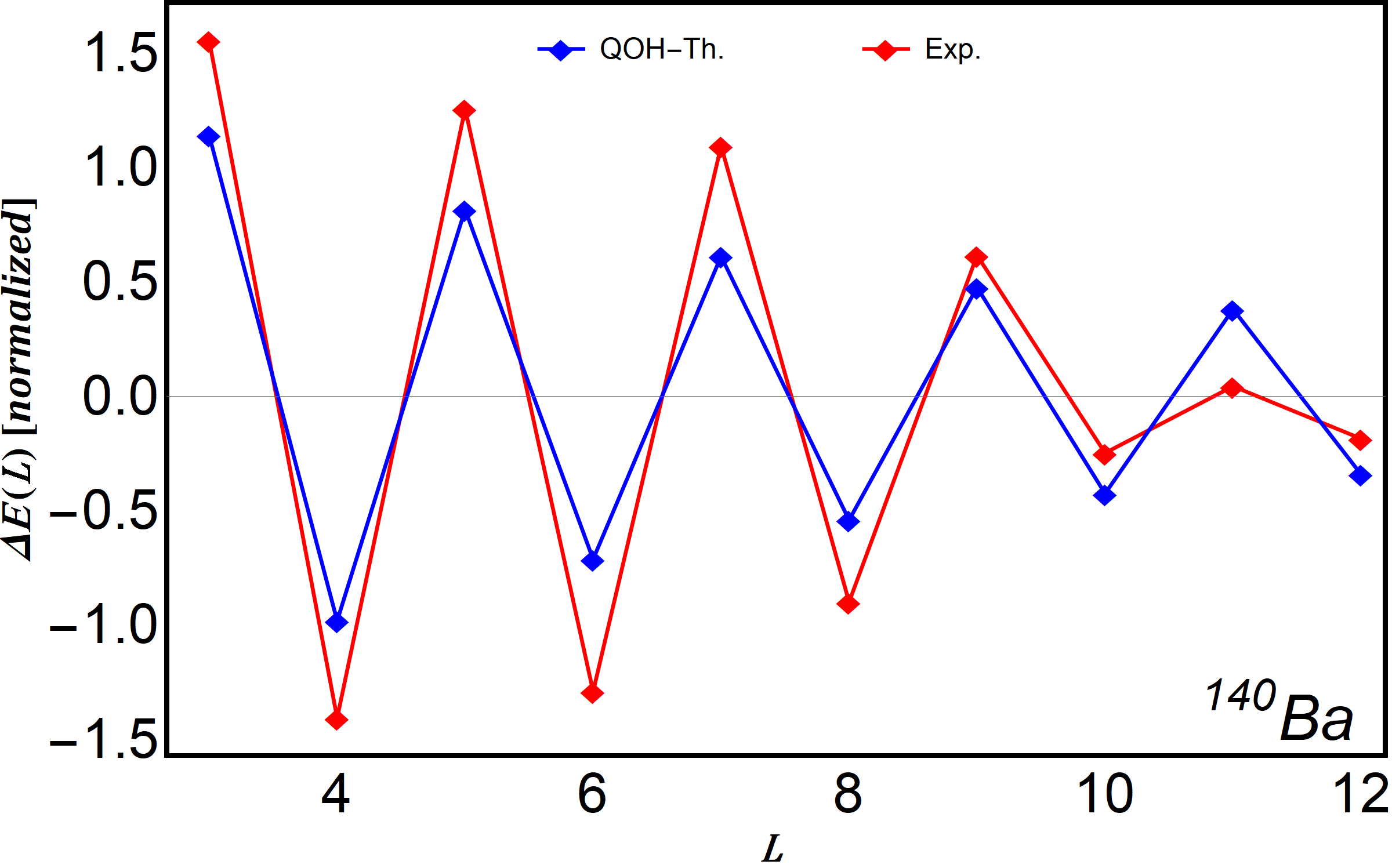}
	\includegraphics[width=.3\textwidth]{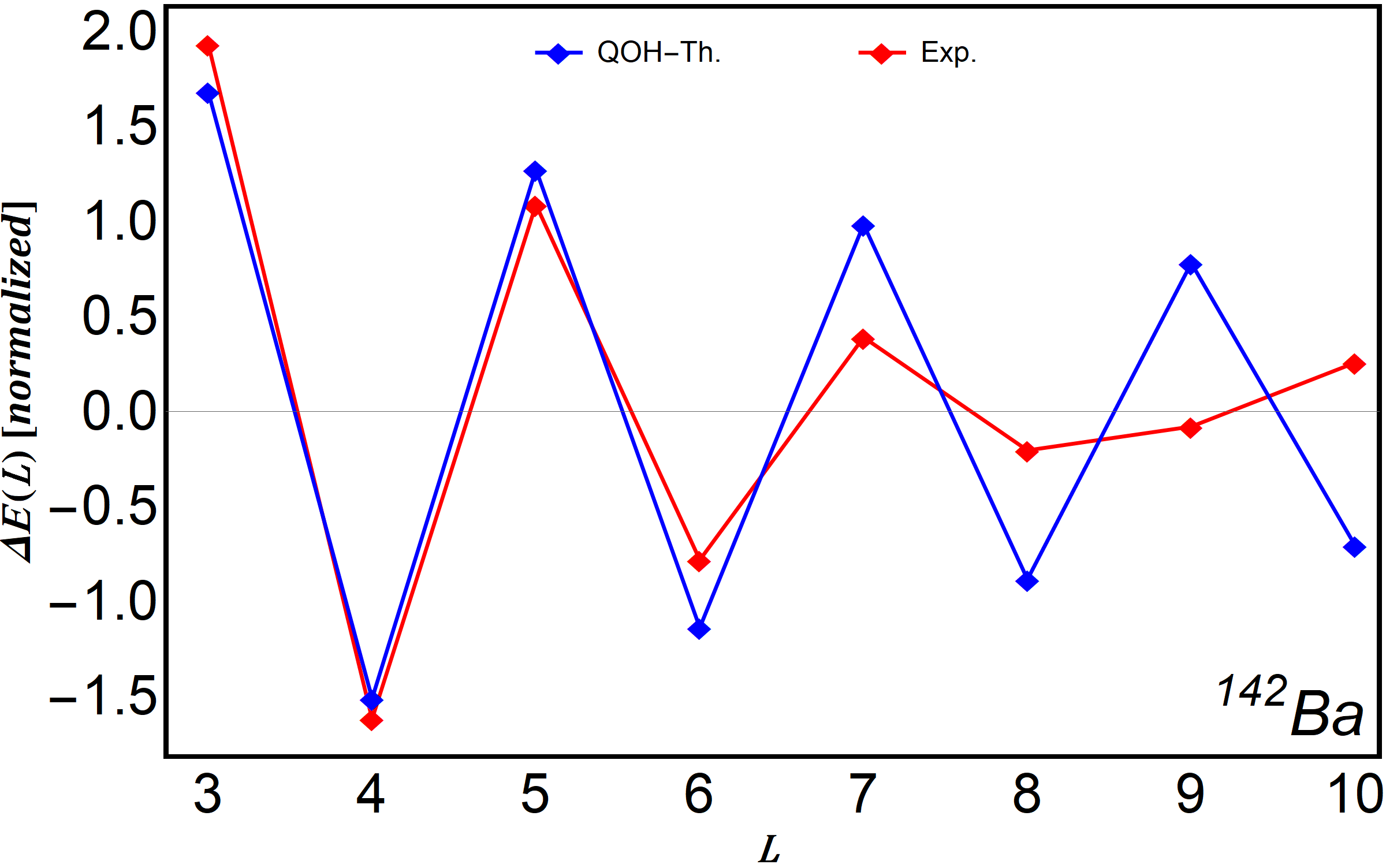}
	\includegraphics[width=.3\textwidth]{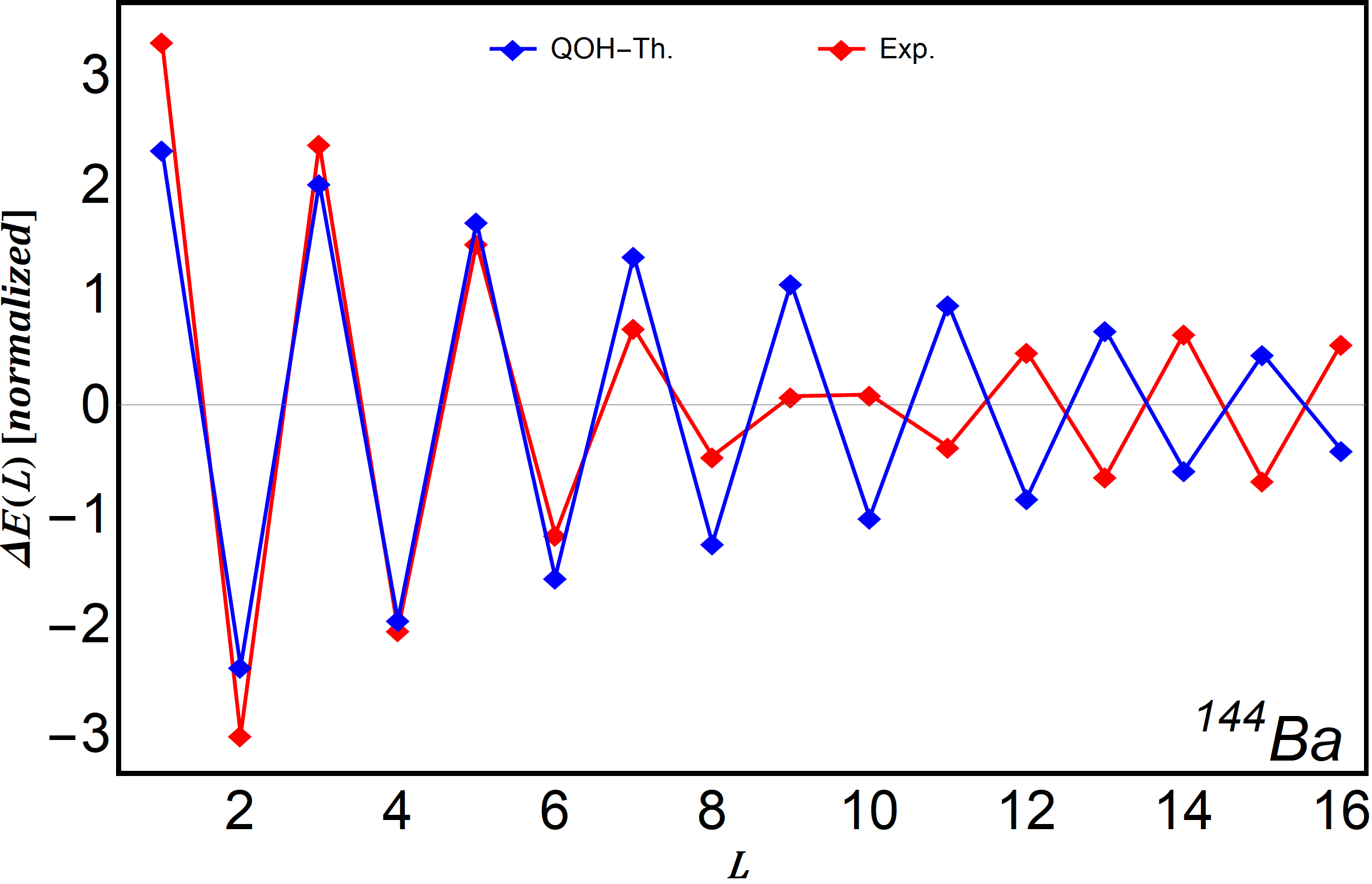}
	
	\caption{Comparison between Experimental and Theoretical Staggering $\Delta E(L)$ Eq. (\ref{stgg}) for vibrational nuclei  ${}^{218-222}$Rn, ${}^{112-114}$Xe, ${}^{150}$Sm, ${}^{144-146}$Ce, ${}^{148}$Nd, ${}^{224-228}$Th and ${}^{140-144}$Ba.}\label{stgS3}
\end{figure*}

\begin{figure*}
	\includegraphics[width=.3\textwidth]{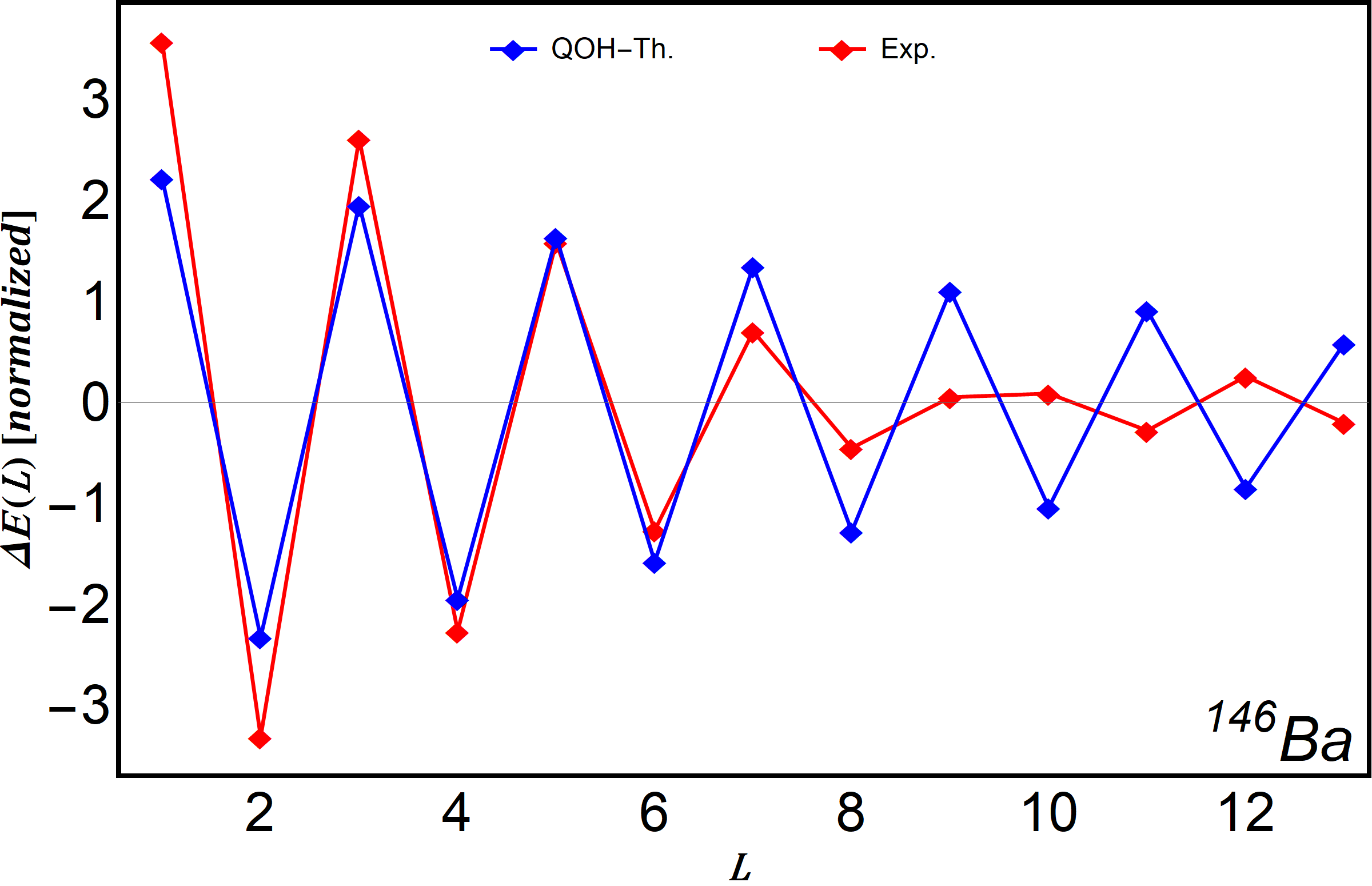}
	\includegraphics[width=.3\textwidth]{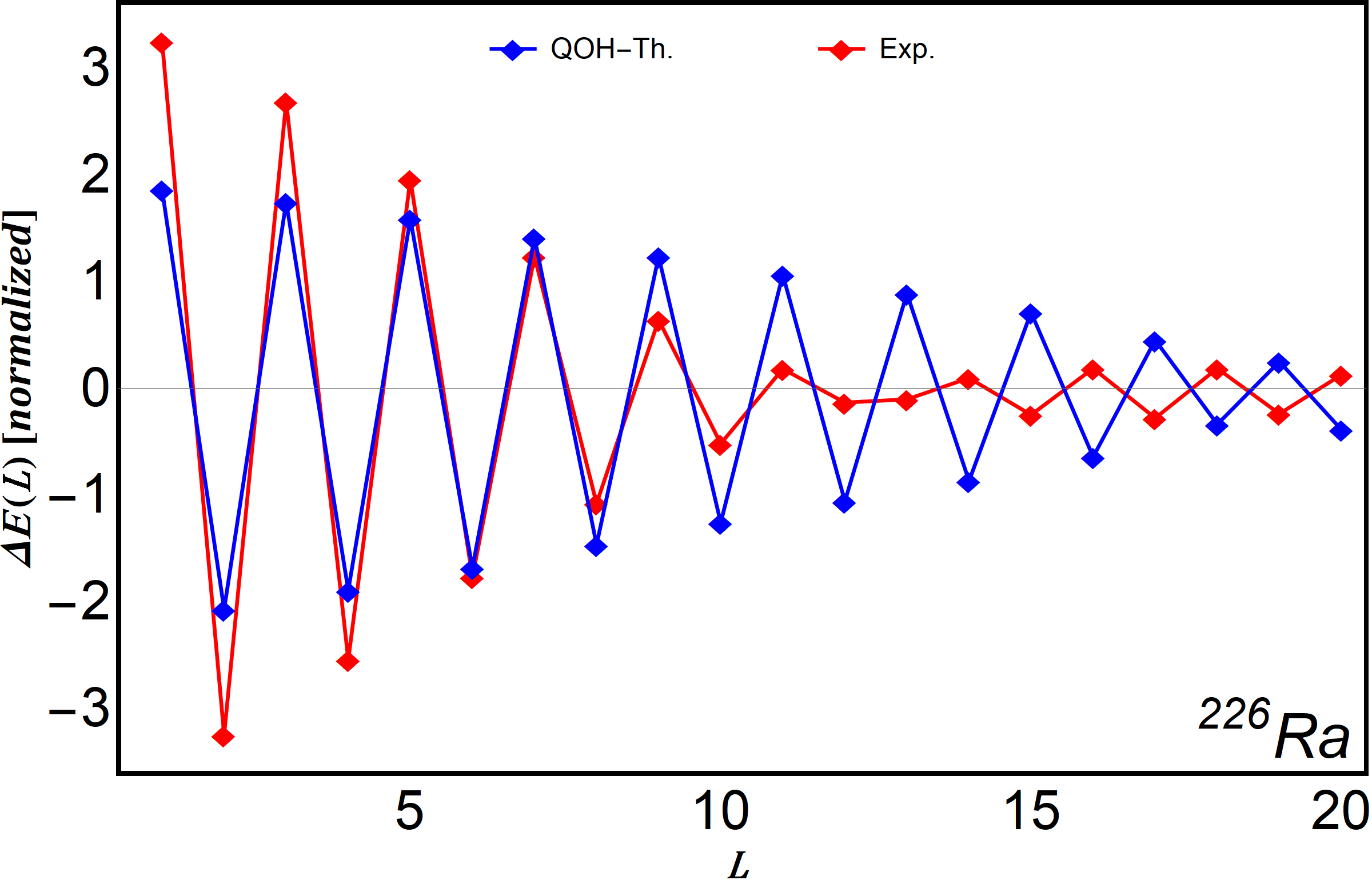}
	\includegraphics[width=.3\textwidth]{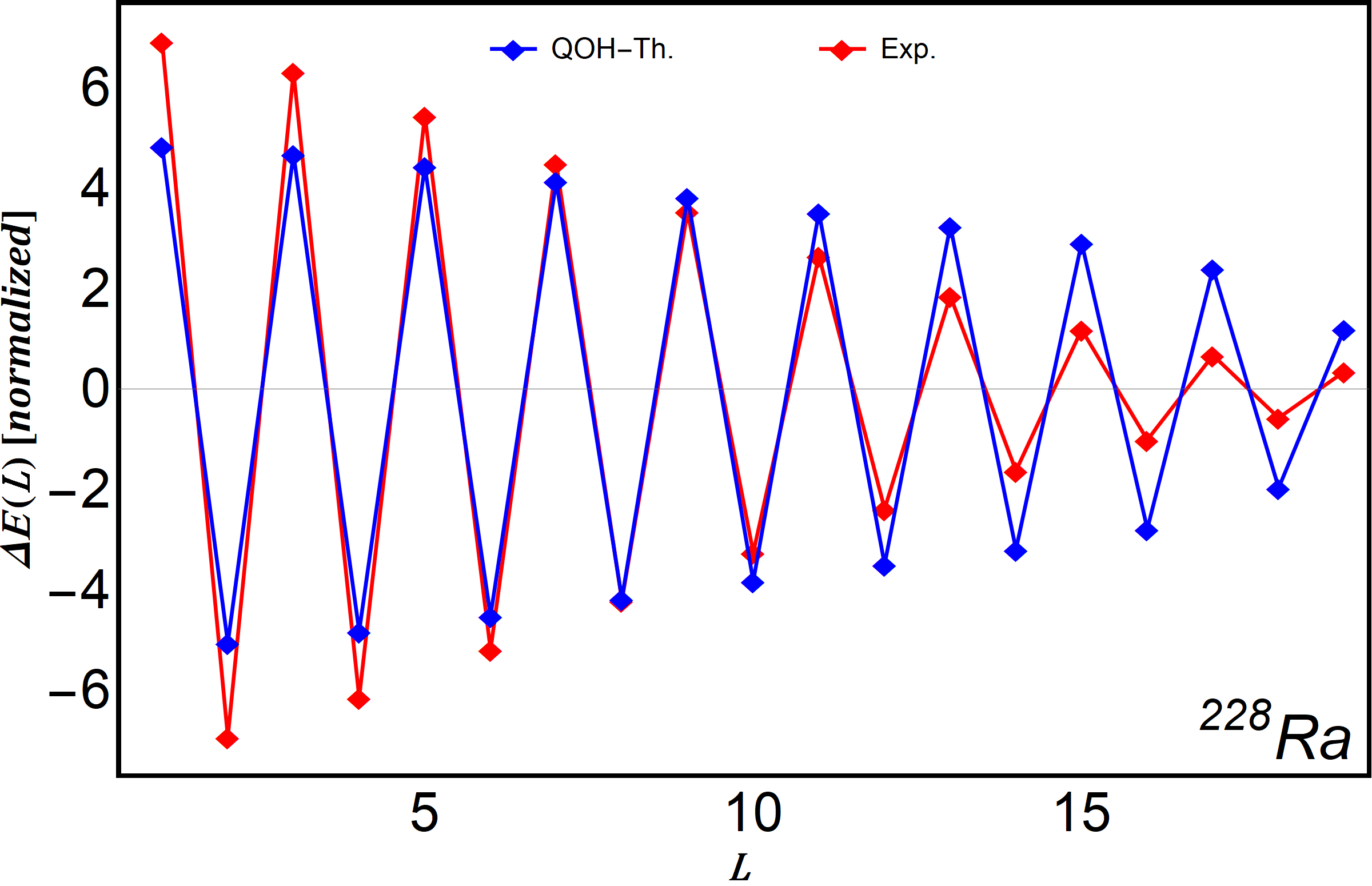}
	
	\includegraphics[width=.3\textwidth]{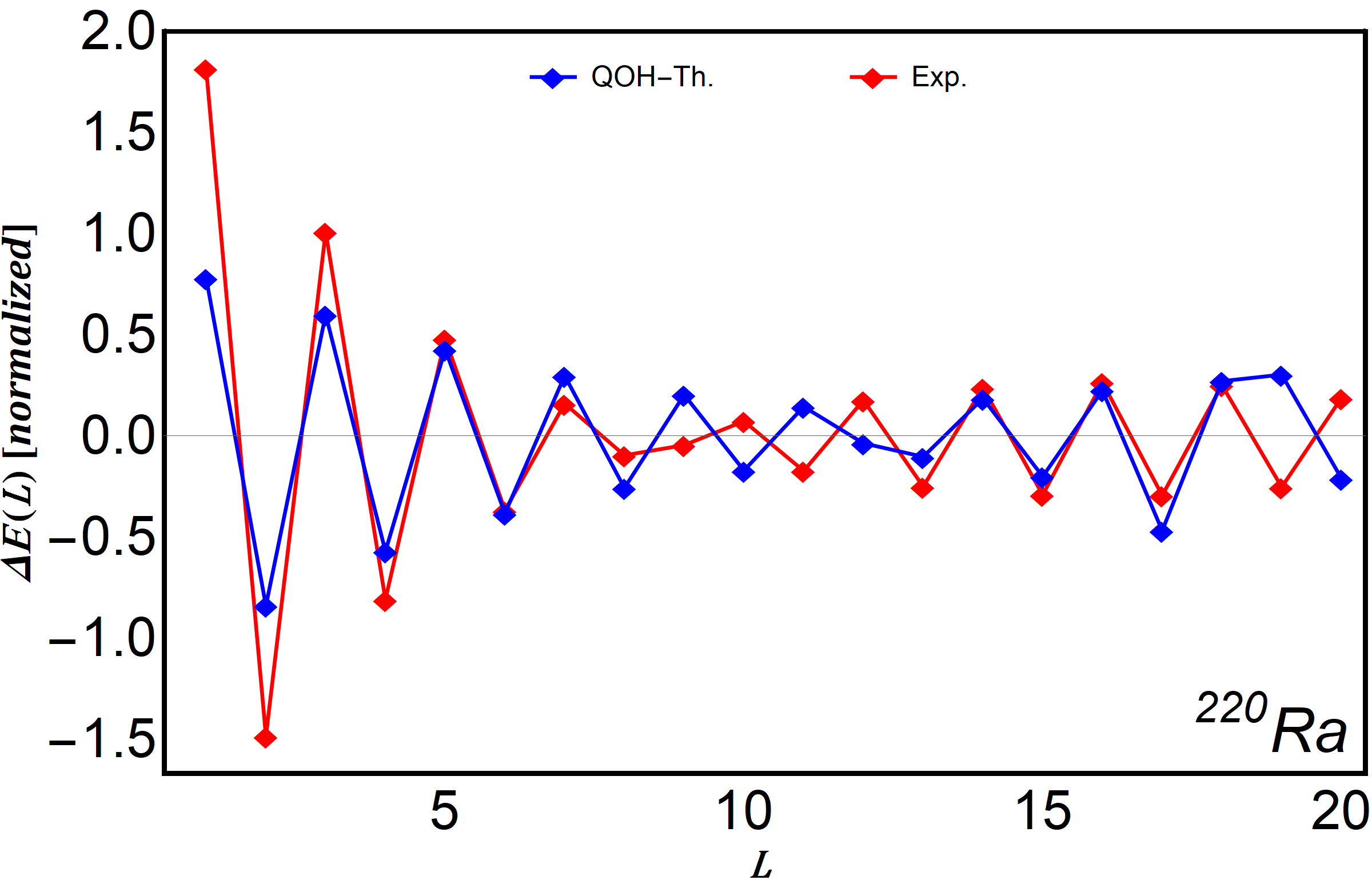}
	\includegraphics[width=.3\textwidth]{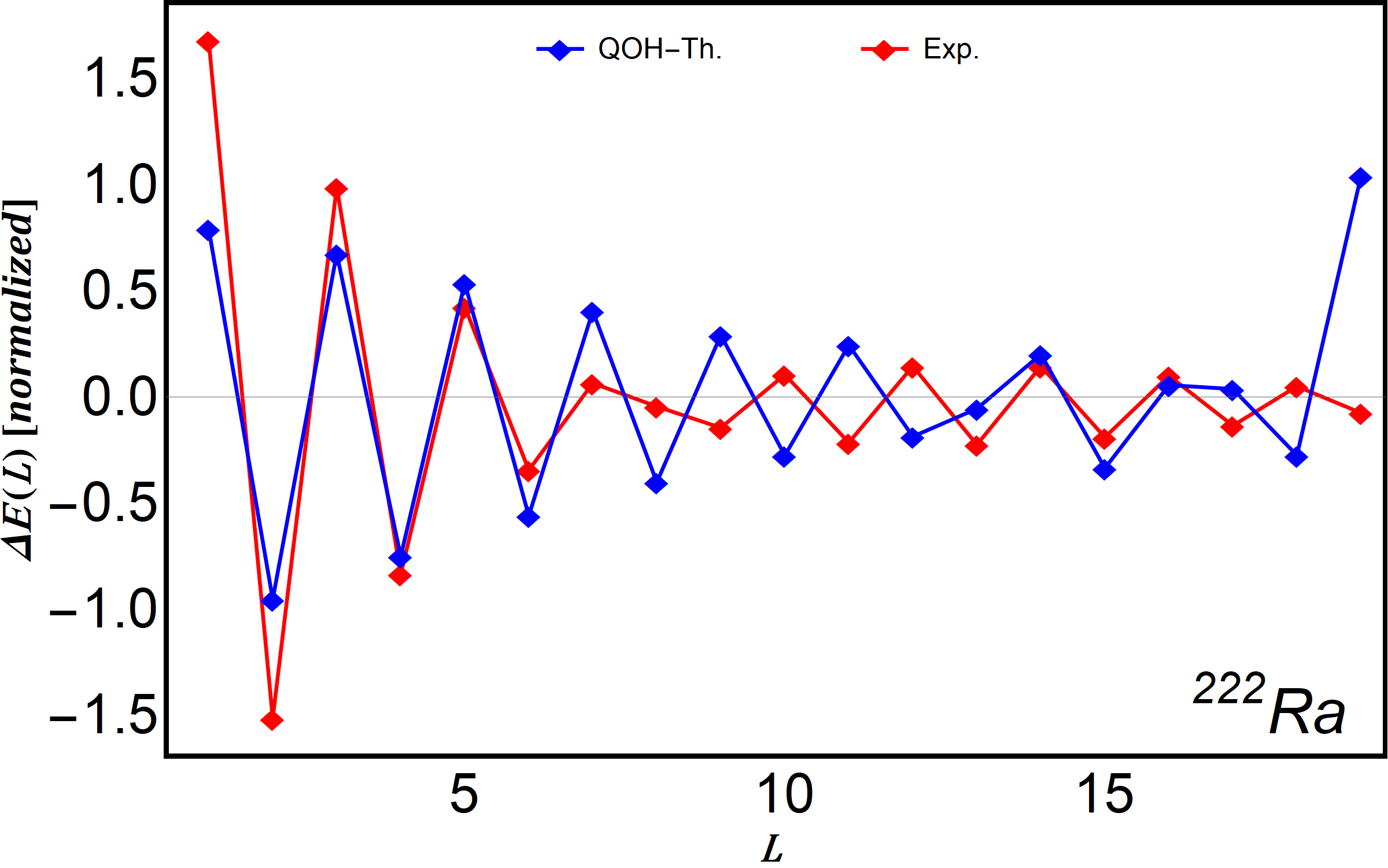}
	\includegraphics[width=.3\textwidth]{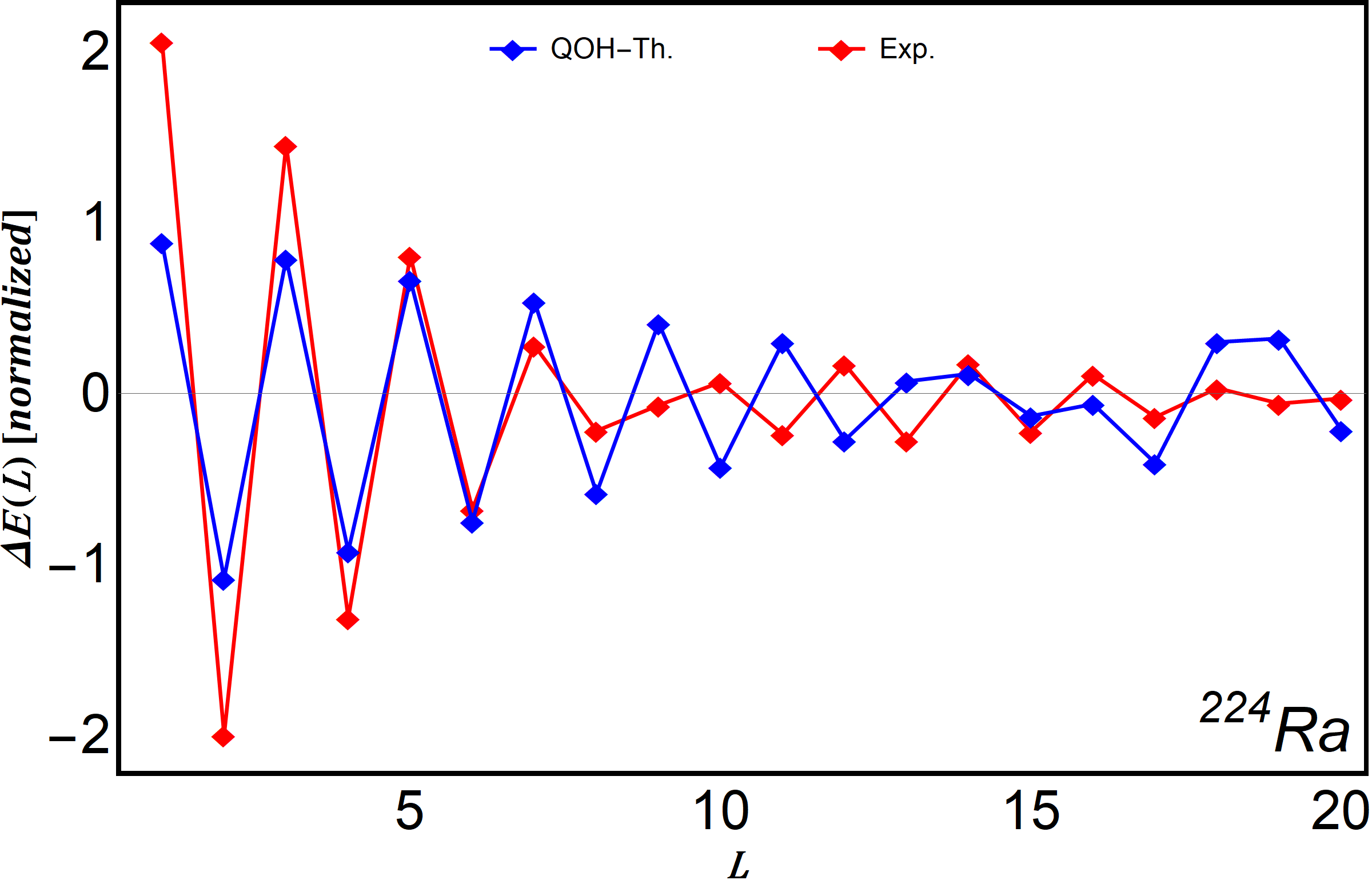}
	
	\caption{Same as in Fig. (\ref{stgS3}), but for	${}^{220-228}$Ra and ${}^{146}$Ba.}\label{stgS4}
\end{figure*}

 For the remaining set of vibrational nuclei, namely ${}^{220-228}$Ra, ${}^{224-228}$Th, ${}^{150}$Sm, ${}^{140-146}$Ba, ${}^{148}$Nd, ${}^{218-222}$Rn, ${}^{112-114}$Xe, and ${}^{144-146}$Ce, shown in Figs. (\ref{stgS3}) and (\ref{stgS4}), where the ground state and the negative parity bands are only slightly separated. Our model provides an approximate reproduction of the staggering patterns, particularly in the low-spin region, suggesting that coupling mechanisms may be necessary for a more complete description \cite{b31}.
\par We observe that the majority of nuclei described by this model exhibit vibrational behaviour. This is linked to the choice of the potential \( v(\phi_1) \), which features a finite barrier with two symmetric minima. Such potential allows for tunnelling and shape oscillations between two reflection-asymmetric shape orientations separated by the barrier \cite{b31}.
\par
\begin{figure*}
	
	\includegraphics[width=.3\textwidth]{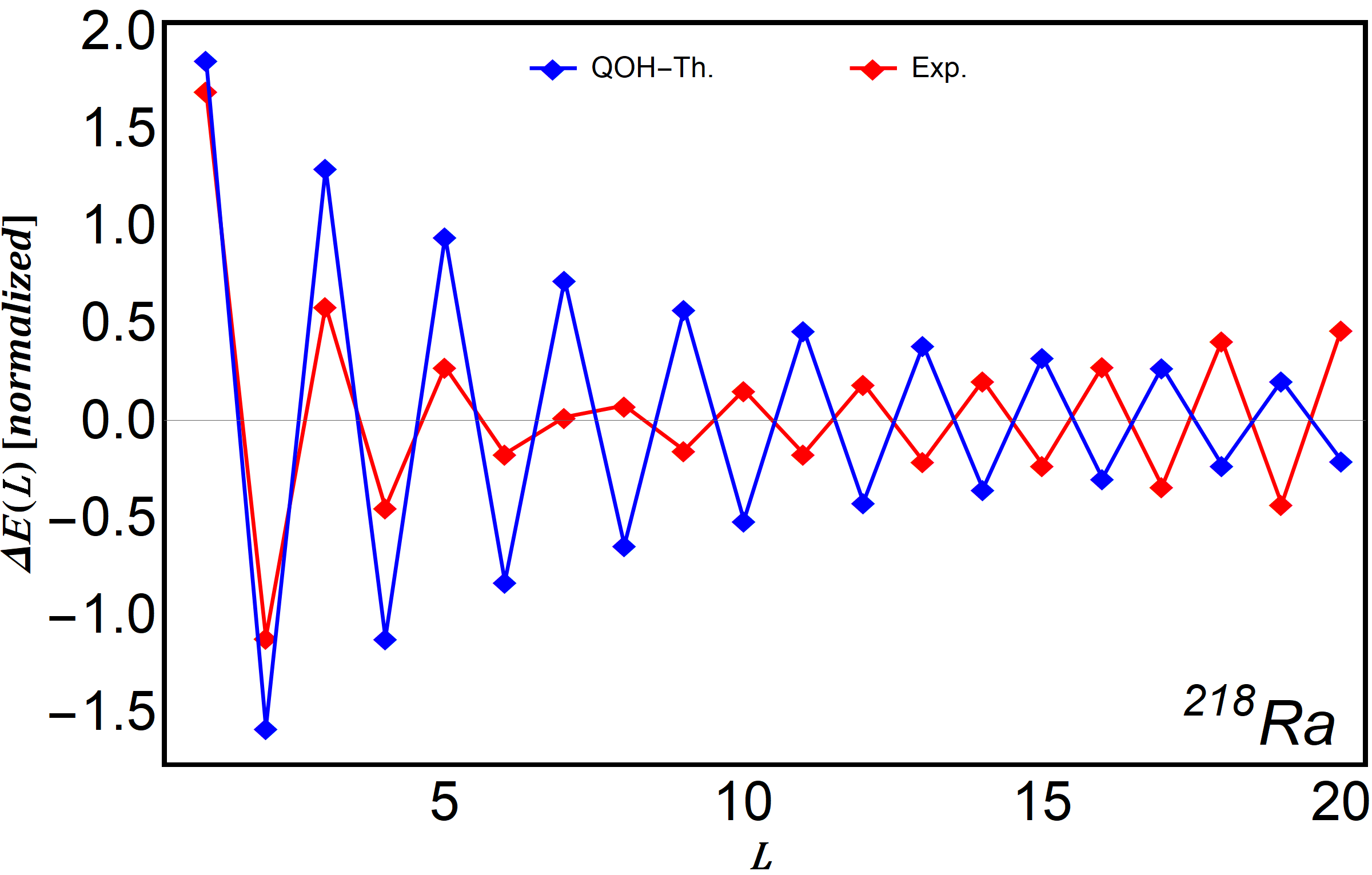}
	\includegraphics[width=.3\textwidth]{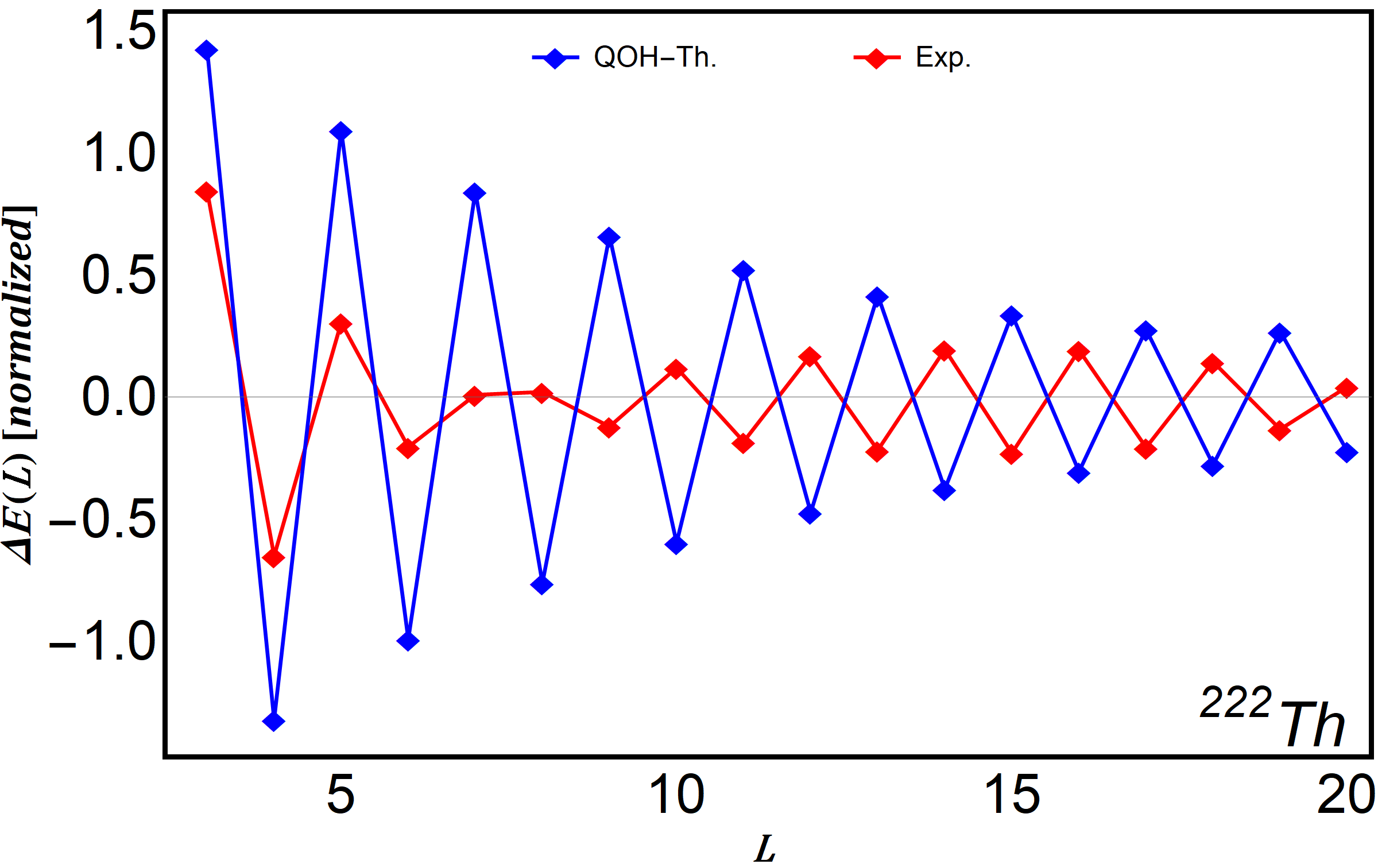}
	\includegraphics[width=.3\textwidth]{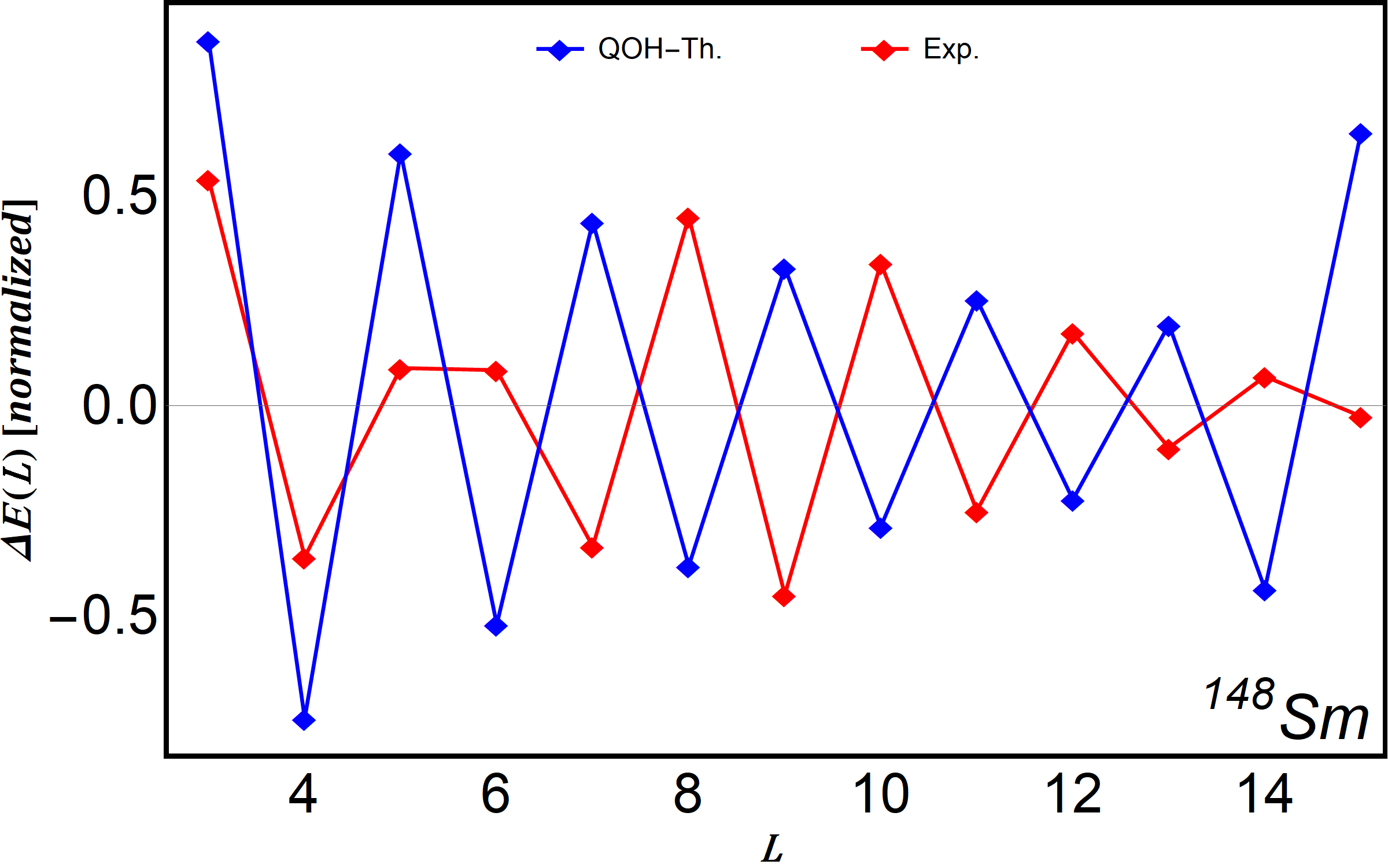}

	\includegraphics[width=.33\textwidth]{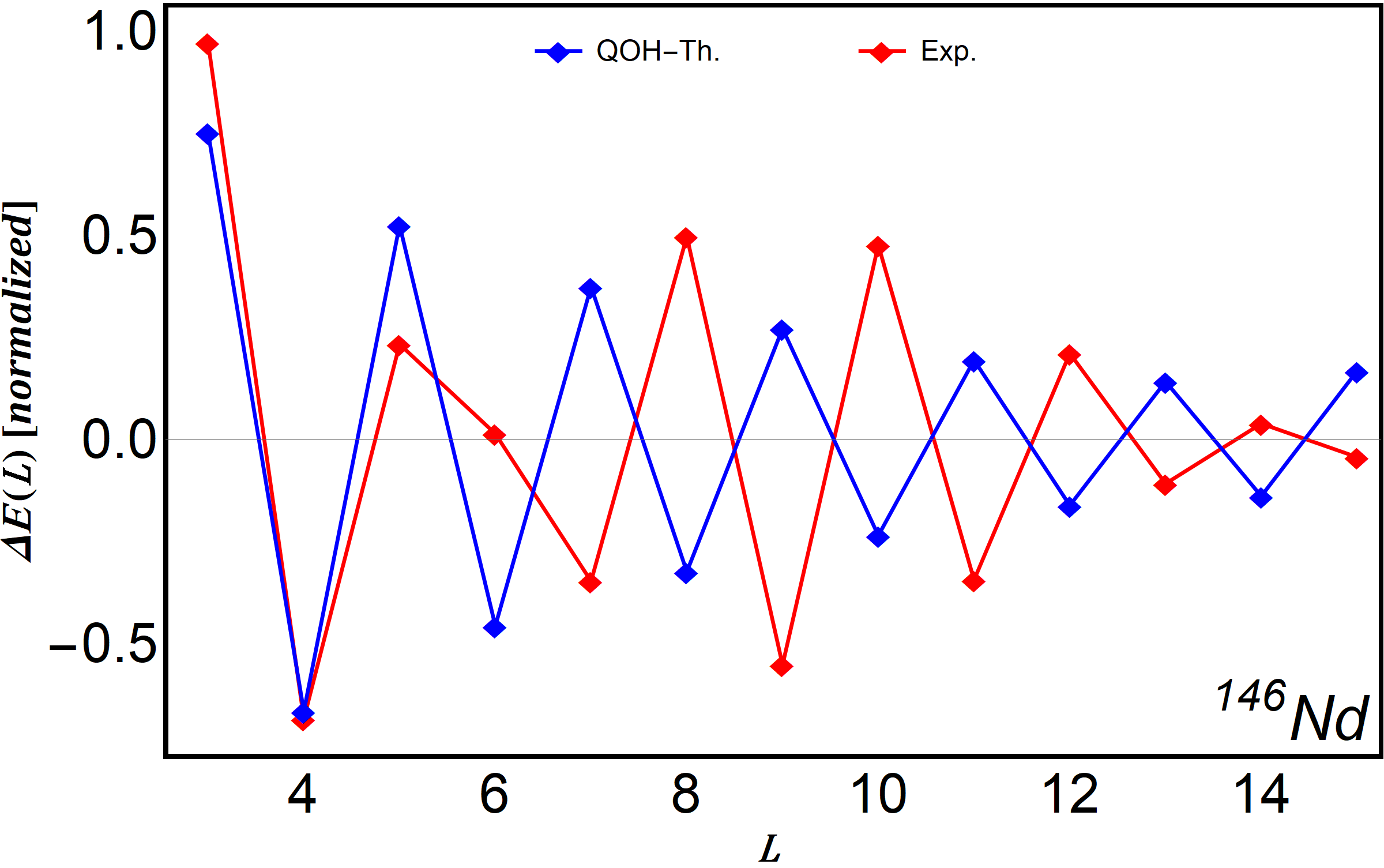}
	\includegraphics[width=.33\textwidth]{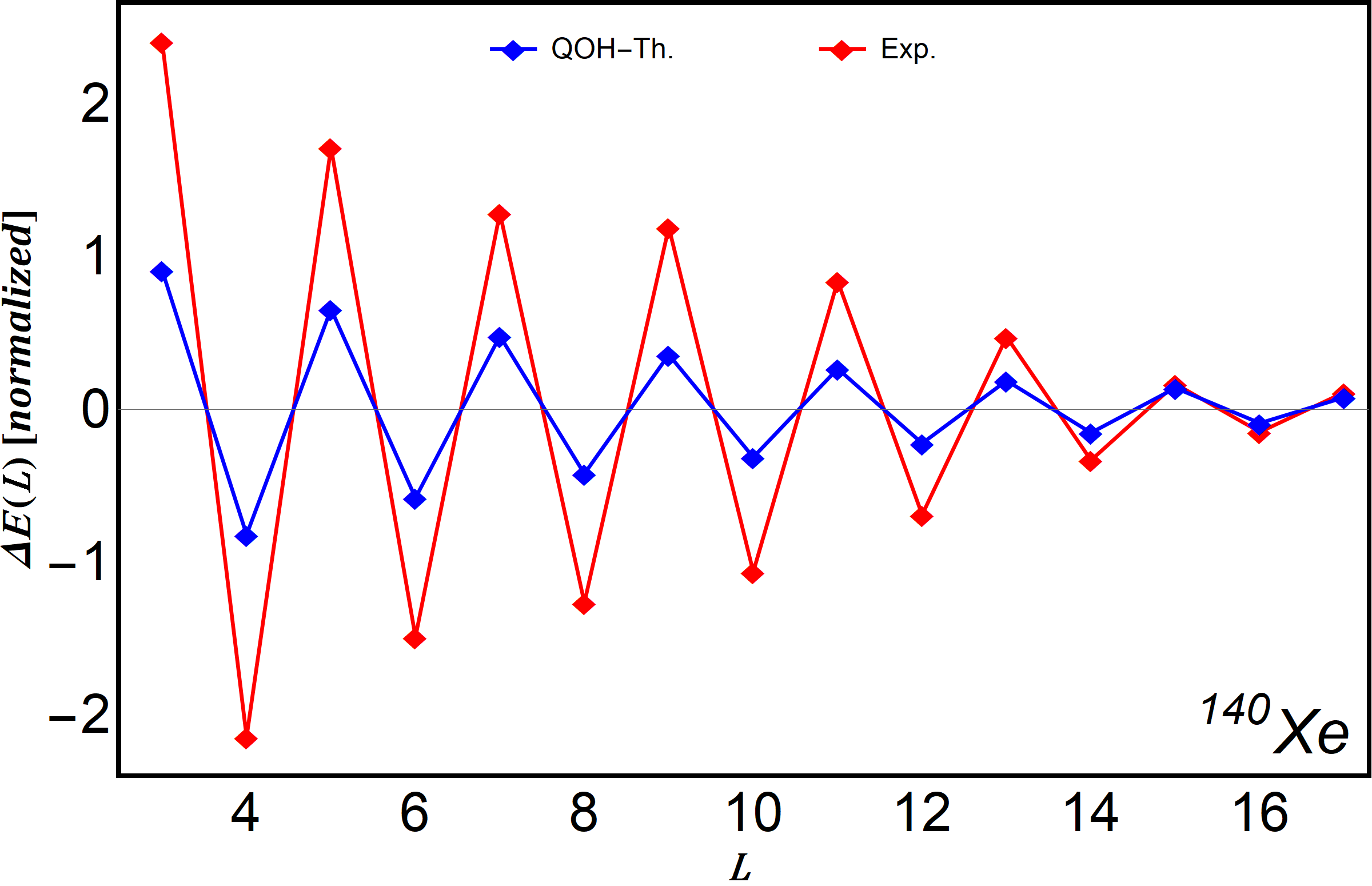}\\

	\caption{Comparison between Experimental and Theoretical Staggering $\Delta E(L)$ Eq. (\ref{stgg}) for ${}^{218}$Ra, ${}^{222}$Th, ${}^{148}$Sm, ${}^{146}$Nd and ${}^{140}$Xe.}\label{stgS5}
\end{figure*}

 However, clear discrepancies are observed for nuclei located in region of octupole deformation, as seen in Figure (\ref{stgS5}), such as ${}^{218}$Ra, ${}^{148}$Sm, ${}^{146}$Nd and ${}^{140}$Xe, as well as for ${}^{222}$Th. Despite these exceptions, the overall agreement between theoretical predictions and experimental data confirms that the model offers a reliable theoretical framework for describing the staggering effects in a wide range of nuclei.
\par Overall, the model captures the essential features of the $\Delta E(L)$ oscillations in nuclei exhibiting strong octupole vibrational behavior, and remains reasonably accurate for cases with weaker staggering, especially at low angular momenta. Nevertheless, clear deviations are observed in nuclei with pronounced octupole deformation, where the experimental staggering amplitude is reduced and not fully reproduced by the calculations. To further clarify the origin of these differences, it is instructive to analyze the behavior of $\Delta E(L)$ in terms of different regimes of octupole collectivity. Large oscillations in $\Delta E(L)$ are characteristic of octupole vibrational nuclei, whereas their suppression signals the onset of stable reflection-asymmetric shapes. This type of analysis has been successfully used to distinguish these regimes, as discussed for example in Ref.\cite{add4}.
\par Within this framework, the present model reproduces well the staggering patterns in vibrational nuclei, but shows limitations in nuclei with static octupole deformation. In the latter case, the staggering amplitude decreases and exhibits noticeable deviations at higher angular momenta. This behavior is not fully reproduced within the current approach, which is based on small oscillations around equilibrium configurations and neglects rotation–vibration coupling. These observations indicate that, while the model captures the essential features of octupole vibrational dynamics, a more complete description of strongly deformed nuclei would require the inclusion of additional coupling mechanisms. In particular, the incorporation of rotation–vibration coupling effects could improve the description of the evolution of staggering patterns at high angular momentum.

\subsubsection{Matrix elements of electric transitions}

\par We now focus on determining the matrix elements of electric transitions using the same values of $c_{2}$, $\phi_{02}$, and $\alpha$ obtained from the rms fit. The values of  the scale factors $F1$, $F1^{\prime}$, $F2$, and $F3$, which appear in the expressions of the matrix elements (\ref{me1}), (\ref{me2}) and (\ref{me3}), resulting from the fit to the available experimental data on transition matrix elements for nuclei such as $^{220,222}$Rn, $^{222,224,228}$Ra, and $^{144,146}$Ba, are presented in Table (\ref{Tab4}). The corresponding matrix elements calculated using these parameters are listed in Tables (\ref{Tab5}) to (\ref{Tab11}). 
The scaling factors $F1, F1^{\prime}, F2$, and $F3$ vary significantly across different nuclei, often spanning several orders of magnitude. This fluctuation reflects their role as effective parameters that absorb structural differences, normalization effects, and collective model limitations. Specifically, the large discrepancies observed in Table (\ref{Tab4}) arise because these mass parameters act as multiplicative factors. Consequently, their values align with the magnitude of the experimental transitions they describe: while $E1$ transitions are typically very weak, $E2$ exhibit intermediate strength, and $E3$ are notably more intense. Furthermore, although electric transition operators depend primarily on quadrupole and octupole coordinates, incorporating the hexadecapole degree of freedom indirectly refines the results. By modifying the collective wave functions and the resulting angular integrals, this inclusion leads to a more accurate agreement with experimental data.


\par A comparative analysis between the experimental transition matrix elements and the theoretical predictions of the QOH model reveals that the model  provides a coherent and accurate description of E1, E2, and E3 transition matrix elements across all examined nuclei $^{220,222}$Rn, $^{222,224,228}$Ra, and $^{144,146}$Ba. It shows a reasonable overall agreement with experimental data for E1 and E3 transitions, while E2 transitions are well reproduced at low angular momentum, with some deviations appearing at higher spin states, indicating potential for improvement. The model is thus a robust framework for investigating alternating parity band structures, effectively capturing the quadrupole-octupole dynamics of the nuclei under consideration.

\begin{table}
	\caption{The values of the scale factors $F1$,  $F1^{\prime}$, $F2$ and $F3$ resulting from the fit for nuclei with available experimental data on transition matrix elements.} \label{Tab4}
	$$
	\begin{array}{|lcccc|}
		\hline \text { \textbf{Nucleus} } & F1 & F1^{\prime} & F2 & F3 \\
		\hline \hline
		& &  &&      \\
		{ }^{222} \mathrm{Ra}& 1.651 \times 10^{-7}&  0.1339  &345&  658.64    \\
		{ }^{224} \mathrm{Ra}& 0.0104& 2.6835 \times 10^{-6}  & 254.722 & 454    \\
		{ }^{228} \mathrm{Ra}& 1.9 \times 10^{-6} & 0.066  & 724.459 & 471      \\
		& &  &&      \\
		{ }^{220} \mathrm{Rn}& 0.0157 & 0.00599  & 377.06 & 760 \\
		{ }^{222} \mathrm{Rn}&0.0888 & 0.0224 & 429.23 & 642.49 \\
		& &  &&      \\
		{ }^{144} \mathrm{Ba}& - & - & 365.25 & 540.93    \\
		{ }^{146} \mathrm{Ba}& 0.00129& 0.00177  &318.54&     361.73 \\
		& &  &&      \\
		\hline
	\end{array}
	$$
\end{table}
\begin{table}[h]    
	\caption{The absolute values of the matrix element $|m.e.|$ of electric transitions in ${ }^{224} \mathrm{Ra}$. The experimental data, in units of $e$ fm, $e$ fm$^2$, $e$ fm$^3$ for E1, E2, E3, respectively, have been taken from Ref. \cite{b32}.}\label{Tab5}
	$$
	\begin{array}{|lcc|}
		\hline\hline
		&  { }^{224} \mathrm{Ra} \qquad \qquad \qquad&   \\
		\hline \text {\textbf{ Matrix element (m.e.)} } & \text { \textbf{Experiment}} & \text {\textbf{QOH - Theory}}\\
		\hline \left\langle I^{\prime} \| E \lambda \| I \right\rangle &  |\textbf{m.e.}|\left(\mathrm{e.fm}^\lambda\right)& |\textbf{m.e.}|\left(\mathrm{e.fm}^\lambda\right)\\
		\hline
		&&\\
		\left\langle 0^{+}\|E 2\| 2^{+}\right\rangle & 199 \pm 3 & \textbf{200.06} \\
		\left\langle 2^{+}\|E 2\| 4^{+}\right\rangle & 315 \pm 6 & 261.23 \\
		\left\langle 4^{+}\|E 2\| 6^{+}\right\rangle & 405 \pm 15 & 251.50 \\
		\left\langle 6^{+}\|E 2\| 8^{+}\right\rangle & 500 \pm 60 & 205.66 \\
		\left\langle 1^{-}\|E 2\| 3^{-}\right\rangle & 230 \pm 11 & \textbf{228.09} \\
		\left\langle 3^{-}\|E 2\| 5^{-}\right\rangle & 410 \pm 60 & 243.02 \\
		\left\langle 0^{+}\|E 2\| 2^{+}_{\small{\tilde{\beta}}}\right\rangle & 23 \pm 4 &  30.30 \\
		&& \\
		\left\langle 0^{+}\|E 3\| 3^{-}\right\rangle & 940 \pm 30 & \textbf{941.67} \\
		\left\langle 2^{+}\|E 3\| 1^{-}\right\rangle & 1370 \pm 140 & 1111.33 \\
		\left\langle 2^{+}\|E 3\| 3^{-}\right\rangle & <4000 & \textbf{1144.19} \\
		\left\langle 2^{+}\|E 3\| 5^{-}\right\rangle & 1410 \pm 190 & \textbf{1498.04} \\
		&&\\
		\left\langle 0^{+}\|E 1\| 1^{-}\right\rangle & <0.018 & \textbf{0.017}  \\
		\left\langle 2^{+}\|E 1\| 1^{-}\right\rangle & <0.03 & \textbf{0.023}  \\
		\left\langle 2^{+}\|E 1\| 3^{-}\right\rangle & 0.026 \pm 0.005 & \textbf{0.026}  \\
		\left\langle 4^{+}\|E 1\| 5^{-}\right\rangle & 0.030 \pm 0.010 & \textbf{0.028}  \\
		\left\langle 6^{+}\|E 1\| 7^{-}\right\rangle & <0.10 & \textbf{0.025}\\
		&&\\
		\hline\hline
	\end{array}
	$$
\end{table}

\begin{table*}[h]
	\caption{The absolute values of the matrix elements $|m.e.|$ for electric transitions in ${}^{222}\mathrm{Ra}$ are presented in units of $e$·fm for E1, $e$·fm$^2$ for E2, and $e$·fm$^3$ for E3 transitions. The experimental values of the E1 transitions are taken from Ref\cite{b33}, while the E2 and E3 values are from Ref\cite{b34}.}\label{Tab6}
	\centering
	\begin{tabular}{|lcc|}
		\hline \hline
		& ${ }^{222} \mathrm{Ra}$ \qquad \qquad \qquad &\\
		\hline \text{ \textbf{Matrix element (m.e.)}} & \text{ \textbf{Experiment}} & \text {\textbf{QOH - Theory}}\\
		\hline $\left\langle I \| E \lambda \| I^{\prime}\right\rangle$ & \textbf{$|m.e.|$} $\left(\mathrm{e.fm}^\lambda\right)$ &  \textbf{$|m.e.|$} $\left(\mathrm{e.fm}^\lambda\right)$ \\
		\hline
		&&\\
		$\left\langle 2^{+}\|E 2\| 2^{+}\right\rangle$ & $ 130 \pm 50$ &  249.92 \\
		$\left\langle 2^{+}\|E 2\| 4^{+}\right\rangle$ &   $298 \pm 15$ & \textbf{294.12} \\
		$\left\langle 4^{+}\|E 2\| 4^{+}\right\rangle$ &  $280 \pm 50$ &  \textbf{258.32} \\
		$\left\langle 4^{+}\|E 2\| 6^{+}\right\rangle$ &  $357 \pm 18$  & 291.33 \\
		$\left\langle 6^{+}\|E 2\| 8^{+}\right\rangle$ &   $415 \pm 23$   & 246.48 \\
		$\left\langle 8^{+}\|E 2\| 10^{+}\right\rangle$ &   $470 \pm 50 $  &168.49\\
		$\left\langle 1^{-}\|E 2\| 3^{-}\right\rangle$ &  $235 \pm 22$   & \textbf{257.69} \\
		$\left\langle 3^{-}\|E 2\| 5^{-}\right\rangle$ &   $310 \pm 40$   & \textbf{280.44}\\
		$\left\langle 5^{-}\|E 2\| 7^{-}\right\rangle$ &   $440 \pm 40$  &251.08 \\
		$\left\langle 7^{-}\|E 2\| 9^{-}\right\rangle$ &  $600 \pm 100$   & 185.75\\
		&&\\
		$\left\langle 0^{+}\|E 3\| 3^{-}\right\rangle$ &   $1130 \pm 90$ & \textbf{1087.29} \\
		$\left\langle 2^{+}\|E 3\| 1^{-}\right\rangle$ &   $850 \pm 240$  & 1294.68 \\
		$\left\langle 2^{+}\|E 3\| 3^{-}\right\rangle$ & $ 900 \pm 500$  & \textbf{1341.85} \\
		$\left\langle 2^{+}\|E 3\| 5^{-}\right\rangle$ &  $1790 \pm 200$  & \textbf{1756.81} \\
		$\left\langle 4^{+}\|E 3\| 1^{-}\right\rangle$ &  $ 2100 \pm 500$  &1519.66\\
		$\left\langle 4^{+}\|E 3\| 3^{-}\right\rangle$ &  $2600_{-900}^{+600}$  & 1565.77\\
		$\left\langle 4^{+}\|E 3\| 5^{-}\right\rangle$ &  $ 1700 \pm 1000$ & \textbf{1759.53} \\
		$\left\langle 4^{+}\|E 3\| 7^{-}\right\rangle$ &  $3300_{-500}^{+3300}$  & 2338.66 \\
		& & \\
		$\left\langle 0^{+}\|E 1\| 1^{-}\right\rangle$ &  $0.082 \pm 0.007$  & \textbf{0.076} \\
		$\left\langle 2^{+}\|E 1\| 1^{-}\right\rangle$ &   $0.114^{+0.025}_{-0.009}$ &  \textbf{0.114}\\
		$\left\langle 2^{+}\|E 1\| 3^{-}\right\rangle$ &  $0.26 \pm 0.05$ & 0.145\\
		$\left\langle 4^{+}\|E 1\| 3^{-}\right\rangle$ &  $0.27 \pm 0.26$ & \textbf{0.180}\\
		$\left\langle 4^{+}\|E 1\| 5^{-}\right\rangle$ &   $0.180_{-0.019}^{+0.048}$ &  \textbf{0.210}\\
		$\left\langle 5^{-}\|E 1\| 6^{+}\right\rangle$ &  $0.277 \pm 0.011$ & \textbf{0.25}\\
		$\left\langle 6^{+}\|E 1\| 7^{-}\right\rangle$ &  $0.273 \pm 0.021$ & \textbf{0.282}\\
		$\left\langle 7^{-}\|E 1\| 8^{+}\right\rangle$ &   $0.317 \pm 0.018$ & \textbf{0.324}\\
		$\left\langle 8^{+}\|E 1\| 9^{-}\right\rangle$ &  $0.48 \pm 0.04$  & 0.361\\
		$\left\langle 9^{-}\|E 1\| 10^{+}\right\rangle$ &  $0.40 \pm 0.05$  & \textbf{0.407}\\
		$\left\langle 10^{+}|| E 1 \mid 11^{-}\right\rangle$   & $0.50 \pm 0.06$  & \textbf{0.44} \\
		$\left\langle 11^{-}\|E 1\| 12^{+}\right\rangle$   & $0.31 \pm 0.04$ & 0.495\\
		&& \\
		\hline \hline
	\end{tabular}
\end{table*}

\begin{table*}[htp]
	\caption{The absolute values of the matrix elements $|m.e.|$ for electric transitions in ${}^{228}\mathrm{Ra}$ are presented in units of $e$·fm for E1, $e$·fm$^2$ for E2, and $e$·fm$^3$ for E3 transitions. The experimental values of the E1 transitions are taken from Ref\cite{b33}, while the E2 and E3 values are from Ref\cite{b34}.}\label{Tab7}
	\centering
	
	\begin{tabular}{|lcc|}
		\hline \hline
		&${ }^{228} \mathrm{Ra}$\qquad \qquad \qquad &\\
		\hline \text{ \textbf{Matrix element (m.e.)}} & \text { \textbf{Experiment}}  & \text {\textbf{QOH - Theory}}\\
		\hline $\left\langle I \| E \lambda \| I^{\prime}\right\rangle$ & \textbf{$|m.e.|$} $\left(\mathrm{e.fm}^\lambda\right)$ &  \textbf{$|m.e.|$} $\left(\mathrm{e.fm}^\lambda\right)$\\
		\hline
		&&\\
		$\left\langle 2^{+}\|E 2\| 2^{+}\right\rangle$&  $30 \pm 170$ &   460.56 \\
		$\left\langle 2^{+}\|E 2\| 4^{+}\right\rangle$&  $387 \pm 19$ &  531.20\\
		$\left\langle 4^{+}\|E 2\| 6^{+}\right\rangle$  & $511 \pm 26$ &  \textbf{470.97} \\
		$\left\langle 6^{+}\|E 2\| 8^{+}\right\rangle$  & $589 \pm 29$ & 312.66\\
		$\left\langle 8^{+}\|E 2\| 10^{+}\right\rangle$  & $750 \pm 40$ &  85.20 \\
		$\left\langle 10^{+}\|E 2\| 12^{+}\right\rangle$  &$710_{-30}^{+50}$ &  197.60 \\
		$\left\langle 1^{-}\|E 2\| 3^{-}\right\rangle$  & $380 \pm 50 $   &\textbf{460.60}\\
		$\left\langle 3^{-}\|E 2\| 5^{-}\right\rangle$ & $390_{-80}^{+40}$ & \textbf{462.45}\\
		$\left\langle 5^{-}\|E 2\| 7^{-}\right\rangle$ & $400 \pm 90$  &\textbf{339.86}\\
		$\left\langle 7^{-}\|E 2\| 9^{-}\right\rangle$  &  $590 \pm 100$  &136.65\\
		&&\\
		$\left\langle 0^{+}\|E 3\| 3^{-}\right\rangle$ & $870 \pm 150$   & 1098.09 \\
		$\left\langle 2^{+}\|E 3\| 1^{-}\right\rangle$& $1360 \pm 230$ &  \textbf{1275.31} \\
		$\left\langle 2^{+}\|E 3\| 3^{-}\right\rangle$ & $60^{+160}_{-230}$  &  1300.65 \\
		$\left\langle 2^{+}\|E 3\| 5^{-}\right\rangle$ & $1710 \pm 230$ &  \textbf{1707.61}\\
		$\left\langle 4^{+}\|E 3\| 1^{-}\right\rangle$ & $400_{-1100}^{+700}$   &1487.47\\
		&&\\
		$\left\langle 0^{+}\|E 1\| 1^{-}\right\rangle$ & $0.043^{+0.032}_{-0.020}$& \textbf{0.044} \\
		$\left\langle 2^{+}\|E 1\| 1^{-}\right\rangle$ & $0.051_{-0.022}^{+0.033}$ & \textbf{0.063} \\
		$\left\langle 2^{+}\|E 1\| 3^{-}\right\rangle$  & $0.09 \pm 0.04$  &\textbf{0.078}\\
		$\left\langle 4^{+}\|E 1\| 3^{-}\right\rangle$ & $0.080_{-0.023}^{+0.036}$ & \textbf{0.093}\\
		$\left\langle 4^{+}\|E 1\| 5^{-}\right\rangle$  & $0.13 \pm 0.06$  & \textbf{0.10}\\
		$\left\langle 5^{-}\|E 1\| 6^{+}\right\rangle$  & $0.074^{+0.018}_{-0.012}$ &0.12 \\
		$\left\langle 7^{-}\|E 1\| 8^{+}\right\rangle$ & $0.13^{+0.10}_{-0.33}$  & \textbf{0.14}\\
		$\left\langle 8^{+}\|E 1\| 9^{-}\right\rangle$ & $0.20^{+0.20}_{-0.09}$  &\textbf{0.16} \\
		&&\\
		\hline \hline
	\end{tabular}
\end{table*}
\begin{table*}[h]
	\caption{The absolute values of the matrix element $|m.e.|$ of electric transitions in ${ }^{220} \mathrm{Rn}$. The experimental data, in units of $e$ fm, $e$ fm$^2$, $e$ fm$^3$ for E1, E2, E3, respectively, have been taken from
		Ref. \cite{b32}.}\label{Tab8}
	\centering
	\begin{tabular}{|lcc|}
		\hline\hline
		&  ${ }^{220} \mathrm{Rn}$ \qquad \qquad \qquad &   \\
		\hline \text {\textbf{ Matrix element (m.e.)} } & \text { \textbf{Experiment}} & \text {\textbf{QOH - Theory}}\\
		\hline $\left\langle I^{\prime} \| E \lambda \| I \right\rangle$ & $ |\textbf{m.e.}|\left(\mathrm{e.fm}^\lambda\right) $&$ |\textbf{m.e.}|\left(\mathrm{e.fm}^\lambda\right)$\\
		\hline
		&&\\
		$<0^{+}\|\mathrm{E} 1\| 1^{-}>$& $<0.10$ & \textbf{0.011}  \\
		$<2^{+}\|\mathrm{E} 1\| 1^{-}>$ & $<0.13$ & \textbf{0.017} \\
		$<2^{+}\|\mathrm{E} 1\| 3^{-}>$ & $<0.18$ & \textbf{0.020}  \\
		$<4^{+}\|\mathrm{E} 1\| 5^{-}>$ & $0.028 \pm 0.009$ & \textbf{0.028}\\
		$<6^{+}\|\mathrm{E} 1\| 7^{-}>$ & $<1.3$ & \textbf{0.033} \\
		&&\\
		$<0^{+}\|\mathrm{E} 2\| 2^{+}>$ & $137 \pm 4$ & \textbf{137.949} \\
		$<1^{-}\|\mathrm{E} 2\| 3^{-}>$ & $180 \pm 60$ & \textbf{190.246}  \\
		$<2^{+} \|$E2 $\| 4^{+}>$ & $212 \pm 4$ & 200.344 \\
		$<3^{-}\|E 2\| 5^{-}>$ & $220 \pm 150$ & \textbf{209.314} \\
		$<4^{+}\|E 2\| 6^{+}>$ & $274 \pm 14$ & 208.812 \\
		&&\\
		$<0^{+}\|\mathrm{E} 3\| 3^{-}>$ & $810 \pm 50$ & \textbf{856.6} \\
		$<2^{+}\|\mathrm{E} 3\| 1^{-}>$ & $<2,600$ & \textbf{1156.86} \\
		$<2^{+}\|\mathrm{E} 3\| 3^{-}>$ & $<5,300$ & \textbf{1182.74}\\
		$<2^{+}\|\mathrm{E} 3\| 5^{-}>$ & $1,700 \pm 400$ & \textbf{1502.27}  \\
		&&\\
		\hline \hline
	\end{tabular}
\end{table*}

\begin{table*}[htp]
	\caption{The absolute values of the matrix element $|m.e.|$ of electric transitions in ${ }^{222} \mathrm{Rn}$. The experimental data, in units of $e$ fm, $e$ fm$^2$, $e$ fm$^3$ for E1, E2, E3, respectively, have been taken from Ref. \cite{b35}.}\label{Tab9}
	\centering
	\begin{tabular}{|lcc|}
		\hline \hline
		&  ${ }^{222} \mathrm{Rn}$ \qquad \qquad \qquad &\\
		\hline \text{ \textbf{Matrix element (m.e.)}} &  \text {\textbf{Experiment}} & \text {\textbf{QOH - Theory}}\\
		\hline $\left\langle I \| E \lambda \| I^{\prime}\right\rangle$  & \textbf{$|m.e.|$} $\left(\mathrm{e.fm}^\lambda\right)$ & \textbf{$|m.e.|$} $\left(\mathrm{e.fm}^\lambda\right)$ \\
		\hline
		&&\\
		$<0^{+}\|\mathrm{E} 1\| 1^{-}>$  & $0.07_{- 0.03}^{ + 0.07}$ & \textbf{0.066}\\[1mm]
		$<1^{-}\|\mathrm{E} 1\| 0^{+}_{\beta}>$ & $0.057_{- 0.020}^{ + 0.020}$& \textbf{0.061}\\
		&&\\
		$<2^{+}\|\mathrm{E} 2\| 2^{+}>$  & $180^{+90}_{-60}$  &\textbf{203.98}\\[1mm]
		$<2^{+}\|\mathrm{E} 2\| 4^{+}>$  & $255_{-10}^{+6}$   &\textbf{243.58}\\[1mm]
		$<4^{+}\|\mathrm{E} 2\| 6^{+}>$ & $352_{-18}^{+13}$   &247.69\\[1mm]
		$<6^{+}\|\mathrm{E} 2\| 8^{+}>$ &$450_{-30}^{+60}$  &201.88\\[1mm]
		$<8^{+}\|\mathrm{E} 2\| 10^{+}>$  & $410_{-90}^{+50}$   &111.65\\[1mm]
		$<1^{-}\|\mathrm{E} 2\| 3^{-}>$  & $210_{-40}^{+40}$  &\textbf{226.09}\\[1mm]
		$<3^{-}\|\mathrm{E} 2\| 5^{-}>$  &$270_{-70}^{+70}$  &\textbf{246.68}\\[1mm]
		$<5^{-}\|\mathrm{E} 2\| 7^{-}>$  &$400_{-90}^{+90}$   &212.44\\[1mm]
		$<7^{-}\|\mathrm{E} 2\| 9^{-}>$  &$550_{-130}^{+210}$   &134.33\\[1mm]
		$<2^{+}\|\mathrm{E} 2\| 0^{+}_{\beta}>$  & $32_{-5}^{+5}$  &78.48\\[1mm]
		&&\\
		$<0^{+}\|\mathrm{E} 3\| 3^{-}>$  & $880_{-80}^{+110}$ & \textbf{816.73}\\[1mm]
		$<2^{+}\|\mathrm{E} 3\| 1^{-}>$   &$500_{-400}^{+200}$ & 1041.09\\[1mm]
		$<2^{+}\|\mathrm{E} 3\| 3^{-}>$  & $\mid$ m.e. $\mid<1500$  &\textbf{1068.16}\\[1mm]
		$<2^{+}\|\mathrm{E} 3\| 5^{-}>$  & $1300_{-300}^{+200}$ &   \textbf{1369.9} \\[1mm]
		$<4^{+}\|\mathrm{E} 3\| 1^{-}>$ & $\mid$ m.e. $\mid<1400$  &   \textbf{1286.39}\\
		&&\\
		\hline \hline
	\end{tabular}
\end{table*}

\begin{table*}[htp]
	\caption{The absolute values of the matrix element $|m.e.|$ of electric transitions in ${ }^{144} \mathrm{Ba}$. The experimental data, in units of $e$ fm$^2$, $e$ fm$^3$ for E2, E3, respectively, have been taken from Ref. \cite{b36}.}\label{Tab10}
	\centering
	$$\begin{array}{|ccc|}
		\hline\hline
		& \qquad \qquad { }^{144} \mathrm{Ba} \qquad \qquad \qquad \qquad & \\
		\hline \text {\textbf{ Matrix element (m.e.)} } & \text { \textbf{Experiment}}  & \text {\textbf{QOH - Theory}}\\
		\hline \left\langle I^{\prime} \| E \lambda \| I \right\rangle &   |\textbf{m.e.}| \left(\mathrm{e.fm}^\lambda\right)&|\textbf{m.e.}|\left(\mathrm{e.fm}^\lambda\right)\\ \hline
		&&\\
		\left\langle 2^{+} \| E 2 \| 0^{+} \right\rangle  & 104.2_{-2200}^{+1700} & \textbf{141.69} \\[1mm]
		\left\langle 4^{+} \| E 2 \| 2^{+} \right\rangle   & 186_{-8100}^{+8600}& \textbf{192.21} \\[1mm]
		\left\langle 6^{+} \| E 2 \| 4^{+} \right\rangle  & 178_{-1000}^{+1200} & \textbf{187.97}\\[1mm]
		\left\langle 8^{+} \| E 2 \| 6^{+} \right\rangle  & 204_{-2300}^{+3500}& \textbf{146.95}\\[1mm]
		\left\langle 3^{-} \| E 3 \| 0^{+} \right\rangle   & 650_{-23000}^{+17000} & \textbf{728.19}\\[1mm]
		\left\langle 5^{-} \| E 3 \| 2^{+} \right\rangle & <1200& \textbf{1172.83}  \\[1mm]
		\left\langle 7^{-} \| E 3 \| 4^{+} \right\rangle & <1600 & \textbf{1584.17}\\
		&  &  \\ \hline
	\end{array}$$
\end{table*}

\begin{table*}[htp]
	\caption{The absolute values of the matrix element $|m.e.|$ of electric transitions in ${ }^{146} \mathrm{Ba}$. The experimental data, in units of $e$ fm, $e$ fm$^2$, $e$ fm$^3$ for E1, E2, E3, respectively, have been taken from Ref. \cite{b37}.}\label{Tab11}
	\centering
	$$\begin{array}{|ccc|}
		\hline\hline
		& \qquad \qquad { }^{146} \mathrm{Ba} \qquad \qquad \qquad \qquad  &  \\
		\hline \text {\textbf{ Matrix element (m.e.)} } &  \text { \textbf{Experiment}}  & \text {\textbf{QOH - Theory}}\\
		\hline \left\langle I^{\prime} \| E \lambda \| I \right\rangle &    |\textbf{m.e.}| \left(\mathrm{e.fm}^\lambda\right)&|\textbf{m.e.}|\left(\mathrm{e.fm}^\lambda\right)\\ \hline
		&&\\
		\left\langle 1^{-} \| E 1 \| 0^{+} \right\rangle   & 0.00223\binom{100}{-80} & \textbf{0.00223}\\[1mm]
		\left\langle 3^{-} \| E 2 \| 1^{-} \right\rangle   & 120(500) & \textbf{182.19}\\[1mm]
		\left\langle 2^{+} \| E 2 \| 0^{+} \right\rangle  & 117(200)  & \textbf{150.10}\\[1mm]
		\left\langle 4^{+} \| E 2 \| 2^{+} \right\rangle   & 197(1400) & \textbf{202.06}\\[1mm]
		\left\langle 6^{+} \| E 2 \| 4^{+} \right\rangle &235\binom{+2000}{-2400}  & \textbf{196.33}\\[1mm]
		\left\langle 8^{+} \| E 2 \| 6^{+} \right\rangle  & 217\binom{+6500}{-3300} & \textbf{153.65}\\[1mm]
		\left\langle 3^{-} \| E 3 \| 0^{+} \right\rangle  &650\binom{+14000}{-20000} &\textbf{583.01} \\[1mm]
		\left\langle 5^{-} \| E 3 \| 2^{+} \right\rangle &1010\binom{+61000}{-20000} &\textbf{933.53}\\[1mm]
		\left\langle 7^{-} \| E 3 \| 4^{+} \right\rangle  & 1250\binom{+85000}{-34000} &\textbf{1248}\\[1mm]
		\left\langle 9^{-} \| E 3 \| 6^{+} \right\rangle &1500\binom{+8000}{-12000} &\textbf{1571.86}\\
		&&\\ \hline
	\end{array}$$
\end{table*}
\subsubsection{Impact of the hexadecapole degree of freedom}
To assess the role of the hexadecapole degree of freedom, we compare the present results with those obtained within quadrupole-octupole models where $\beta_4$ is absent\cite{b15,b19}.
We observe that the inclusion of the $\phi_2$ variable leads to:
\begin{itemize}
	\item[(i)] an improved description of low-lying negative-parity states (notably $1^-$ and $3^-$), essentially achieved through the introduction of the sextic potential $v(\phi_1)$, characterized by a finite barrier with two symmetric minima,
	\item[(ii)]a refined reproduction of the staggering amplitude in vibrational nuclei, and
	\item[(iii)]a systematic reduction of the RMS deviation for a large subset of nuclei.
\end{itemize}
The parameter $\phi_{02}$ can be interpreted as a measure of the relative contribution of the hexadecapole deformation, while $c_2$ governs the stiffness of oscillations around this equilibrium configuration. These results demonstrate that the $\lambda=4$ degree of freedom provides a non-negligible correction to the collective dynamics, even in cases where static hexadecapole deformation is not present. Note that even if it is static (and not dynamic) it should not be dominant.
	\section{CONCLUSIONS}
We have introduced the analytic quadrupole-octupole-hexadecapole axially symmetric model (QOH) as a unified framework for nuclear collective motion up to $\lambda = 4$. Extending previous AQOA models \cite{b15,b19,b38,b39}, the QOH model incorporates the hexadecapole degree of freedom through three parameters ($c_{2}, \phi_{02}, \alpha$) from the centered harmonic oscillator potential for $\phi_{2}$ and the sextic potential for $\phi_{1}$. The model successfully describes both octupole vibrations and stable octupole deformations across a wide range of nuclei, including ${ }^{218-230} \mathrm{Ra}$, ${ }^{222-234} \mathrm{Th}$, ${ }^{148-154} \mathrm{Sm}$, ${ }^{120} \mathrm{Ba}$, ${ }^{140-148} \mathrm{Ba}$, ${ }^{230-240} \mathrm{U}$, ${ }^{154-160} \mathrm{Gd}$, ${ }^{146-152} \mathrm{Nd}$, ${ }^{218-226} \mathrm{Rn}$, ${ }^{236-244} \mathrm{Pu}$, ${ }^{112-114} \mathrm{Xe}$, ${ }^{140-142} \mathrm{Xe}$, and ${ }^{144-146} \mathrm{Ce}$. The hexadecapole component proves relevant in specific regions, and the model reliably reproduces staggering patterns and electric transition matrix elements (notably E1 and E3), providing a practical tool for studying the evolution of octupole correlations and the fine structure of alternating parity bands.
\appendix
\section{Normalization Constant}
\label{app:A}
We present, here, the derivation of the normalization constants of the collective wave functions associated with the angular variable $\phi_{2}$. So, for small vibrations from  $\phi_{02}$, the volume element becomes \textbf{$d\tau \propto  \sin \phi_{02}\cdot d \phi_{2}$}. We define the expression of the normalization constant $N_{n_{\phi_{2}}}$ for the bands considered in this paper, corresponding to $n_{\phi_{2}}=0$, where the Hermite polynomials are unity. We know that Eq. 3.322.2 of Ref.\cite{b40} is expressed as follow
\begin{equation}
	\int_0^{+ \infty} e^{-x^2-2 b x} \mathrm{~d} x=\frac{\sqrt{\pi}}{2} \times e^{b^2}  [1-\operatorname{Erf}(b)]\label{Eq20}
\end{equation}
where $\text{Erf}(x)$ is the error function, having the property $\text{Erf}(-x)$=-\,$\text{Erf}(x)$. Using Eq.(\ref{Eq20}), we finally obtain the normalization constant
\begin{align}\label{Norm}
	N_{n_{\phi_{2}}}= \sqrt{\frac{2\,a}{\sqrt{\pi } \sin (\phi_{02}) \left(1+\text{Erf}\left(a \phi_{02}-\frac{\cot (\phi_{02})}{2 a}\right)\right) }}  e^{-\frac{\cot ^2(\phi_{02} )}{8 a^2}}
\end{align}
with $a=\left(c_{2} \over 2 \, \sin^2 \phi_{02} \right)^{1/4}$.
The normalization constants $N_{n_{\phi_{1}}}^{(M_s)}$ are determined from the condition that the ansatz functions (\ref{Eq28}) be normalized to unity with the integration measure $d\phi_{1}$. These constants can be determined either numerically or analytically by evaluating integrals of the form :
\begin{align}
	I_A=\int_{-\infty}^{\infty} y^A \mathrm{e}^{-\frac{y^4}{2}-\alpha y^2} \mathrm{~d} y= 2^{-\frac{A}{4}-\frac{5}{4}} \left(1+(-1)^A\right) \nonumber\\ \times \Gamma \left(\frac{A+1}{2}\right) U\left(\frac{A+1}{4},\frac{1}{2},\frac{\alpha ^2}{2}\right) 
\end{align} 
\section{Radial and angular Integrals}
This appendix is devoted to the explicit expressions of the radial integrals
$I_{\tilde{\beta}}^{(EL)}(n_i,n_f)$ and the angular integrals
$I_{\phi_{1}}^{(EL)}$ and $I_{\phi_{2}}^{(EL)}$, which enter the evaluation of the electric transition matrix elements:
\label{app:B}
\begin{itemize}
\item$$I_{\tilde \beta}^{(E2)}(0,0)=\frac{\Gamma\left(\frac{a_{i}^{\pi}}{2}+\frac{a_{f}^{\pi}}{2}+\frac{3}{2}\right)}
	{\sqrt{\Gamma(a_{i}^{\pi}+1)\Gamma(a_{f}^{\pi}+1)}},\quad \quad \pi=\pm \,.$$
\item$$I_{\tilde \beta}^{(E2)}(1,0)=\frac{\Gamma\left(\frac{a_{i}^{\pi}}{2}+\frac{a_{f}^{\pi}}{2}+\frac{3}{2}\right) \Gamma\left(\frac{a_{i}^{\pi}}{2}-\frac{a_{f}^{\pi}}{2}+\frac{1}{2}\right)}
	{\sqrt{\Gamma(a_{i}^{\pi}+2)\Gamma(a_{f}^{\pi}+1)} \Gamma\left(\frac{a_{i}^{\pi}}{2}-\frac{a_{f}^{\pi}}{2}-\frac{1}{2}\right)},$$
\item$$I_{\tilde \beta}^{(E2)}(0,1)=\frac{\Gamma\left(\frac{a_{f}^{\pi}}{2}+\frac{a_{i}^{\pi}}{2}+\frac{3}{2}\right) \Gamma\left(\frac{a_{f}^{\pi}}{2}-\frac{a_{i}^{\pi}}{2}+\frac{1}{2}\right)}
	{\sqrt{\Gamma(a_{f}^{\pi}+2)\Gamma(a_{i}^{\pi}+1)} \Gamma\left(\frac{a_{f}^{\pi}}{2}-\frac{a_{i}^{\pi}}{2}-\frac{1}{2}\right)},$$
\item$$I_{\tilde \beta}^{(E1)}(0, 0)=\frac{\Gamma\left(\frac{a_{i}^{\pi}}{2}+\frac{a_{f}^{\pi}}{2}+2\right)}
	{\sqrt{\Gamma(a_{i}^{\pi}+1)\Gamma(a_{f}^{\pi}+1)}}, $$
\item$$I_{\tilde \beta}^{(E1)}(0, 1)=\frac{\Gamma\left(\frac{a_{i}^{\pi}}{2}+\frac{a_{f}^{\pi}}{2}+2\right)\Gamma\left(\frac{a_{f}^{\pi}}{2}-\frac{a_{i}^{\pi}}{2}\right)}
	{\sqrt{\Gamma(a_{i}^{\pi}+1)\Gamma(a_{f}^{\pi}+2)}\Gamma\left(\frac{a_{f}^{\pi}}{2}-\frac{a_{i}^{\pi}}{2}-1\right)}.$$
\item$$I_{\phi_{1}}^{(E2)} =\int_{-\infty}^{\infty} \chi_{n_{\phi_{1}},\, L_i}^{\;(M_s)}(\phi_{1})\; \textcolor{black}{\cos \phi_{1}}  \; \chi_{n_{\phi_{1}},\, L_f}^{\;(M_s)}(\phi_{1})\; \mathrm{~d} \phi_{1},$$
\item$$I_{\phi_{1}}^{(E3)} =\int_{-\infty}^{\infty} \chi_{n_{\phi_{1}},\, L_i}^{\;(M_s)}(\phi_{1})\; \textcolor{black}{\sin \phi_{1}}  \; \chi_{n_{\phi_{1}},\, L_f}^{\;(M_s)}(\phi_{1})\; \mathrm{~d} \phi_{1}, $$
\item$$I_{\phi_{1}}^{(E1)}=  \int_{-\infty}^{\infty} \chi_{n_{\phi_{1}},\, L_i}^{\;(M_s)}(\phi_{1})\; \textcolor{black}{\frac{\sin 2 \phi_{1}}{2}}  \; \chi_{n_{\phi_{1}},\, L_f}^{\;(M_s)}(\phi_{1})\; \mathrm{~d} \phi_{1} ,$$
\item$$ I_{\phi_{1}}^{(E1)\prime}= \int_{-\infty}^{\infty} \chi_{n_{\phi_{1}},\, L_i}^{\;(M_s)}(\phi_{1})\; \textcolor{black}{\sin \phi_{1}}  \; \chi_{n_{\phi_{1}},\, L_f}^{\;(M_s)}(\phi_{1})\; \mathrm{~d} \phi_{1}.$$
\item$$	I_{\phi_{2}}^{(E2)}=I_{\phi_{2}}^{(E3)}= 
	N_{n_{\phi_{2}}}^{\prime} \times  \int^{+\infty}_{0} e^{- \beta \phi_{2}^2 - \gamma \phi_{2}} \cdot \textcolor{black}{\sin\phi_2}\; d  \phi_{2}
	$$
\item$$I_{\phi_{2}}^{(E1)}=  N_{n_{\phi_{2}}}^{\prime} \times  \int^{+\infty}_{0} e^{- \beta \phi_{2}^2 - \gamma \phi_{2}} \cdot \textcolor{black}{\sin^{2} \phi_{2}}\; d  \phi_{2}, $$ 
	
\item$$ I_{\phi_{2}}^{(E1)\prime}= N_{n_{\phi_{2}}}^{\prime}\times  \int^{+\infty}_{0} e^{- \beta \phi_{2}^2 - \gamma \phi_{2}} \cdot \textcolor{black}{{\sin 2\phi_{2} \over 2}}\; d  \phi_{2}  $$
	
	with : \quad $\beta= a^2 $ , $\gamma=\left( \cot (\phi_{02}) - 2 a^2 \phi_{02}\right)$ and $N_{n_{\phi_{2}}}^{\prime} = N_{n_{\phi_{2}}}^{2} \cdot \sin \phi_{02} \cdot e^{- a^2 \phi_{02}^2+ \phi_{02} \cot (\phi_{02})}$.
\end{itemize}

\end{document}